# Coordination corrected *ab initio* formation enthalpies

Rico Friedrich,[1,2] Demet Usanmaz,[1,2] Corey Oses,[1,2] Andrew Supka,[3] Marco Fornari,[2,3] Marco Buongiorno Nardelli,[2,4] Cormac Toher,[1,2] and Stefano Curtarolo[2,5,6,*]

[1] *Department of Mechanical Engineering and Materials Science, Duke University, Durham, North Carolina 27708, USA*
[2] *Center for Materials Genomics, Duke University, Durham, North Carolina 27708, United States*
[3] *Department of Physics and Science of Advanced Materials Program, Central Michigan University, Mount Pleasant, Michigan 48859, USA*
[4] *Department of Physics and Department of Chemistry, University of North Texas, Denton, TX 76203, USA*
[5] *Materials Science, Electrical Engineering, Physics and Chemistry, Duke University, Durham NC, 27708, USA*
[6] *Fritz-Haber-Institut der Max-Planck-Gesellschaft, 14195 Berlin-Dahlem, Germany*

The correct calculation of formation enthalpy is one of the enablers of *ab-initio* computational materials design. For several classes of systems (*e.g.* oxides) standard density functional theory produces incorrect values. Here we propose the "Coordination Corrected Enthalpies" method (CCE), based on the number of nearest neighbor cation-anion bonds, and also capable of correcting relative stability of polymorphs. CCE uses calculations employing the Perdew, Burke and Ernzerhof (PBE), Local Density Approximation (LDA) and Strongly Constrained and Appropriately Normed (SCAN) exchange correlation functionals, in conjunction with a quasiharmonic Debye model to treat zero-point vibrational and thermal effects. The benchmark, performed on binary and ternary oxides (halides), shows very accurate room temperature results for all functionals, with the smallest mean absolute error of 27 (24) meV/atom obtained with SCAN. The zero-point vibrational and thermal contributions to the formation enthalpies are small and with different signs — largely cancelling each other.

## I. INTRODUCTION

The accurate prediction of the thermodynamic stability of a compound — crucial in computational materials design [1] — mostly relies on the calculation of the formation enthalpy: the enthalpy change with respect to elemental reference phases. Using Density Functional Theory (DFT), the formation energy, neglecting pressure-volume contributions, is routinely computed *ab initio*. For systems where elements and compounds are metallic, *i.e.* chemically similar, accurate results are usually obtained by using standard (semi)local approximations to DFT [2, 3]. They include the Local Density Approximation (LDA) [4, 5] or the Generalized Gradient Approximation (GGA), for instance PBE [6]. In this way, formation energies for millions of metal alloys have already been calculated in materials databases such as AFLOW [7–10], the Materials Project [11, 12] and OQMD [13, 14].

When the compound and the elements have a different chemical character, as for example in case of oxides, nitrides or sulfides, the situation is less favorable. For oxides, the compound is typically an ionic insulator while the elements are metals or semiconductors and a diatomic gas. When comparing to experimental enthalpies [15–18], standard approximations of DFT lead to Mean Absolute Errors (MAEs) of the order of several hundred meV/atom. For reaction energies between binary and ternary oxides, within a similar chemical realm, a smaller average error of about 24-35 meV/atom has been observed [19].

**Correcting DFT.** Different attempts have been made to calculate more accurate formation energies *ab initio*. A modified version of PBE was proposed by Sarmiento-Pérez *et al.* [20]: three functional parameters were optimized, improving results by about a factor of two. The hybrid functional HSE06 yields only a slight improvement for transition metal oxides [21]. The recently developed Strongly Constrained and Appropriately Normed (SCAN) meta-generalized-gradient approximation [22] has an accuracy limited to about 100 meV/atom [23, 24].

**Beyond DFT.** Non self-consistent EXact eXchange plus Random Phase Approximation (EXX+RPA) calculations can lead to more accurate formation energies by about a factor of two to three compared to PBE [25, 26]. The renormalized adiabatic PBE method improves the results based on RPA for 19 main group and two transition metal oxides by about a factor of two [27]. A Bayesian Error Estimation Functional (mBEEF) systematically improves PBE results reaching an MAE of about 120 meV/atom for a test set of 24 compounds [28]. Applying a correction method on top of the functional could reduce the MAE to 90 meV/atom, which is 20-60 meV/atom less than if the correction is applied on top of other functionals. Unfortunately, such computationally expensive approaches are not suitable for screening large materials sets and do not, in general, reach the necessary chemical accuracy of 1 kcal/mol (≈ 40 meV/atom).

**Empirical corrections.** Several empirical correction schemes have been established for formation energies calculated with DFT by comparing to experimentally measured formation enthalpies. Wang *et al.* [29] suggested an oxygen correction of 1.36 eV per $O_2$ to be subtracted from formation energies calculated with PBE. The approach was extended to $H_2$, $N_2$, $F_2$ and $Cl_2$ for different functionals [30]. For sulfides, a different correction is found depending on whether the anion is $S^{2-}$ or $S_2^{2-}$ [31]. Jain

* stefano@duke.edu

*et al.* suggested an empirical scheme for mixing GGA and GGA+$U$ calculations to compute formation enthalpies for compounds containing transition metal elements [32]. An MAE of 45 meV/atom was achieved for a test set of 49 ternary oxides with respect to experimental values [32]. A local environment dependent GGA+$U$ method based on the GGA/GGA+$U$ mixing scheme was also developed [33]. It introduced significantly more parameters and achieved an MAE of 19 meV/atom for a test set of 52 transition metal oxides. In the Fitted Elemental-phase Reference Energies (FERE) method [34, 35], element specific corrections were used to optimize the error cancellation when calculating total energy differences between chemically dissimilar materials. Corrected formation energies calculated for a test set of 55 ternary compounds gave an MAE of 48 meV/atom [35]. In conclusion, existing correction schemes and advanced theoretical approaches do not, in general, reach an accuracy of the order of the thermal energy at room temperature (∼25 meV) for formation enthalpies.

**Topological corrections: coordination corrected enthalpies.** Here, we propose a physically motivated correction scheme — Coordination Corrected Enthalpies (CCE), based on the number of bonds between each cation and surrounding anions. Compared to previous approaches, it leads to systematically more accurate results. The smallest MAE of 27 (24) meV/atom for a test set of ternary oxides (halides) is reached when starting from SCAN calculations. Contrary to earlier approaches, the ansatz also allows correction of the relative stability of polymorphs with different number of cation-anion bonds.

The article mainly focuses on oxides because of: **i.** high technological relevance, **ii.** abundance of experimental thermochemical data, especially for ternary oxides and **iii.** generally low error bars of the experimental values allowing accurate corrections and predictions. Calculated room temperature formation enthalpies for a set of 79 binary and 71 ternary oxides are presented employing the three main approximations to the DFT exchange-correlation functional: LDA, PBE and SCAN.

In other schemes, temperature effects have been completely neglected [32, 34, 35], or room temperature experimental values were interpolated to 0 K using a Debye model parameterized with the measured room temperature heat capacities and entropies [19, 33]. Here, the thermal contributions to the formation enthalpy are calculated via a quasiharmonic Debye model [36–40]. Our approach includes the contribution due to zero-point vibrational energies.

The methodology of calculating coordination corrected room temperature formation enthalpies is presented in Section II. The DFT derived and CCE results are discussed in Section III. Conclusions are summarized in Section IV. Additional comparisons are given in Appendices A to D. Tables with structure data, values of the corrections, of calculated, corrected and experimental formation enthalpies/energies as well as the vibrational contributions are listed in Appendix F.

## II. METHODOLOGY

**Room temperature formation enthalpies.** The formation enthalpy includes contributions due to the pressure-volume term (*e.g.* for $O_2$). The formation energy takes into account only internal energy contributions. The formalism, introduced for oxides, works equivalently for other polar systems.

From DFT, an approximate formation energy $\Delta_{\mathrm{f}}E^{0,\mathrm{DFT}}$ of an oxide $A_{x_1}B_{x_2}\ldots\mathrm{O}_{x_n}$ at zero $T$ and $p$, without zero-point vibrational energies, can be calculated:

$$\Delta_{\mathrm{f}}E^{0,\mathrm{DFT}}_{A_{x_1}\ldots\mathrm{O}_{x_n}} = U^{0,\mathrm{DFT}}_{A_{x_1}\ldots\mathrm{O}_{x_n}} - \left[\sum_{i=1}^{n-1} x_i U_i^{0,\mathrm{DFT}} + \frac{x_n}{2} U^{0,\mathrm{DFT}}_{\mathrm{O}_2}\right], \tag{1}$$

where $U^{0,\mathrm{DFT}}_{A_{x_1}\ldots\mathrm{O}_{x_n}}$, $U_i^{0,\mathrm{DFT}}$ and $U^{0,\mathrm{DFT}}_{\mathrm{O}_2}$ are the total energies of the compound per formula unit, the $i$-element reference phase per atom, and $O_2$, respectively, and $x_1, ..., x_n$ are stoichiometries.

The tabulated experimentally measured standard formation enthalpy at the reference temperature $T_{\mathrm{r}}$=298.15 K, $\Delta_{\mathrm{f}}H^{\circ,T_{\mathrm{r}},\mathrm{exp}}$, corresponds to:

$$\Delta_{\mathrm{f}}H^{\circ,T_{\mathrm{r}},\mathrm{exp}}_{A_{x_1}\ldots\mathrm{O}_{x_n}} = H^{\circ,T_{\mathrm{r}}}_{A_{x_1}\ldots\mathrm{O}_{x_n}} - \left[\sum_{i=1}^{n-1} x_i H_i^{\circ,T_{\mathrm{r}}} + \frac{x_n}{2} H^{\circ,T_{\mathrm{r}}}_{\mathrm{O}_2}\right], \tag{2}$$

where $H^{\circ,T_{\mathrm{r}}}_{A_{x_1}\ldots\mathrm{O}_{x_n}}$, $H_i^{\circ,T_{\mathrm{r}}}$ and $H^{\circ,T_{\mathrm{r}}}_{\mathrm{O}_2}$ are the standard enthalpies of the compound per formula unit, the $i$-element reference phase per atom and $O_2$, respectively, all at $T_{\mathrm{r}}$.

Using $H = U + pV$ and neglecting the $pV$ terms for the compound and the elements (less $O_2$), the formation enthalpy becomes:

$$\Delta_{\mathrm{f}}H^{\circ,T_{\mathrm{r}},\mathrm{exp}}_{A_{x_1}\ldots\mathrm{O}_{x_n}} \approx U^{T_{\mathrm{r}}}_{A_{x_1}\ldots\mathrm{O}_{x_n}} - \left[\sum_{i=1}^{n-1} x_i U_i^{T_{\mathrm{r}}} + \frac{x_n}{2} H^{\circ,T_{\mathrm{r}}}_{\mathrm{O}_2}\right]. \tag{3}$$

Generally, neglecting $pV$ is a very good approximation: pressures are small and the molar volumes of condensed systems are typically three orders of magnitude smaller than gases — the contribution to the formation enthalpy is expected to be well below 1 meV/atom.

Writing the total energies and the standard enthalpy of $O_2$ at $T_{\mathrm{r}}$ as the value at 0 K plus the difference between $T_{\mathrm{r}}$ and 0 K, and separating the zero-point vibrational energy for each system, gives:

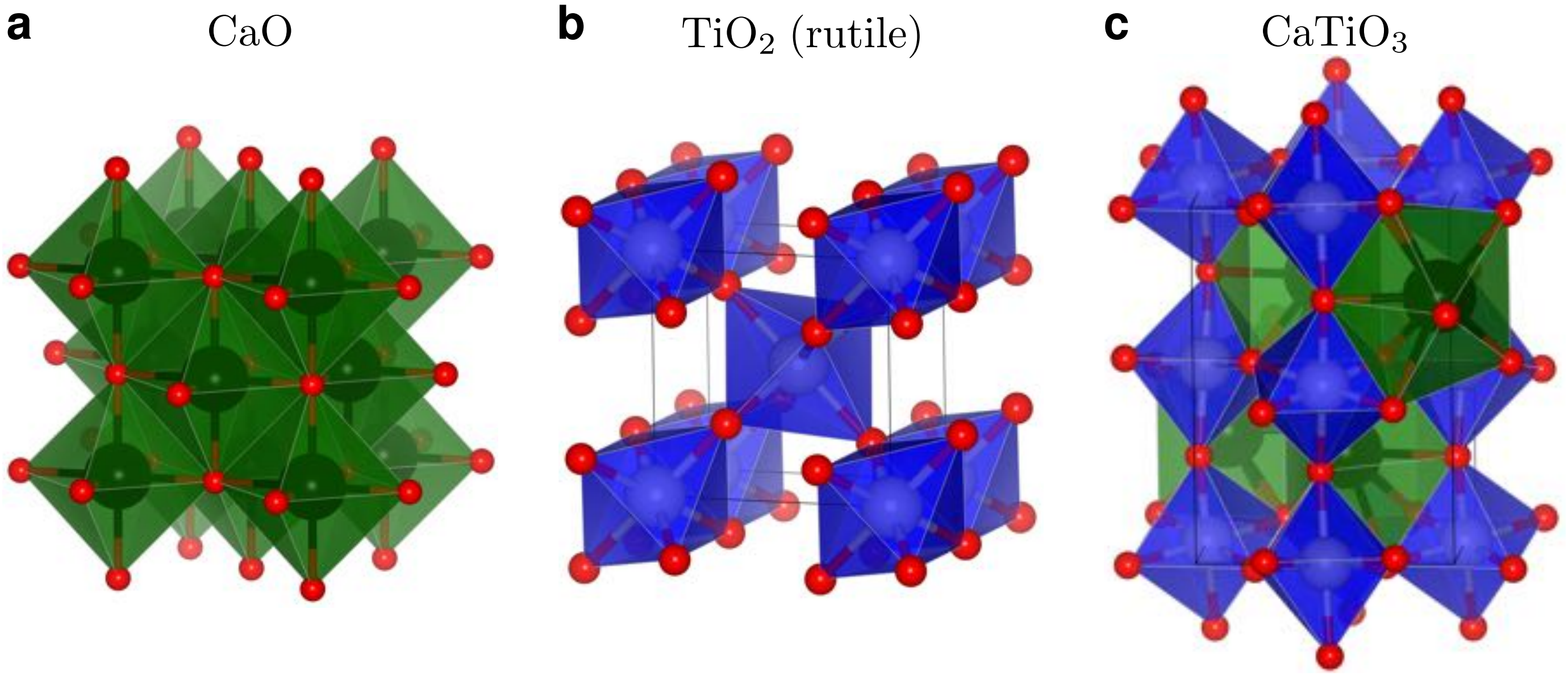

FIG. 1. **Coordination change.** Crystal structures of (**a**) CaO, (**b**) rutile $TiO_2$, and (**c**) $CaTiO_3$ (perovskite). The coordination polyhedra of Ca and Ti are shown green and blue, respectively. Note: Ca is six-fold (octahedrally) coordinated with oxygen in CaO and eight-fold coordinated in $CaTiO_3$, requiring coordination corrections. Colors: Ca black, Ti light gray, and O red [41].

$$\begin{aligned}\Delta_{\mathrm{f}} H^{\circ,T_{\mathrm{r}},\mathrm{exp}}_{A_{x_1}\ldots\mathrm{O}_{x_n}} &\approx U^{0}_{A_{x_1}\ldots\mathrm{O}_{x_n}} + U^{\mathrm{ZP}}_{A_{x_1}\ldots\mathrm{O}_{x_n}} + \Delta U^{T_{\mathrm{r}}-0\mathrm{K}}_{A_{x_1}\ldots\mathrm{O}_{x_n}} - \left[\sum_{i=1}^{n-1} x_i \left(U^0_i + U^{\mathrm{ZP}}_i + \Delta U^{T_{\mathrm{r}}-0\mathrm{K}}_i\right) + \frac{x_n}{2}\left(U^0_{\mathrm{O}_2} + U^{\mathrm{ZP}}_{\mathrm{O}_2} + \Delta H^{\circ,T_{\mathrm{r}}-0\mathrm{K}}_{\mathrm{O}_2}\right)\right] \\ &\approx \Delta_{\mathrm{f}} E^{0}_{A_{x_1}\ldots\mathrm{O}_{x_n}} + \Delta_{\mathrm{f}} E^{\mathrm{ZP}}_{A_{x_1}\ldots\mathrm{O}_{x_n}} + \Delta_{\mathrm{f}} H^{\mathrm{TC}}_{A_{x_1}\ldots\mathrm{O}_{x_n}} \approx \Delta_{\mathrm{f}} H^{\circ,T_{\mathrm{r}},\mathrm{cal}}_{A_{x_1}\ldots\mathrm{O}_{x_n}}, \end{aligned} \quad (4)$$

where $U^{\mathrm{ZP}}_{A_{x_1}\ldots\mathrm{O}_{x_n}}$, $U^{\mathrm{ZP}}_i$ and $U^{\mathrm{ZP}}_{\mathrm{O}_2}$ are the zero-point vibrational energies of the compound, the $i$-element reference and $O_2$, respectively. $\Delta_{\mathrm{f}} H^{\circ,T_{\mathrm{r}},\mathrm{cal}}_{A_{x_1}\ldots\mathrm{O}_{x_n}}$ stands for the calculated standard formation enthalpy at $T_{\mathrm{r}}$. The terms are:

$$\Delta_{\mathrm{f}} E^{0}_{A_{x_1}\ldots\mathrm{O}_{x_n}} \equiv U^{0}_{A_{x_1}\ldots\mathrm{O}_{x_n}} - \left[\sum_{i=1}^{n-1} x_i U^0_i + \frac{x_n}{2} U^0_{\mathrm{O}_2}\right] \quad (5)$$

is the internal energy contribution excluding vibrational effects;

$$\Delta_{\mathrm{f}} E^{\mathrm{ZP}}_{A_{x_1}\ldots\mathrm{O}_{x_n}} \equiv U^{\mathrm{ZP}}_{A_{x_1}\ldots\mathrm{O}_{x_n}} - \left[\sum_{i=1}^{n-1} x_i U^{\mathrm{ZP}}_i + \frac{x_n}{2} U^{\mathrm{ZP}}_{\mathrm{O}_2}\right] \quad (6)$$

collects all Zero-Point (ZP) contributions;

$$\Delta_{\mathrm{f}} H^{\mathrm{TC}}_{A_{x_1}\ldots\mathrm{O}_{x_n}} \equiv \Delta U^{T_{\mathrm{r}}-0\mathrm{K}}_{A_{x_1}\ldots\mathrm{O}_{x_n}} - \left[\sum_{i=1}^{n-1} x_i \Delta U^{T_{\mathrm{r}}-0\mathrm{K}}_i + \frac{x_n}{2} \Delta H^{\circ,T_{\mathrm{r}}-0\mathrm{K}}_{\mathrm{O}_2}\right] \quad (7)$$

is the overall Thermal Contribution (TC).

The internal energy contribution to $\Delta_{\mathrm{f}} H^{\circ,T_{\mathrm{r}},\mathrm{exp}}_{A_{x_1}\ldots\mathrm{O}_{x_n}}$ can be identified with $\Delta_{\mathrm{f}} E^{0,\mathrm{DFT}}_{A_{x_1}\ldots\mathrm{O}_{x_n}}$ calculated with DFT according to Eq. (1). The pressure dependence is negligible at the standard value of 1 bar.

For the thermal contribution, the internal energy differences between 0 K and $T_{\mathrm{r}}$ are almost entirely due to vibrations. The quantity is estimated by using the AFLOW Automatic GIBBS Library (AGL) via a quasiharmonic Debye model [36–40] with default parameters (28 strained structures, 1% lattice strain increments [39]). The approach is tested by comparing the calculated internal energy difference between 0 K and $T_{\mathrm{r}}$ with experimental enthalpy differences as illustrated in Fig. A.6 (Appendix A), indicating good agreement for both compounds and references.

The AGL calculations also provide a zero-point vibrational energy, which is used to treat the zero-point contribution. Notably, *e.g.* for BeO the energy is calculated to be 0.11 eV/atom for all three functionals, which agrees *exactly* with the value reported in Ref. [42] obtained from more expensive phonon calculations. In the rest of the article, the sum of the zero-point and thermal contributions is denoted as the vibrational contribution.

For $O_2$, the enthalpy difference between 0 K and $T_{\mathrm{r}}$ can be estimated from a perfect diatomic gas with five degrees of freedom where the bond-stretching vibrational mode is not excited at $T_{\mathrm{r}}$, leading to 90 meV/$O_2$ [43]. The value agrees exactly with the tabulated enthalpy difference from the NIST-JANAF thermochemical tables [16]. For $F_2$, $Cl_2$, $BF_3$ and $SiF_4$, the enthalpy differences from NIST-JANAF corresponding to 91, 95, 121 and 159 meV are taken, respectively. The zero-point vibrational energy of $O_2$ is calculated, using the experimental oxygen vibrational frequency of 1580.1932 $\mathrm{cm}^{-1}$ [16], to be 98 meV/$O_2$. For $F_2$, $Cl_2$, $BF_3$ and $SiF_4$, the calculated zero-point energies are 55, 35, 339 and 346 meV. Similarly, for Hg the total energy at 0 K is calculated for the low temperature rhombohedral structure, with the zero-point vibrational energy obtained from AGL. The experimental enthalpy difference from 0 K to $T_\mathrm{r}$ of 97 meV/Hg atom from the NIST-JANAF tables [16], including fusion at 234.29 K, is used to account for thermal effects.

**Coordination corrected enthalpies scheme.** The remaining deviation between calculated and measured room temperature formation enthalpies is almost entirely due to the internal energy contribution $\Delta_\mathrm{f} E^{0,\mathrm{DFT}}_{A_{x_1}\ldots \mathrm{O}_{x_n}}$ obtained with DFT. Compounds with strong polar bonds are chemically different from elements — mostly metallic plus a diatomic gas. As already noted by Lany [34] and Stevanović *et al.* [35], this leads to an incomplete error cancellation when calculating total energy differences — standard semilocal functionals do not allow calculation of accurate total energies.

Since a reliable description of the bonding in a material is central for capturing its properties, it seems reasonable to assume in first approximation that DFT makes errors *per bond*. As such, the CCE scheme considers the number of nearest neighbor bonds (coordination number) formed between the cation and oxygen. The approach enables accounting for coordination changes, as illustrated in Fig. 1 for the case of CaO, rutile $TiO_2$ and perovskite $CaTiO_3$. For the binary oxides, Ca is six-fold (octahedrally) coordinated by O in the rocksalt structure of CaO, while Ti is six-fold in rutile $TiO_2$. For Ti, the coordination number remains the same in $CaTiO_3$, but the number of nearest neighbor Ca-O bonds changes to eight. The phenomenon is quite common for several elements when going from binary to ternary oxides, and can be captured within CCE.

The corrections per bond $\delta H^{A^{+\alpha}}_{A-\mathrm{O}}$ are extracted from the deviation between the calculated and experimental formation enthalpies of binary oxides $A_{x_1}\mathrm{O}_{x_2}$ for each functional:

$$\Delta_\mathrm{f} H^{\circ,T_\mathrm{r},\mathrm{cal}}_{A_{x_1}\mathrm{O}_{x_2}} - \Delta_\mathrm{f} H^{\circ,T_\mathrm{r},\mathrm{exp}}_{A_{x_1}\mathrm{O}_{x_2}} = x_1 N_{A-\mathrm{O}} \delta H^{A^{+\alpha}}_{A-\mathrm{O}}, \quad (8)$$

where $N_{A-\mathrm{O}}$ is the number of nearest neighbor $A-\mathrm{O}$ bonds of element $A$ in oxidation state $+\alpha$. CCE is constructed to be dependent on $+\alpha$: the energetic position of the bonding states and hence also the correction are expected to be oxidation state specific. In AFLOW, oxidation numbers can be determined by a Bader analysis [44, 45], while ensuring that the sum over all atoms equals zero. When counting bonds for distorted or low symmetry environments, a length variation up to 0.5 Å is allowed. After trying different tolerances, this value is found to lead to the best results. In the case of $CaTiO_3$ (see Fig. 1(c)) the nearest neighbor Ca-O bond length varies between 2.36 and 2.69 Å for the relaxed PBE structure.

As mentioned before, DFT errors do not only originate from the inaccurate treatment of the bonding in the compound, but also from the lack of error cancellation with the different reference phases. CCE corrections per bond implicitly include those of the elemental references — for a given bonded pair of atoms, reference phases are constant and the lack of error cancellation is then "absorbed" into corrections per bond. It especially applies to the molecular $O_2$ reference, for which the atomization energy is known to be poorly described in DFT [29].

The energy corrections extracted from binary oxides are then applied to the test-set of ternary oxides $A_{x_1}B_{x_2}\mathrm{O}_{x_3}$ to calculate the corrected formation enthalpies:

$$\Delta_\mathrm{f} H^{\circ,T_\mathrm{r},\mathrm{cor}}_{A_{x_1}B_{x_2}\mathrm{O}_{x_3}} = \Delta_\mathrm{f} H^{\circ,T_\mathrm{r},\mathrm{cal}}_{A_{x_1}B_{x_2}\mathrm{O}_{x_3}} - \sum_{i=1,2} x_i N_{i-\mathrm{O}} \delta H^{i^{+\alpha}}_{i-\mathrm{O}}, \quad (9)$$

where $N_{i-\mathrm{O}}$ is the number of nearest neighbor bonds between the cation $i$-species and oxygen.

Compared to other approaches [29, 32–35], it is important to note that at fixed composition, CCE is capable of correcting the relative stability of polymorphs with different coordination numbers.

Comparisons are performed with a quasi-FERE approach following the ideas of Refs. [34, 35]. A least-squares problem for all binary oxides in the fitting set is solved for the element specific corrections $\delta H^{\mathrm{qFERE}}_i$:

$$\Delta_\mathrm{f} H^{\circ,T_\mathrm{r},\mathrm{exp}}_{A_{x_1}\mathrm{O}_{x_2}} = \Delta_\mathrm{f} H^{\circ,T_\mathrm{r},\mathrm{cal}}_{A_{x_1}\mathrm{O}_{x_2}} - \sum_{i=1,2} x_i \delta H^{\mathrm{qFERE}}_i. \quad (10)$$

The corrections are then added to the calculated reference enthalpies used to calculate the corrected formation enthalpies. Contrary to the original FERE [34, 35], here **i.** no Hubbard-$U$ term is used, **ii.** only oxides are considered in the fitting set, **iii.** the corrections are determined and applied with respect to the calculated room temperature formation enthalpies rather than DFT formation energies, and **iv.** in part different experimental data are used.

**Principal thermodynamic considerations.** There is also another *caveat.* Corrections depending linearly on the concentration (like the previously proposed renormalization of the chemical potential of one or more species) are equivalent to tilting the whole Gibbs landscape, and might — in some cases — lead to thermodynamic paradoxes. For example, consider the case of non-ideal activity *vs.* concentration, differing from the Raoult's law with a negative(positive) deviation at low(high) concentration [46]. Any linear interpolation tends to balance the deviations and erroneusly correct the chemical potential by decreasing its non-ideal behavior. This is a rare scenario. Yet, phase diagrams having a very-high monotectoid and very-low eutectoid do exist, and the accuracy of calculated critical temperatures would be reduced with unappropropriately corrected enthalpies. The problem can be solved only by including more information in the DFT correction, introducing non linearity and/or considering topology and oxidation states like in the case of CCE.

***Ab-initio* calculations.** Calculations are performed using the AFLOW framework [7–10, 45, 47–50] leveraging the Vienna *Ab-initio* Simulation Package (VASP) [51, 52] with projector-augmented-wave pseudopotentials [53] of version 5.4. The exchange-correlation functionals LDA [4, 5], PBE [6] and SCAN [22] are employed. The parameters of the structural relaxation and static calculations largely follow the AFLOW Standard for entries from the ICSD library [47] with the internal VASP precision set to ACCURATE. No Hubbard-$U$ term is used, and for the elements Li, Be, Na and W, pseudopotentials with the labels Li, Be, Na_pv and W_sv are taken, respectively. For calculating total energy differences between a compound and its references, the kinetic energy cutoff is set to be 40% larger than the highest value recommended among all pseudopotentials for the compound but to at least 560 eV (oxygen cutoff). For magnetic systems, spin-polarized calculations are performed with all possible ferro-, ferri- and antiferromagnetic configurations initialized for five different sizes of the induced magnetic moments in the primitive unit cell. For computational efficiency, for $Ti_4O_7$, $Ti_5O_9$ and $Ti_6O_{11}$, only four different ferromagnetic configurations were initialized. The final magnetic state with the lowest total energy is considered for the formation enthalpy.

All room temperature structures are obtained from the AFLOW-ICSD online library [7, 9, 10, 47]. The selection is based on the structure information in the Kubaschewski *et al.* tables [15]. If it is insufficient, it is taken from the Springer Materials database [54]. The ICSD numbers, space groups and Pearson symbols are listed in Table A.V and A.VI (Appendix F). Spacegroups and Pearson symbols are calculated with AFLOW-SYM [55]. For $SiO_2$, both the $\alpha$-quartz (space group $P3_121$ #152; Pearson symbol hP9; AFLOW prototype `A2B_hP9_152_c_a` [49, 50, 56]) and $\alpha$-cristobalite ($P4_12_12$ #92; tP12; `A2B_tP12_92_b_a` [49, 50, 57]) prototypes are considered. $TiO_2$ is calculated in the rutile ($P4_2/mnm$ #136; tP6; `A2B_tP6_136_f_a` [49, 50, 58]) and anatase ($I4_1/amd$ #141; tI12; `A2B_tI12_141_e_a` [49, 50, 59]) structures. $Al_2SiO_5$ is represented in the kyanite ($P\bar{1}$ #2; aP32) and andalusite ($Pnnm$ #58; oP32) structures. $CaSiO_3$ is treated as wollastonite ($P\bar{1}$ #2; aP30) and pseudowollastonite ($C2/c$ #15; mS60). For $O_2$, $F_2$, $Cl_2$, $BF_3$ and $SiF_4$, a $10 \times 10 \times 10$ Å$^3$ cubic box is used, the intermolecular bond length is relaxed until the forces are smaller than 10 meV/Å, and the Brillouin zone is sampled only at the $\Gamma$-point.

**Selection of experimental data.** The accuracy of experimental data used is crucial. For oxides and halides, several reliable thermochemical libraries do exist, and here, we rely on the collections of Kubaschewski *et al.* [15], NIST-JANAF [16], Barin [17] and NBS [18].

For the validation of the experimental room temperature enthalpies, a procedure similar to Hautier *et al.* [19] is applied. Each $\Delta_f H^{\circ,T_r,\text{exp}}$ of Kubaschewski *et al.* [15] is first compared to the values from the NIST-JANAF database [16], which is believed to be the most accurate [19]. If the deviation exceeds 5 meV/atom, the value from Ref. [16] is used. For the oxides with no corresponding entry in NIST-JANAF, the formation enthalpies are compared with the Barin ones. If the values differ by more than 10 meV/atom, Barin's $\Delta_f H^{\circ,T_r,\text{exp}}$ is used. $NaCrO_2$ is an exception: the Kubaschewski formation enthalpy is taken, since the Barin value deviates by 0.15 eV/atom from the Kubaschewski and NBS data. Both Hautier *et al.* [19] and Aykol & Wolverton [33] used the $\Delta_f H^{\circ,T_r,\text{exp}}$ from Kubaschewski and obtained good agreement with the calculated reaction energies and formation enthalpies — this would not have been possible with the Barin value. In general, the NBS collection might not be considered as a suitable source for comparisons: When compared to all others, it exhibits several examples with significant deviations (Appendix B). This might be at least partially due to the special consistency requirements within NBS [18]. Oxides from Kubaschewski with no corresponding formation enthalpy in Barin are therefore excluded. For halides, the procedure is relaxed for $NaBF_4$ and $Na_2SiF_6$ due to the scarcity of experimental data for polar ternaries other than oxides. In these two cases, the Kubaschewski formation enthalpy is taken, which could only be verified by NBS.

## III. RESULTS AND DISCUSSION

### A. Room temperature DFT+AGL results

The difference between calculated DFT+AGL and experimental room temperature formation enthalpies for 79 binary and 71 ternary oxides for the three functionals employed are illustrated in Fig. 2. The vibrational (zero-point + thermal) contribution is shown in the lower panels of panels (a,b). MAEs are included in Table I. The calculated formation enthalpies for each functional, together with the experimental values, are included in Tables A.VII and A.IX (Appendix F). The vibrational, zero-point and thermal contributions are listed in Tables A.XII and A.XIV.

**Vibrational contribution.** In general the vibrational term is very small (lower panels of Fig. 2(a,b)), and decreases with increasing atomic number of the non-O elements. The maximum value of 23 (23) meV/atom is reached for $Al_2O_3$ (kyanite-$Al_2SiO_5$) with SCAN. The minimum of −22 (−4) meV/atom occurs for HgO ($PbWO_4$) with LDA (PBE and SCAN). For HgO, this is due to the heat of fusion of Hg being about 24 meV/atom at 234.29 K [16]. On average, the absolute vibrational value for binaries (ternaries) is very small: 5, 7 and 6 (7, 9 and 8) meV/atom for PBE, LDA and SCAN, respectively, due to partial cancellations of the zero-point and thermal contributions (Section III of Appendix C provides

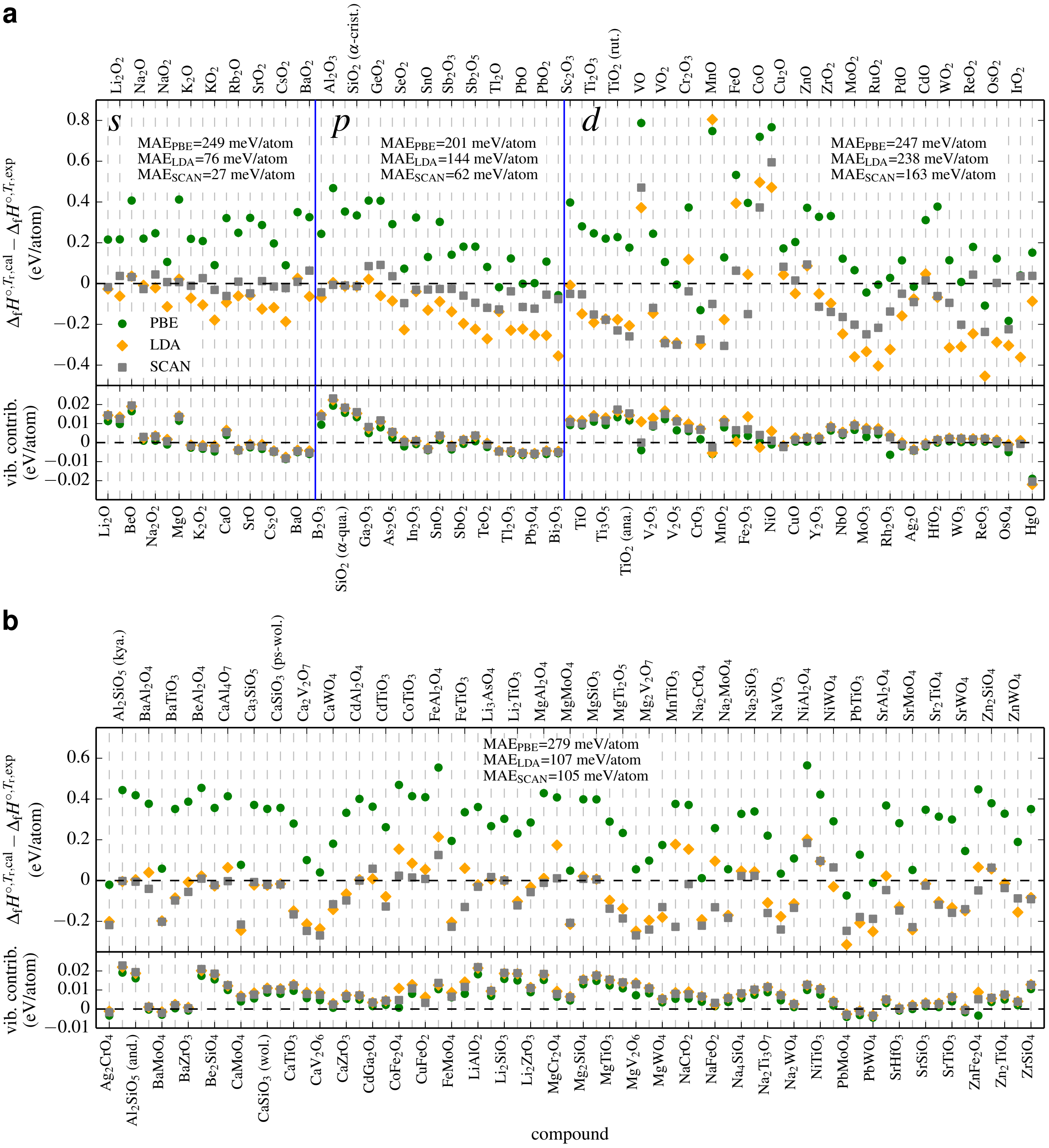

FIG. 2. **Uncorrected enthalpies.** Differences between calculated (Eq. (4)) and experimental room temperature formation enthalpies of binary oxides ((**a**) upper panel) and vibrational (zero-point + thermal) contribution to the calculated formation enthalpy ((**a**) lower panel). Vertical blue lines separate the different $l$-blocks with respect to the position of the non-O element of the compound in the periodic table. Differences between calculated and experimental room temperature formation enthalpies of ternary oxides ((**b**) upper panel) and vibrational contribution to the calculated formation enthalpy ((**b**) lower panel).

additional insights). Unless stated otherwise, our PBE, LDA and SCAN formation enthalpies include vibrational contributions, which, despite the often negligible values, consistently improve the MAEs of LDA and SCAN for binaries and ternaries by 2-5 meV/atom (Table I). For PBE, the MAE increases when including the vibrational value

TABLE I. **MAEs of uncorrected and corrected enthalpies.** MAEs of the uncorrected room temperature DFT+AGL, CCE and quasi-FERE corrected formation enthalpies for both binary and ternary oxides with respect to the experimental values. The numbers in brackets denote the MAEs of the calculated and corrected formation energies when no vibrational contribution is considered. Note that for the binary oxides CCE is basically exact by construction. All values in meV/atom.

| calculation type | binaries | | | ternaries | | |
|---|---|---|---|---|---|---|
| | PBE | LDA | SCAN | PBE | LDA | SCAN |
| plain DFT+AGL | 235 (234) | 176 (178) | 105 (107) | 279 (273) | 107 (109) | 105 (110) |
| CCE corrected | 5 (5) | 4 (4) | 3 (3) | 38 (38) | 29 (30) | 27 (27) |
| quasi-FERE corrected | 53 (54) | 44 (44) | 48 (48) | 43 (42) | 35 (36) | 44 (44) |

— most likely an artifact for the functional having the largest errors.

**Comparison of calculated and experimental results.** In Figure 2(a), the compounds are grouped according to the $l$-block of the non-O element in the periodic table. Materials are ordered with respect to increasing atomic number of the non-O element. PBE tends to underestimate the formation enthalpy leading to the largest deviations from the experimental values (MAE 235 meV/atom). Both LDA and SCAN show an increasingly better performance with total MAEs of 176 and 105 meV/atom, respectively. The findings are in agreement with previous reports [23, 35] including similar compounds. LDA was found to systematically yield better formation energies than PBE [34] for a much smaller set of 13 (9 binary, 4 ternary) oxides.

Results indicate a pronounced dependence on the $l$-character of the non-O element. For $s$-oxides, SCAN gives very accurate formation enthalpies with an MAE of 27 meV/atom, with LDA and PBE showing increasing deviations. For $p$-oxides, all functionals display a decreasing trend in $\Delta_{\mathrm{f}}H^{\circ,T_{\mathrm{r}},\mathrm{cal}}$ with respect to $\Delta_{\mathrm{f}}H^{\circ,T_{\mathrm{r}},\mathrm{exp}}$ with increasing atomic number of the non-O species, the trend being weakest for SCAN. Spin-orbit coupling could be the culprit, although often the effect largely cancels out when calculating formation energies [35, 60]. Instead, the trend might be caused by an increasing degree of covalency. MAEs for the combined set of all $s$- and $p$- (main group) oxides of 223, 113 and 46 meV/atom are obtained for PBE, LDA and SCAN, respectively. The values are in good agreement with Ref. [23], where a largely similar set of main group oxides was investigated. For transition metal, *i.e.* $d$-oxides, all functionals show large errors of several hundred meV/atom, with SCAN having the smallest MAE of 163 meV/atom. For the ternary oxides, deviations similar to the binaries are shown in Fig. 2(b): MAEs are 279, 107 and 105 meV/atom for PBE, LDA and SCAN.

Further improvements on a semilocal DFT level might be difficult considering that SCAN already fulfills all known constraints required for the exact functional [22]. A promising direction might be provided by the recently developed size-extensive self-interaction correction scheme [61–64] potentially leading to more accurate formation enthalpies.

### B. Coordination corrected enthalpies

This section compares the two correction schemes described in the Methods section: CCE and the quasi-FERE approach. The oxygen correction introduced by Wang *et al.* [29] is not considered as it shows a strong dependency on the fitting set when $p$-oxides are included (see Appendix D).

CCE uses the deviation between calculated and experimental room temperature formation enthalpies of single-valence binary oxides to obtain corrections per cation-O bond for each functional. They are then applied to the calculated formation enthalpies of ternary and mixed-valence binary oxides. The quasi-FERE method uses the binary data to obtain element specific corrections, optimizing the systematic error cancellation between the total energies/enthalpies of the references with respect to the compound [34, 35].

**Corrected binary results.** For the binary fit set, CCE gives almost exact solutions, as indicated by the small MAEs of 5, 4 and 3 meV/atom for PBE, LDA and SCAN (Table I). The corrections per bond are included in Table II and in Table A.VII (Appendix F). The quality is not surprising: the scheme is constructed to reproduce the experimental formation enthalpies of the single-valence binary oxides. The few other cases include mixed-valence compounds, multiple polymorphs at the same composition and per- as well as superoxides, leading to non-zero MAE for the binary set (Table I) and allowing assessment of CCE reliability. For $SbO_2$, the corrections obtained from $Sb_2O_3$ and $Sb_2O_5$ are used. $Pb_3O_4$ is refined based on PbO and $PbO_2$, and for $Ti_3O_5$ the corrections from $Ti_2O_3$ and $TiO_2$ (rutile) are taken. For $SiO_2$ ($\alpha$-cristobalite) and $TiO_2$ (anatase), the $\delta H_{A-\mathrm{O}}^{A+\alpha}$ determined from $SiO_2$ ($\alpha$-quartz) and $TiO_2$ (rutile) are applied, respectively. The results are included in Table A.VII (Appendix F). For all cases, the corrected values agree well with the experimental data — typically within 20 meV/atom — and a maximum systematic deviation of about 50 meV/atom is observed for $Pb_3O_4$.

The per- and superoxides cannot be corrected exactly, since their structure incorporates bonds between the cation and O, as well as an internal O-O bond. The values are corrected based on the assumption that for the cation-O bond the correction of the normal ($O^{2-}$) oxide can be taken, and the O-O bond correction is transferable between (su)peroxides: O-O correction for peroxides is obtained from $Li_2O_2$; the O-O correction for superoxides is derived from $KO_2$. The two values are listed in Table A.VII (Appendix F). All other per- and superoxides are corrected based on these values. In general, the procedure leads to good agreement with experiment, and the largest absolute deviation occurs for the corrected PBE value of $NaO_2$: 124 meV/atom (the absolute deviations for LDA and SCAN are 17 and 45 meV/atom, respectively).

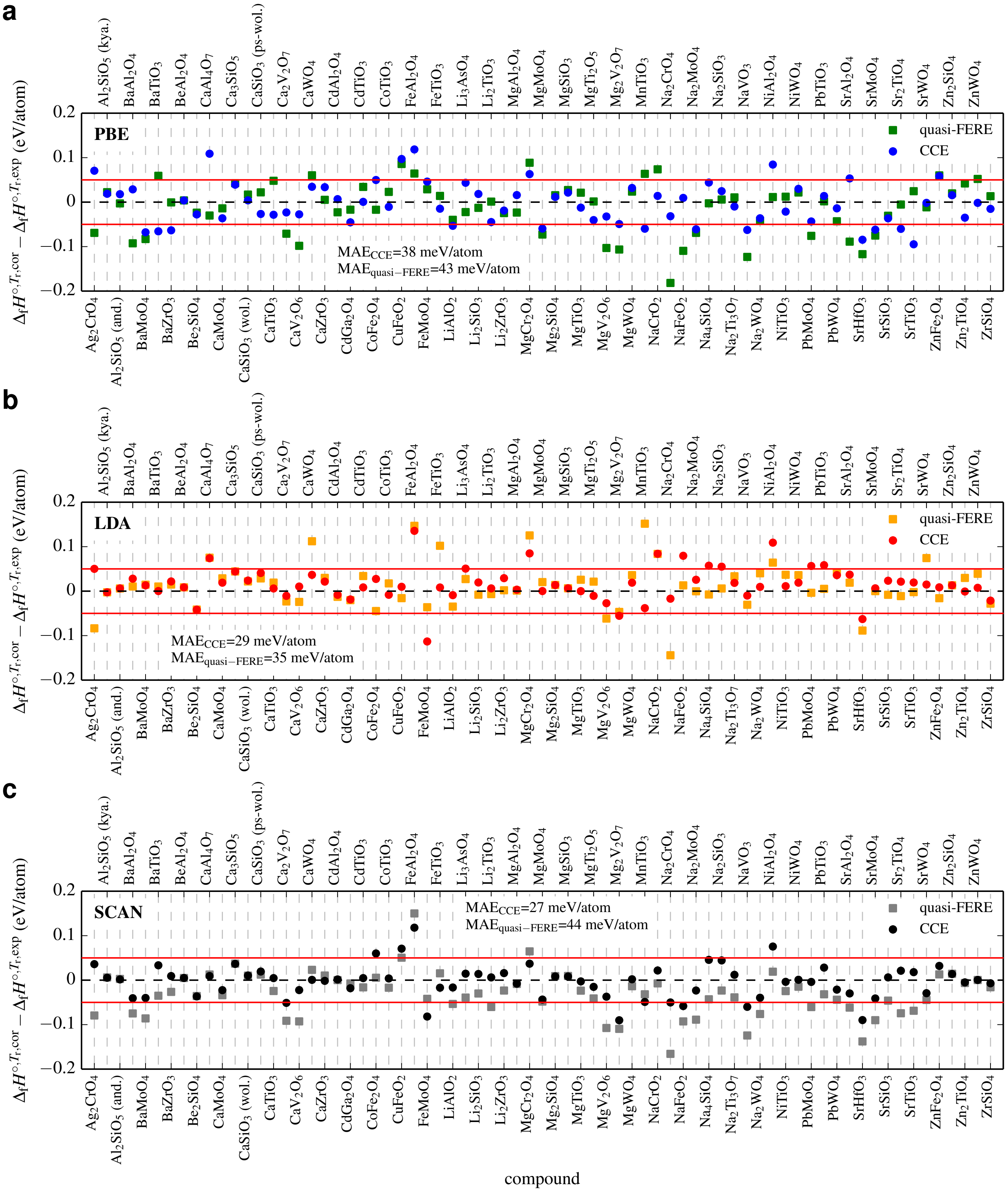

FIG. 3. **Corrected enthalpies.** Differences between corrected and experimental room temperature formation enthalpies of the test set of 71 ternary oxides for the CCE and quasi-FERE correction schemes on top of PBE (**a**), LDA (**b**) and SCAN (**c**). Note the different energy scale compared to Fig. 2. The red lines at ±50 meV/atom indicate the typical MAE of previous correction schemes [32, 35].

For quasi-FERE, MAEs of 53, 44 and 48 meV/atom for PBE, LDA and SCAN are obtained for the binary fit set. They well agree with the MAE of 54 meV/atom reported for the original FERE in Ref. [35] using PBE for the fit set of binary compounds. Our calculations indicate that even better agreement is obtainable when using LDA or SCAN in the DFT calculations.

FERE tends to yield large deviations if multivalent $p$-oxides such as SnO, $SnO_2$, $Sb_2O_3$, $SbO_2$, $Sb_2O_5$, $Tl_2O$, $Tl_2O_3$, PbO, $PbO_2$ and $Pb_3O_4$ are considered [35]. Indeed, for these systems, errors for the quasi-FERE corrected values partly exceeding 100 meV/atom are observed, in agreement with Ref. [35]. CCE circumvents the problem through its explicit dependence on the oxidation state of the cation according Eq. (8).

**Corrected ternary results.** The differences between CCE and quasi-FERE corrected and experimental room temperature formation enthalpies are displayed in Fig. 3 for the test set of 71 ternary oxides calculated with PBE (panel (a)), LDA (panel (b)) and SCAN (panel (c)). MAEs are included in Table I and the formation enthalpies are listed in Tables A.IX and A.XI (Appendix F). The importance of using *ab-initio* data as input for CCE is discussed in Appendix E. CCE predicts accurate results for almost all ternary compounds: MAE is 38, 29 and 27 meV/atom with PBE, LDA and SCAN, respectively. Compared to plain DFT+AGL, the errors are decreased by about a factor of 4-7. The mean deviations are significantly smaller than 45 and 48 meV/atom predicted by the GGA/GGA+$U$ mixing and FERE corrections of Refs. [32] and [35]. For the quasi-FERE method on the same set of compounds, MAEs of 43, 35 and 44 meV/atom are obtained for the corrected values of PBE, LDA and SCAN [65]. CCE consistently yields more accurate results than quasi-FERE for all three functionals. The MAEs of CCE are slightly larger than the 19 meV/atom of the local environment dependent GGA+$U$ method [33]. The latter scheme, however, uses about a factor two more parameters and is constructed for transition metal compounds. On the contrary CCE is applicable to all systems. CCE is simpler and more intuitive.

The largest single absolute deviation over the whole set is also higher for the quasi-FERE method — 182, 152 and 166 meV/atom for PBE ($Na_2CrO_4$), LDA ($MnTiO_3$) and SCAN ($Na_2CrO_4$) — compared to CCE — 118, 135 and 118 meV/atom for PBE, LDA and SCAN (always $FeAl_2O_4$).

TABLE II: **CCE corrections per bond.** Corrections per bond $\delta H_{A-\mathrm{O}}^{A^{+\alpha}}$ of the CCE method for each cation species $A$ in oxidation states $+\alpha$ obtained from calculated room temperature formation enthalpies of binary oxides. The numbers in brackets denote the corrections derived from the calculated DFT formation energies when no vibrational contribution is considered. The corrections for Si and Ti in oxidation state +4 are obtained from $\alpha$-quartz and rutile, respectively. All corrections in eV/bond.

| cation species $A$ | $+\alpha$ | $\delta H_{A-\mathrm{O}}^{A^{+\alpha}}$ PBE+AGL | (PBE) | $\delta H_{A-\mathrm{O}}^{A^{+\alpha}}$ LDA+AGL | (LDA) | $\delta H_{A-\mathrm{O}}^{A^{+\alpha}}$ SCAN+AGL | (SCAN) |
|---|---|---|---|---|---|---|---|
| Li | +1 | 0.0809 | (0.0766) | −0.0100 | (−0.0154) | −0.0065 | (−0.0118) |
| Be | +2 | 0.2035 | (0.1953) | 0.0180 | (0.0083) | 0.0160 | (0.0060) |
| B | +3 | 0.2030 | (0.1952) | −0.0572 | (−0.0693) | −0.0357 | (−0.0472) |
| Na | +1 | 0.0826 | (0.0823) | −0.0033 | (−0.0043) | −0.0101 | (−0.0113) |
| Mg | +2 | 0.1373 | (0.1335) | 0.0072 | (0.0025) | 0.0023 | (−0.0023) |
| Al | +3 | 0.1950 | (0.1869) | 0.0020 | (−0.0073) | −0.0028 | (−0.0124) |
| Si ($\alpha$-qua.) | +4 | 0.2648 | (0.2530) | −0.0098 | (−0.0233) | −0.0070 | (−0.0208) |
| K | +1 | 0.0821 | (0.0830) | −0.0269 | (−0.0263) | −0.0041 | (−0.0035) |
| Ca | +2 | 0.1070 | (0.1057) | −0.0308 | (−0.0332) | −0.0203 | (−0.0222) |
| Sc | +3 | 0.1656 | (0.1618) | −0.0034 | (−0.0083) | −0.0212 | (−0.0257) |
| Ti | +2 | 0.1169 | (0.1131) | −0.0619 | (−0.0667) | −0.0221 | (−0.0265) |
| Ti | +3 | 0.1025 | (0.0980) | −0.0796 | (−0.0855) | −0.0633 | (−0.0688) |
| Ti (rut.) | +4 | 0.1138 | (0.1072) | −0.0882 | (−0.0965) | −0.1150 | (−0.1237) |
| V | +2 | 0.2623 | (0.2637) | 0.1240 | (0.1203) | 0.1568 | (0.1568) |
| V | +3 | 0.1018 | (0.0984) | −0.0608 | (−0.0661) | −0.0498 | (−0.0531) |
| V | +4 | 0.0528 | (0.0467) | −0.1413 | (−0.1497) | −0.1462 | (−0.1537) |
| V | +5 | −0.0037 | (−0.0082) | −0.2033 | (−0.2118) | −0.2101 | (−0.2179) |
| Cr | +3 | 0.1553 | (0.1528) | 0.0495 | (0.0454) | −0.0159 | (−0.0189) |
| Cr | +6 | −0.1305 | (−0.1323) | −0.2980 | (−0.3053) | −0.2745 | (−0.2813) |
| Mn | +2 | 0.2492 | (0.2513) | 0.2682 | (0.2700) | −0.0333 | (−0.0325) |
| Mn | +4 | 0.0640 | (0.0600) | −0.0885 | (−0.0943) | −0.1528 | (−0.1582) |
| Fe | +2 | 0.1775 | (0.1763) | 0.1312 | (0.1310) | 0.0210 | (0.0188) |
| Fe | +3 | 0.1648 | (0.1633) | 0.0187 | (0.0130) | −0.0625 | (−0.0655) |
| Co | +2 | 0.2398 | (0.2397) | 0.1655 | (0.1662) | 0.1243 | (0.1230) |
| Ni | +2 | 0.2555 | (0.2558) | 0.1572 | (0.1552) | 0.1982 | (0.1980) |
| Cu | +1 | 0.1293 | (0.1310) | 0.0328 | (0.0340) | 0.0618 | (0.0635) |

TABLE II. (*continued*)

| cation species $A$ | $+\alpha$ | $\delta H_{A-\mathrm{O}}^{A+\alpha}$ PBE+AGL | (PBE) | $\delta H_{A-\mathrm{O}}^{A+\alpha}$ LDA+AGL | (LDA) | $\delta H_{A-\mathrm{O}}^{A+\alpha}$ SCAN+AGL | (SCAN) |
|---|---|---|---|---|---|---|---|
| Cu | +2 | 0.1018 | (0.1015) | −0.0245 | (−0.0258) | 0.0075 | (0.0068) |
| Zn | +2 | 0.1858 | (0.1853) | 0.0433 | (0.0423) | 0.0468 | (0.0455) |
| Ga | +3 | 0.2034 | (0.2009) | 0.0105 | (0.0068) | 0.0427 | (0.0386) |
| Ge | +4 | 0.2030 | (0.1992) | −0.0300 | (−0.0357) | 0.0457 | (0.0397) |
| As | +5 | 0.2039 | (0.2022) | −0.0599 | (−0.0636) | 0.0251 | (0.0212) |
| Se | +4 | 0.0730 | (0.0750) | −0.2267 | (−0.2277) | −0.0960 | (−0.0963) |
| Rb | +1 | 0.0934 | (0.0950) | −0.0229 | (−0.0215) | 0.0035 | (0.0049) |
| Sr | +2 | 0.1073 | (0.1082) | −0.0195 | (−0.0192) | −0.0160 | (−0.0155) |
| Y | +3 | 0.1363 | (0.1358) | −0.0209 | (−0.0219) | −0.0474 | (−0.0483) |
| Zr | +4 | 0.1419 | (0.1393) | −0.0416 | (−0.0451) | −0.0597 | (−0.0631) |
| Nb | +2 | 0.0610 | (0.0593) | −0.1235 | (−0.1263) | −0.0820 | (−0.0845) |
| Mo | +4 | 0.0327 | (0.0292) | −0.1797 | (−0.1843) | −0.1008 | (−0.1053) |
| Mo | +6 | −0.0440 | (−0.0470) | −0.3335 | (−0.3410) | −0.2490 | (−0.2558) |
| Ru | +4 | −0.0027 | (−0.0048) | −0.2020 | (−0.2057) | −0.1085 | (−0.1118) |
| Rh | +3 | 0.0115 | (0.0141) | −0.1347 | (−0.1363) | −0.0572 | (−0.0583) |
| Pd | +2 | 0.0568 | (0.0578) | −0.0793 | (−0.0793) | −0.0250 | (−0.0245) |
| Ag | +1 | −0.0115 | (−0.0083) | −0.0568 | (−0.0540) | −0.0683 | (−0.0653) |
| Cd | +2 | 0.1037 | (0.1042) | 0.0158 | (0.0162) | 0.0050 | (0.0053) |
| In | +3 | 0.1349 | (0.1353) | −0.0163 | (−0.0167) | −0.0127 | (−0.0130) |
| Sn | +2 | 0.0650 | (0.0670) | −0.0653 | (−0.0638) | −0.0148 | (−0.0130) |
| Sn | +4 | 0.1512 | (0.1505) | −0.0442 | (−0.0460) | −0.0133 | (−0.0153) |
| Sb | +3 | 0.1177 | (0.1207) | −0.1150 | (−0.1135) | −0.0218 | (−0.0198) |
| Sb | +5 | 0.1056 | (0.1052) | −0.1304 | (−0.1323) | −0.0551 | (−0.0573) |
| Te | +4 | 0.0610 | (0.0630) | −0.2035 | (−0.2033) | −0.0893 | (−0.0885) |
| Cs | +1 | 0.0983 | (0.1008) | −0.0588 | (−0.0567) | −0.0073 | (−0.0050) |
| Ba | +2 | 0.1167 | (0.1183) | 0.0085 | (0.0098) | 0.0028 | (0.0042) |
| Hf | +4 | 0.1617 | (0.1617) | −0.0290 | (−0.0296) | −0.0263 | (−0.0269) |
| W | +4 | 0.0570 | (0.0567) | −0.1575 | (−0.1587) | −0.0473 | (−0.0483) |
| W | +6 | 0.0052 | (0.0050) | −0.2063 | (−0.2078) | −0.1347 | (−0.1362) |
| Re | +4 | 0.0898 | (0.0897) | −0.1230 | (−0.1242) | 0.0218 | (0.0212) |
| Re | +6 | −0.0722 | (−0.0727) | −0.3027 | (−0.3040) | −0.1582 | (−0.1597) |
| Os | +4 | 0.0613 | (0.0617) | −0.1438 | (−0.1443) | 0.0015 | (0.0012) |
| Os | +8 | −0.2288 | (−0.2225) | −0.3805 | (−0.3793) | −0.2803 | (−0.2773) |
| Ir | +4 | 0.0198 | (0.0202) | −0.1808 | (−0.1813) | 0.0177 | (0.0180) |
| Hg | +2 | 0.1515 | (0.1700) | −0.0870 | (−0.0650) | 0.0375 | (0.0580) |
| Tl | +1 | −0.0090 | (−0.0065) | −0.0687 | (−0.0665) | −0.0635 | (−0.0612) |
| Tl | +3 | 0.0513 | (0.0536) | −0.0955 | (−0.0936) | −0.0159 | (−0.0140) |
| Pb | +2 | −0.0005 | (0.0028) | −0.1115 | (−0.1088) | −0.0575 | (−0.0548) |
| Pb | +4 | 0.0538 | (0.0568) | −0.1272 | (−0.1250) | −0.0273 | (−0.0250) |
| Bi | +3 | −0.0286 | (−0.0258) | −0.1775 | (−0.1752) | −0.0381 | (−0.0356) |

TABLE III: **CCE corrections per bond for halides.** Corrections per bond $\delta H_{A-X}^{A+\alpha}$ ($X$ =F,Cl) of the CCE method for each cation species $A$ in oxidation states $+\alpha$ obtained from calculated room temperature formation enthalpies of binary halides. The numbers in brackets denote the corrections derived from the calculated DFT formation energies when no vibrational contribution is considered. The values below (above) the horizontal line refer to chlorides (fluorides). The corrections for B in oxidation state +3 and for Si in oxidation state +4 are obtained from the gaseous molecular systems $BF_3$ and $SiF_4$. All corrections in eV/bond.

| cation species $A$ | $+\alpha$ | $\delta H_{A-X}^{A+\alpha}$ PBE+AGL | (PBE) | $\delta H_{A-X}^{A+\alpha}$ LDA+AGL | (LDA) | $\delta H_{A-X}^{A+\alpha}$ SCAN+AGL | (SCAN) |
|---|---|---|---|---|---|---|---|
| Li | +1 | 0.0788 | (0.0748) | 0.0120 | (0.0070) | −0.0482 | (−0.0532) |
| Na | +1 | 0.0833 | (0.0807) | 0.0258 | (0.0225) | −0.0473 | (−0.0503) |
| K | +1 | 0.0718 | (0.0702) | 0.0083 | (0.0060) | −0.0472 | (−0.0490) |
| Be | +2 | 0.2073 | (0.2008) | 0.0563 | (0.0480) | −0.1215 | (−0.1300) |
| B | +3 | 0.2447 | (0.2093) | 0.1587 | (0.1247) | −0.1640 | (−0.1987) |
| Al | +3 | 0.2488 | (0.2353) | 0.0572 | (0.0415) | −0.1208 | (−0.1367) |
| Si | +4 | 0.3135 | (0.2833) | 0.1750 | (0.1450) | −0.1525 | (−0.1825) |
| Na | +1 | 0.1000 | (0.0972) | 0.0568 | (0.0537) | 0.0180 | (0.0152) |
| K | +1 | 0.0938 | (0.0913) | 0.0488 | (0.0460) | 0.0168 | (0.0142) |
| Ca | +2 | 0.1608 | (0.1552) | 0.0742 | (0.0680) | 0.0227 | (0.0167) |
| Al | +3 | 0.1933 | (0.1845) | 0.0498 | (0.0400) | 0.0485 | (0.0388) |

TABLE III. (*continued*)

| cation species $A$ | $+\alpha$ | $\delta H_{A-X}^{A+\alpha}$ PBE+AGL | (PBE) | $\delta H_{A-X}^{A+\alpha}$ LDA+AGL | (LDA) | $\delta H_{A-X}^{A+\alpha}$ SCAN+AGL | (SCAN) |
|---|---|---|---|---|---|---|---|

When CCE predicts a similar value for all three functionals with a large deviation with respect to the experimental data, the measured $\Delta_{\mathrm{f}} H^{\circ,T_{\mathrm{r}},\mathrm{exp}}$ might be inaccurate. The conclusion is further confirmed if the quasi-FERE corrected values predict a similar trend. Based on the analysis, the experimental data of $FeAl_2O_4$ and $NiAl_2O_4$ might be too low (*i.e.* too negative) by about 120-140 and 80-110 meV/atom, respectively. $SrHfO_3$ might be too high by about 60-90 meV/atom.

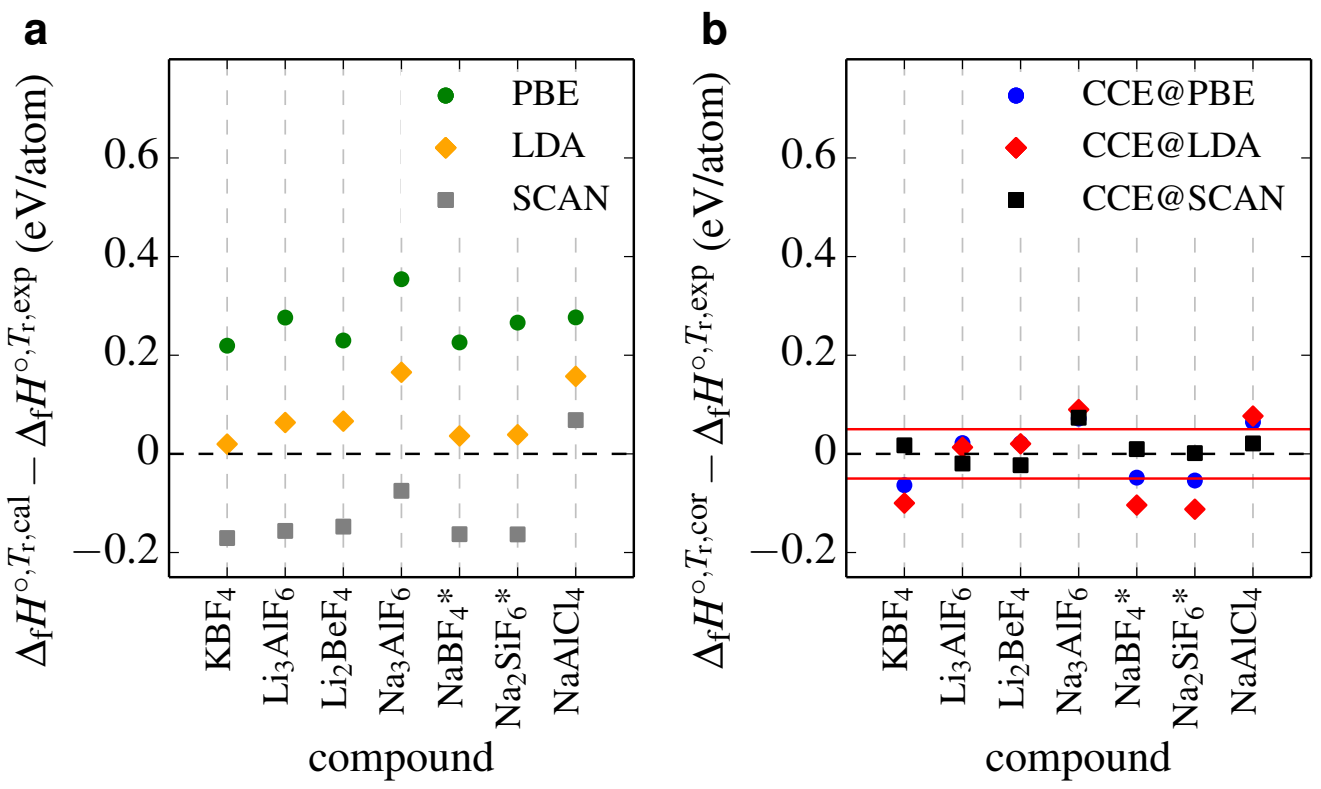

FIG. 4. **Uncorrected** *vs.* **corrected enthalpies for halides.** Differences between calculated (**a**) as well as corrected (**b**) and experimental room temperature formation enthalpies for seven ternary halides. For the compounds marked with "*" the experimental formation enthalpy from Kubaschewski *et al.* [15] can only be verified by NBS [18]. The red lines at ±50 meV/atom indicate the typical MAE of previous correction schemes [32, 35].

The scarcity of reliable experimental data for polar ternary systems other than oxides restricts the number of compounds available to demonstrate the generality of CCE. In Figure 4, uncorrected and corrected results are presented for a set of seven ternary halides. The formation enthalpies are listed in Tables A.VIII, A.X, A.XVII and A.XIX with the vibrational, zero-point and thermal contributions in Tables A.XIII and A.XV (Appendix F). Corrections are given in Tables III, A.VIII and A.XVII. It is difficult to ascribe a statistical significance to a set of only seven entries: MAEs amount to 264, 78 and 135, as well as 49, 74 and 24 meV/atom for the uncorrected and corrected results of PBE, LDA and SCAN, respectively. CCE guarantees a significant improvement in all cases. For $Na_3AlF_6$, and potentially also $NaAlCl_4$, the experimental value might be too low by about 70-90 and 20-80 meV/atom. The accuracy of the corrected results for $KBF_4$, $NaBF_4$ and $Na_2SiF_6$ is interesting — in these cases part of the corrections are obtained from the gaseous molecular $BF_3$ and $SiF_4$ phases and applied to solid ternaries. For PBE and LDA, the corrected results show rather large errors revealing that for these functionals the corrections per bond are not well transferable from molecules to solids. This biases the MAE particularly for the corrected LDA values. For SCAN, however, the corrected results are accurate, further showcasing the better suitability of this functional for CCE. The functionals' different behaviors agree with previous reports [66, 67].

The vibrational (zero-point + thermal) contribution to the formation enthalpy can be largely included in the corrections without explicit calculation, being mostly element specific. For example, for binary oxides the vibrational term is highest for $Al_2O_3$, BeO and $SiO_2$ ($\alpha$-quartz), ranging from 16 to 23 meV/atom (depending on the functional). For ternaries, the largest value is found for kyanite $Al_2SiO_5$ with 19 to 23 meV/atom. MAEs of the corrected formation energies obtained without vibrational contribution in both the binary-fit and ternary-test sets (as in Refs. [34, 32, 35]) are included in brackets in Table I. They deviate no more than 1 meV/atom from the MAEs of the corrected DFT+AGL results. Thus, $\Delta_{\mathrm{f}} E^{0,\mathrm{DFT}}$ corrections can be reliably based on only $\Delta_{\mathrm{f}} H^{\circ,T_{\mathrm{r}},\mathrm{exp}}$. In addition, the vibrational term usually does not lead to significant differences between two structures at the same composition. This has already been seen with machine learning analysis [68]. Therefore, the following discussion is based on results directly obtained with DFT.

**Relative stability.** CCE can also correct the relative stability of same stoichiometry structures with different number of nearest neighbor cation-O bonds. $Al_2SiO_5$ is an example: kyanite is the experimental ground state and andalusite is higher in energy. PBE falsely predicts kyanite to be 19 meV/atom above andalusite (−2.937 *vs.* −2.956 eV/atom). CCE correctly gives kyanite to be lower by 4 meV/atom (−3.343 *vs.* −3.339 eV/atom), in good agreement with the experimental values (−3.361 *vs.* −3.358 eV/atom).

The situation is more evident with polymorphs having large energy differences. Experimentally, MnO and CoO have rocksalt ground states. In Ref. [23], it was reported that PBE and SCAN predict other ground states for both systems with only 4 cation-O bonds, in disagreement with the experimental finding: 6. CCE solves the issue. We take the DFT ground states provided in Ref. [23], relax and primitivize them. PBE/SCAN for MnO and SCAN for CoO find zincblende (space group $F\bar{4}3m$ #216; Pearson symbol cF8; AFLOW prototype `AB_cF8_216_c_a` [49, 50, 69]). With PBE the final CoO structure is body-centered tetragonal ($I\bar{4}m2$ #119; tI4; `AB_tI4_119_c_a` [49, 50, 70]). For CoO, PBE and SCAN erroneously give the energies of the relaxed geometries to be 164 and 103 meV/atom below rocksalt. CCE solves the dilemma. When corrected, they become 76 and 20 meV/atom above the experimental ground state. For MnO the PBE structure is corrected from being 5 meV/atom more-stable to 246 meV/atom less-stable than the experimental re-port. The MnO structure given by SCAN is already 44 meV/atom higher than rocksalt. CCE reduces its difference to 11 meV/atom without changing the correct experimental order. CCE succeeds in all examples. Any scheme dealing only with stoichiometry (such as FERE) would not be able to disentangle the relative stability.

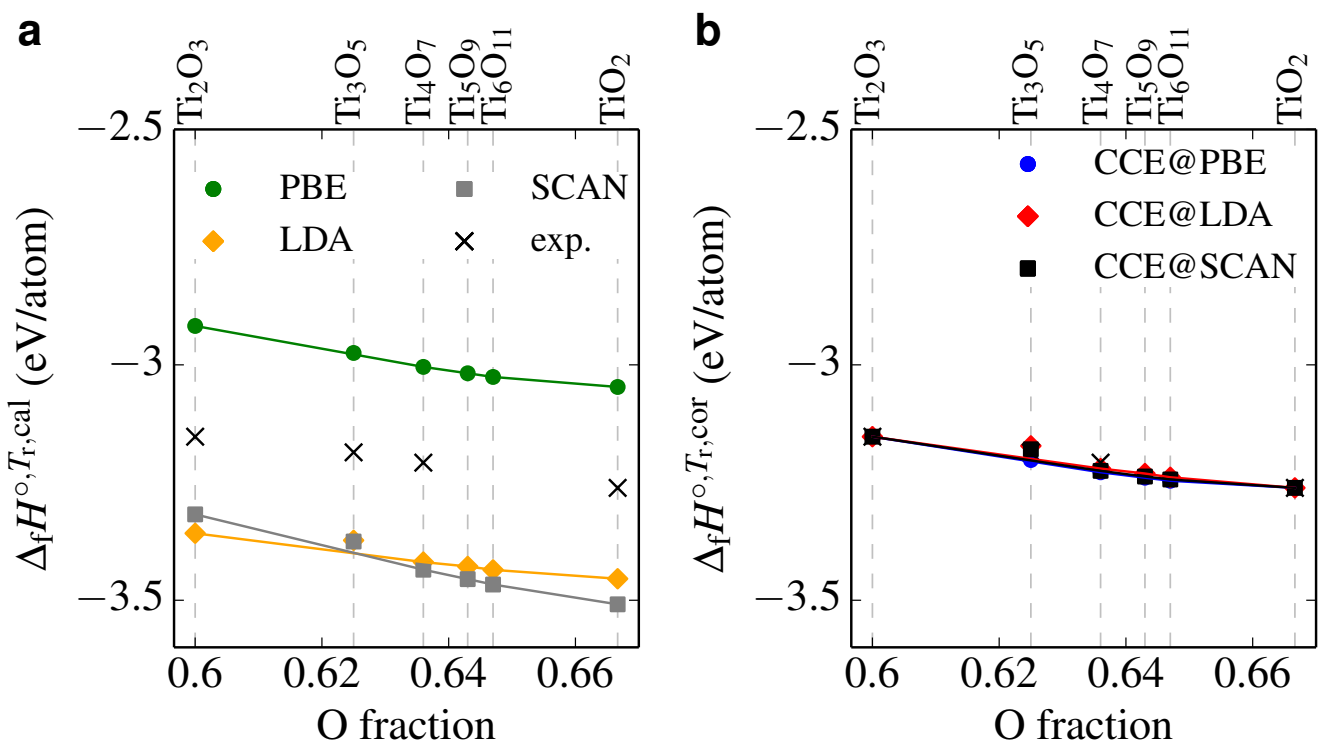

FIG. 5. **Uncorrected *vs.* corrected convex hull.** Section of the convex hull between $Ti_2O_3$ and $TiO_2$ (rutile) for the Ti-O system from plain DFT (**a**) and corrected by CCE (**b**).

**Application to Ti-O systems.** To test whether CCE will also yield quantitatively reliable results for defect energies, the method is applied to Ti-O. The corrections are obtained from $Ti_2O_3$ and rutile $TiO_2$, and are then applied to predict the enthalpies of other oxides, including crystallographic shear compounds (Magnéli phases) $Ti_nO_{2n-1}$. In Figure 5, the section of the convex hull phase diagram [71] between $Ti_2O_3$ and $TiO_2$ is presented for both uncorrected and CCE corrected results. Plain DFT captures well the position of all structures with respect to the individual convex hull for each functional, but yields quantitative errors of the order of several 100 meV/atom in all three cases. When corrected by CCE, all three functionals produce formation enthalpies within 10-20 meV/atom of experiments. Note that for all functionals (corrected and uncorrected) and from the experimental data, $Ti_3O_5$ is found to be above the stability hull by up to about 30 meV/atom.

## IV. CONCLUSIONS

We have introduced a Coordination Corrected Enthalpies (CCE) scheme based on the number of nearest neighbor cation-anion bonds. 71 (7) ternary oxides (halides) are used as a test set. CCE gives very accurate corrected formation enthalpies with MAEs of 38 (49), 29 (74) and 27 (24) meV/atom for PBE, LDA and SCAN, respectively. Zero-point and finite temperature vibrational contributions are treated within a quasiharmonic Debye model and are found to largely cancel out. Errors are significantly smaller than previous approaches [32, 34, 35]. Because CCE considers bonding connectivity and topology, it can also correct the relative stability of different structures at a given composition.

Correction schemes for formation enthalpies are the steps in a ladder of approximations:
**i.** The oxygen correction of Ref. [29] applies a constant energy shift per $O_2$; it can be seen as a $0^{\mathrm{th}}$ order step: one parameter for all oxides. The approach typically leads to mean absolute errors of 100 meV/atom or larger, and can be combined with the GGA/GGA+$U$ mixing scheme for improved accuracy [32].
**ii.** The FERE method [34, 35] corrects the elemental reference energy of each species of the compound; it is a $1^{\mathrm{st}}$ order approximation: one parameter per element. FERE's accuracy is typically limited to about 40-50 meV/atom. Improvements require considering the characteristics of the compounds.
**iii.** CCE leverages the topology of nearest neighbor shells. CCE yields accurate formation enthalpies with an average absolute error as small as 20-30 meV/atom. The method is simple and easy to extend to other materials classes, *e.g.* nitrides, phosphides or sulfides. It can be used to predict a wide variety of properties relying on accurate formation enthalpies such as battery voltages, defect energies and the formation of high-entropy materials [72].

## V. DATA AVAILABILITY.

All the *ab-initio* data are freely available to the public as part of the AFLOW online repository and can be accessed through AFLOW.org following the REST-API interface [45] and AFLUX search language [10].

## VI. ACKNOWLEDGMENTS

We thank Ohad Levy, Frisco Rose, Eric Gossett, David Hicks and Denise Ford for fruitful discussions. The authors acknowledge support by DOD-ONR (N00014-15-1-2863, N00014-15-1-2266, N00014-17-1-2090, N00014-16-1-2326, N00014-17-1-2876). R.F. acknowledges support from the Alexander von Humboldt foundation under the Feodor Lynen research fellowship. C.O. acknowledges support from the National Science Foundation Graduate Research Fellowship under Grant No. DGF-1106401. S.C. acknowledges financial support from the Alexander von Humboldt foundation.

## VII. AUTHOR CONTRIBUTIONS

R.F. and S.C. proposed the formation enthalpy correction. All authors — R.F., D.U., C.O., A.S., M.F., M.B.N., C.T., S.C. — discussed the results and contributed to the writing of the article.

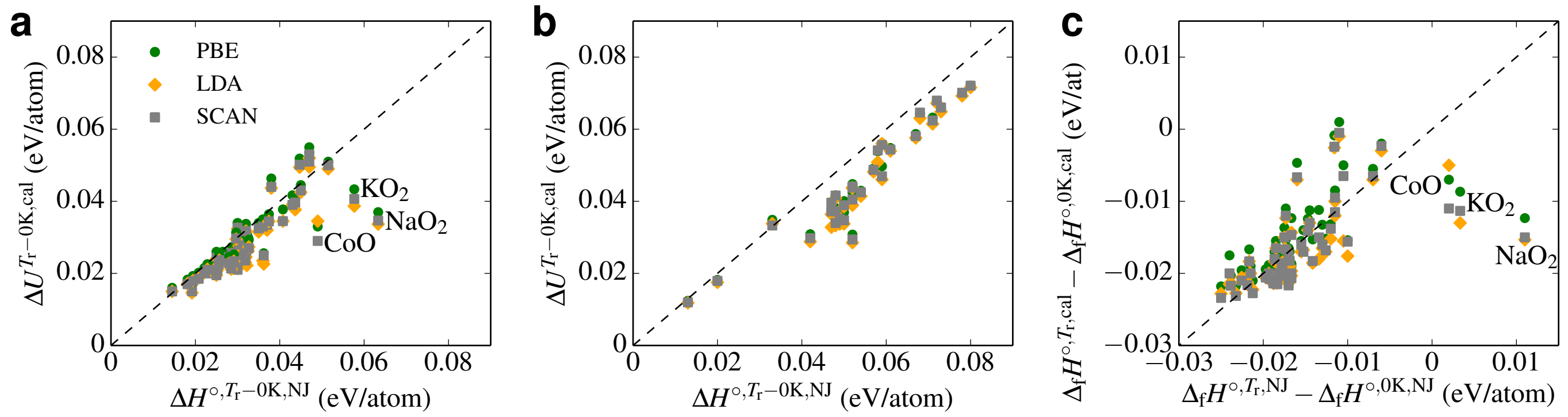

FIG. A.6. **Internal energy *vs.* enthalpy differences.** Calculated internal energy differences *vs.* experimental enthalpy differences from the NIST-JANAF (NJ) collection [16] between 0 and $T_{\mathrm{r}}$=298.15 K for (**a**) binary and ternary oxides and (**b**) elemental reference phases. A comparison of the calculated *vs.* experimental formation enthalpy differences for the same set of compounds as in (**a**) is presented in (**c**). Points showing significant deviations are labeled with the compound in (**a**,**c**).

### Appendix A: Comparing experimental enthalpy differences with calculated energy differences

In Figure A.6(a) the calculated internal energy differences from AGL are plotted *vs.* the experimental enthalpy differences [16] between 0 K and $T_{\mathrm{r}}$ for 52 (36 binary and 16 ternary) oxides. The calculated values approximate the measured data in general very well, with a slight tendency to underestimate them. The MAE is 4, 5 and 5 meV/atom for calculations using PBE, LDA and SCAN, respectively. There are three cases with major deviations: $NaO_2$, $KO_2$ and CoO. The largest difference between calculated and experimental values occurs for $NaO_2$: 30 meV/atom for LDA. Superoxides might be difficult to describe within the AGL Debye model. For CoO the deviation of 15-20 meV/atom can be assigned to the general difficulty of describing the properties of the material as noted before for the calculated formation enthalpy in Fig. 2, also recognized previously [35].

In Figure A.6(b) an equivalent plot is presented for 29 mostly metallic elements. The MAE is 8, 10 and 9 meV/atom for PBE, LDA and SCAN. In general, the calculations slightly underestimate the experimental enthalpy difference. This can be understood as the AGL approach neglects electronic contributions to the internal energy change. The errors are significantly smaller than the MAEs of CCE in Fig. 3, which are of the order of 30 meV/atom.

The combined effect of the thermal excitations for the compounds and elements on the formation enthalpy is illustrated in Fig. A.6(c). It depicts the difference $(\Delta_{\mathrm{f}}H^{\circ,T_{\mathrm{r}},\mathrm{cal}} - \Delta_{\mathrm{f}}H^{\circ,0\mathrm{K},\mathrm{cal}})$ *vs.* $(\Delta_{\mathrm{f}}H^{\circ,T_{\mathrm{r}},\mathrm{NJ}} - \Delta_{\mathrm{f}}H^{\circ,0\mathrm{K},\mathrm{NJ}})$ for the same set of compounds as in Fig. A.6(a). The MAEs are 3, 4 and 4 meV/atom for PBE, LDA and SCAN, indicating that the underestimation of individual enthalpy differences in Figs. A.6(a,b) is partially canceled in the formation enthalpy. The three systems showing large deviations (up to 26 meV/atom for $NaO_2$) are the same as in Fig. A.6(a): the only compounds with a positive experimental difference between 0 K and $T_{\mathrm{r}}$. In conclusion, the AGL approach treats thermal contributions reliably at low computational cost.

### Appendix B: Comparing experimental formation enthalpies from different sources

Comparing the data from different experimental sources gives insight on the exact formation enthalpy. Figure A.7 shows the difference between the four collections [15–18] for binary (Figs. A.7(a,b)) and ternary (Figs. A.7(c,d)) oxides. Table A.IV includes the mean absolute deviations. Not all oxides are available in all collections. The number of compounds reported is therefore added in brackets after the name of the collection.

TABLE A.IV. **Comparison of experimental values.** Mean absolute deviations for the experimental formation enthalpies from different sources for binary (above the diagonal) and ternary (below the diagonal) oxides: Kub (Kubaschewski) [15], NJ [16], Barin [17] and NBS [18]. The numbers in brackets after the abbreviations, in the first row for binary and in the first column for ternary oxides, denote the number of oxides for which formation enthalpies were available. The numbers in brackets in the second column are the mean absolute deviation when $MgTi_2O_5$ is excluded from the set. All values in meV/atom.

| | Kub (79) | NJ (41) | Barin (79) | NBS (76) |
|---|---|---|---|---|
| Kub (71) | - | 4 | 8 | 14 |
| NJ (16) | 20 (7) | - | 1 | 8 |
| Barin (71) | 17 (14) | 3 | - | 10 |
| NBS (59) | 19 (15) | 8 | 8 | - |

In general, most of the values for one oxide agree well. However, in certain cases, deviations significantly exceeding ±50 meV/atom are observed, for which the respective compound is labeled. For the binary oxides in Figs. A.7(a,b), significant deviations always occur when data from the NBS tables are compared to another source as also indicated by the mean absolute deviations in Ta-ble A.IV. The largest values of 14, 10 and 8 meV/atom are found between NBS and the other collections. This might be due to the special internal consistency requirements inside the NBS tables [18].

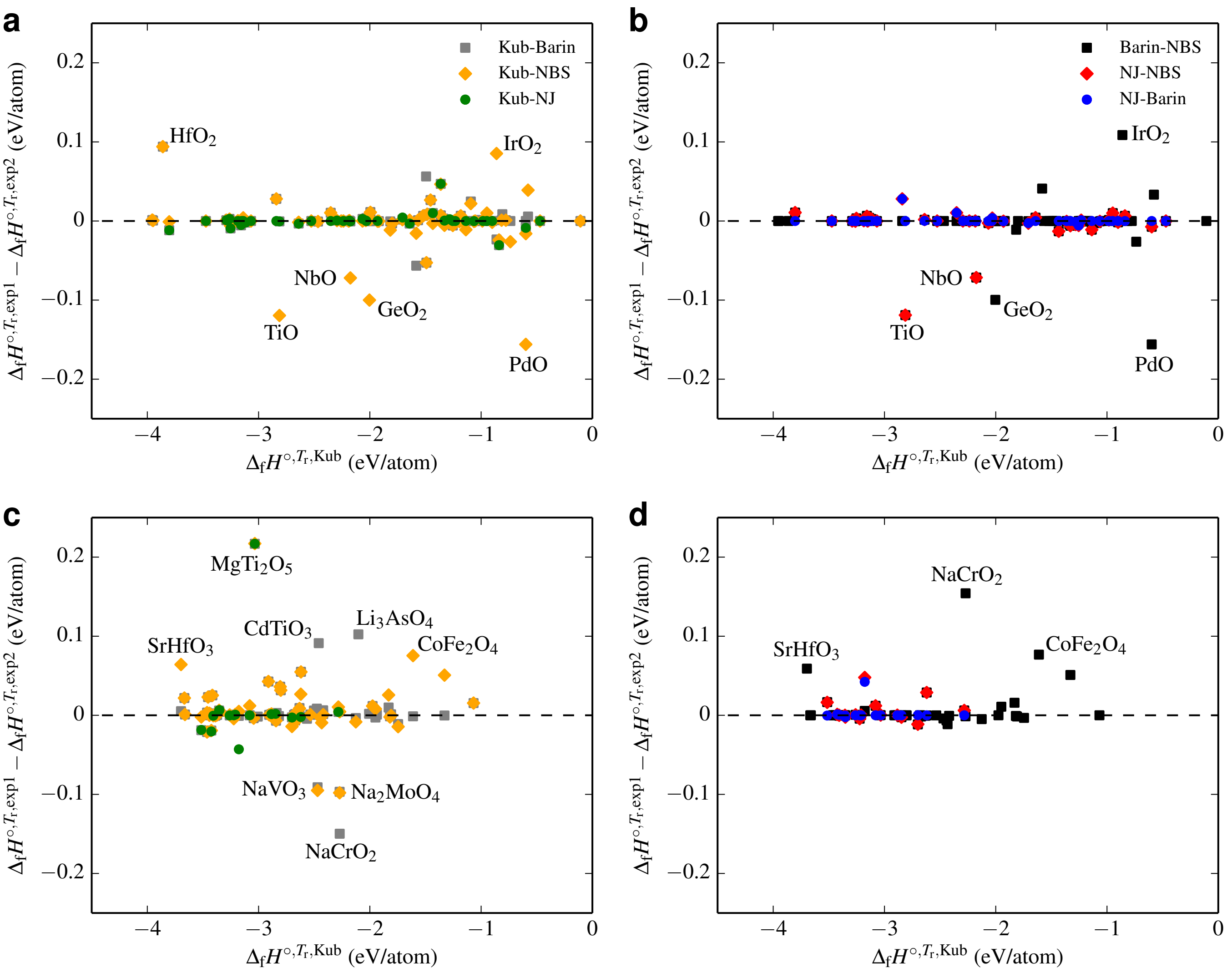

FIG. A.7. **Comparison of experimental values.** Comparison of the experimental room temperature formation enthalpies from different sources for binary (**a**,**b**) as well as ternary (**c**,**d**) oxides. Points showing deviations $\gg \pm 50$ meV/atom are labeled with the compound. Kub refers to Ref. [15], NJ to Ref. [16], Barin to Ref. [17] and NBS to Ref. [18]. Despite corrections can be done to bring MAE to 20-50 meV/atom, there is substantial scatter in experimental measurements.

For ternary oxides significant deviations are seen between Kubaschewski *et al.* and Barin as well as between Kubaschewski *et al.* and NBS in Fig. A.7(c). Some of the formation enthalpies for ternary oxides might not be very accurate in Kubaschewski. For $MgTi_2O_5$, Kubaschewski $\Delta_f H^{\circ,T_r,\mathrm{exp}}$ is clearly inaccurate since all three other collections suggest the same value, which deviates by more than 200 meV/atom. It is a bias to the mean absolute deviation, especially when compared to NIST-JANAF (NJ) as there are only 16 common ternary oxides within the two collections. Therefore, in the second column in Table A.IV the numbers in brackets were calculated with $MgTi_2O_5$ excluded. The largest mean absolute deviations are observed when NBS is involved with the exception of the comparison between Barin and Kubaschewski. In Figure A.7(d) three significant deviations are also found between the Barin and NBS collections. Therefore, the NBS data might not be suited for comparisons with calculated results.

## Appendix C: Comparison of the zero-point and thermal contributions to the formation enthalpy

Figures A.8(a,b) depict the individual zero-point and thermal contributions to the formation enthalpy according to Eq. (4) for binary and ternary oxides, respectively. The ZPC is usually larger. Both contributions are typically about two orders of magnitude smaller than the total values of the formation enthalpies. The maximum ab-

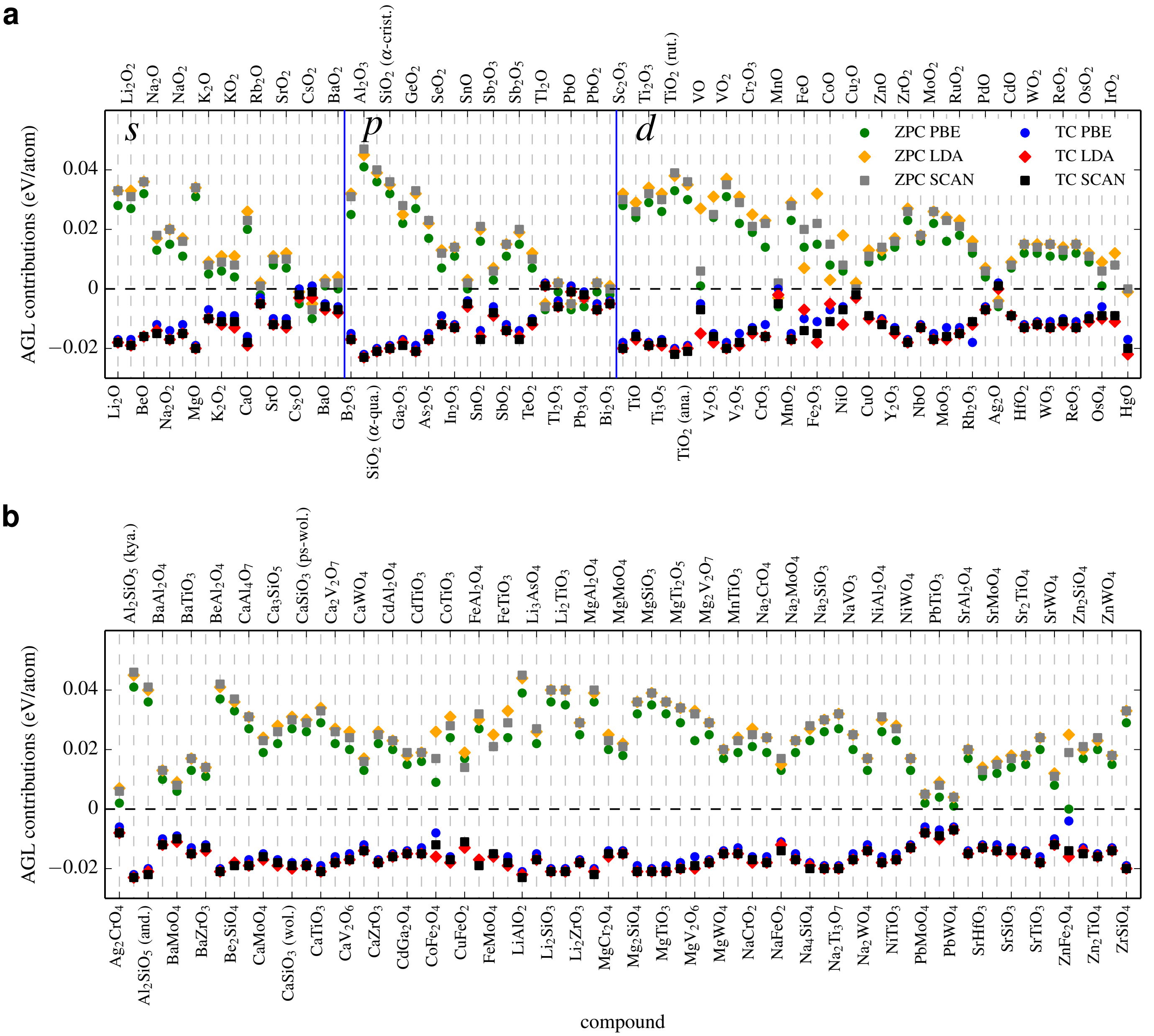

FIG. A.8. **Zero-point *vs.* thermal contribution.** Comparison of the zero-point (ZPC) and thermal (TC) contributions to the room temperature formation enthalpies for binary (**a**) and ternary (**b**) oxides.

solute values of the zero-point contribution (thermal contribution) of 47 (23) and 46 (23) meV/atom are reached for $Al_2O_3$ and kyanite $Al_2SiO_5$ for binary and ternary oxides with SCAN, respectively. For binaries, the zero-point contribution is almost always positive (or minimally −10 meV/atom for $CsO_2$ for PBE), while the thermal is basically always negative (or maximally 2 meV/atom for $Ag_2O$ and $Tl_2O$ for PBE). For ternaries the contributions always have opposite sign leading to a partial cancellation, involving two effects: **i.** the bonds for the ionic compound (for especially light elements) are rather stiff, leading to a large zero-point vibrational energy with respect to the references, giving a positive contribution to the formation enthalpy; **ii.** for systems with stiff bonds, only a small part of the vibrational spectrum is accessible to thermal excitations, giving a negative contribution to the formation enthalpies, as the enthalpies of the elements increase more strongly with temperature than the compounds.

### Appendix D: Dependence of the oxygen correction on the fitting set

According to Wang *et al.* [29], DFT formation energies of oxides can be corrected by plotting the difference between calculated and experimental values for non-transition metal compounds, normalized to $O_2$ in the formula unit (Fig. A.9). Here, the calculated room tem-perature formation enthalpies are used. In contrast, the original work considered formation energies directly calculated with DFT. The differences are negligible given the small size of the vibrational contribution. As proposed by Wang *et al.*, [29] the $\Delta_f H^{\circ,T_r,\mathrm{cal}}$ should have a rather constant shift from $\Delta_f H^{\circ,T_r,\mathrm{exp}}$, which was estimated to be 1.36 eV/$O_2$ for PBE, based on a fit of six non-transition metal ($s$- and $p$-) oxides. The calculated formation enthalpies could then be corrected by subtracting the value.

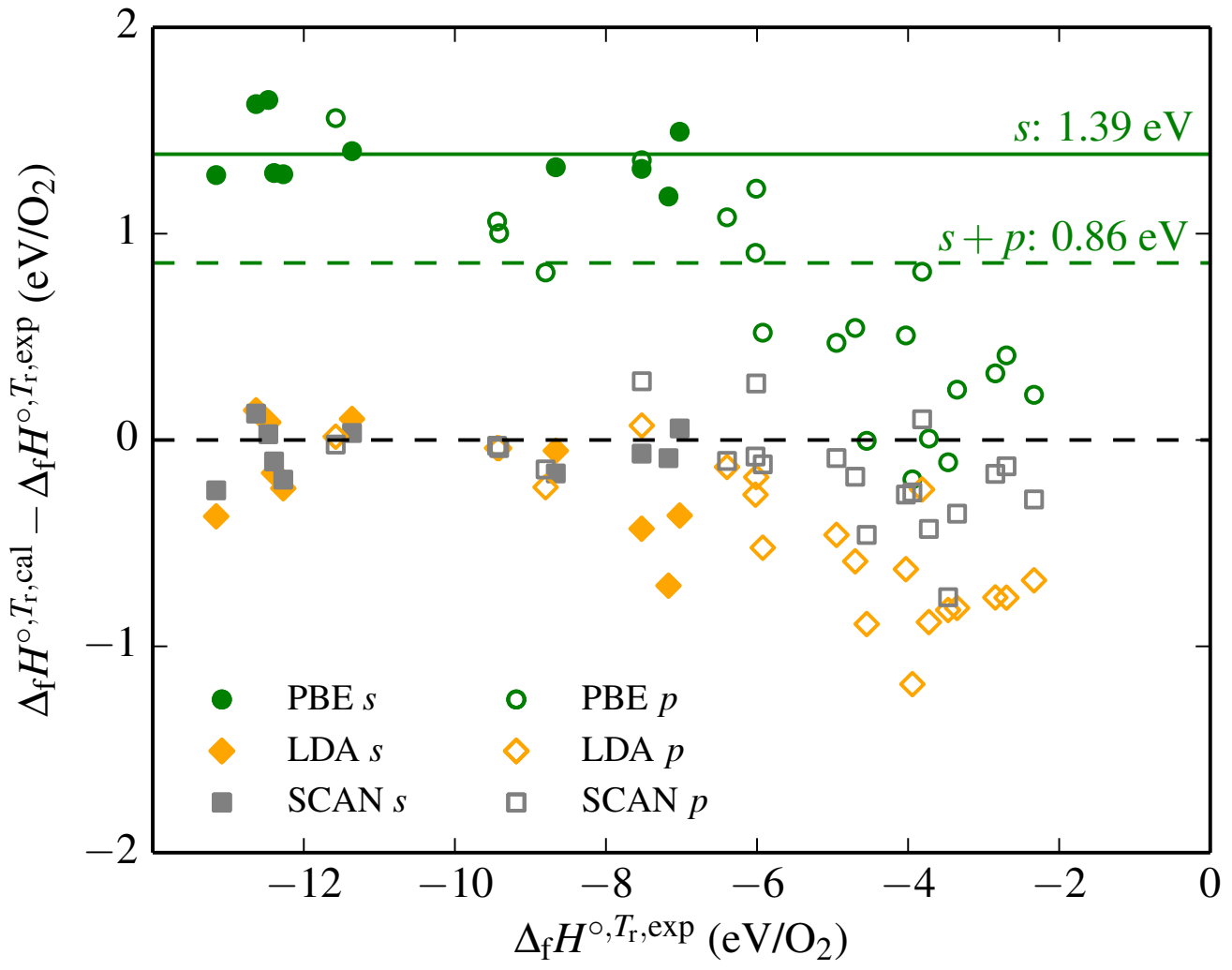

FIG. A.9. **Variation of the oxygen correction.** Dependence of the oxygen correction on the set of compounds used for the fitting. The filled (open) symbols correspond to $s$-oxides ($p$-oxides). For PBE, the fitting curves are drawn with only the $s$-oxides used to obtain the correction (solid green line) and all $s$- and $p$-oxides are used (dashed green line). The scale is normalized to eV/$O_2$.

This is a good approximation for binary $s$-oxides ($O^{2-}$ anions) and PBE (filled green symbols in Fig. A.9): the oxygen correction fitted to the data amounts to 1.39 eV/$O_2$, in close agreement with the initially proposed value. However, when all $p$-oxides are included (open green symbols in Fig. A.9), a constant shift is not adequate, leading to a very different correction of 0.86 eV. Also, for LDA and SCAN, when the formation enthalpies of $p$-oxides are included, the scatter increases towards lower values. The trend is less pronounced than for PBE.

When fitting the oxygen correction to the formation enthalpies of the $s$-oxides calculated for SCAN, a small value 0.06 eV is obtained. The SCAN binding energy for $O_2$ (−5.59 eV) is also closer to experiment (−5.12 eV) [16] than PBE (−6.05 eV) and LDA (−7.49 eV), in good agreement with previous reports [22, 29, 34, 62, 73]. As such, the better description of oxides by SCAN *vs.* PBE is mostly due to its improved treatment of $O_2$.

## Appendix E: Importance of *ab-initio* data for CCE

To investigate the importance of the information included in the *ab-initio* data on the accuracy of CCE, we derive the corrections per bond from the experimental binaries and apply to the ternaries without using DFT data (CCE@exp). In Figure A.10, the results are compared to the CCE corrected SCAN data (CCE@SCAN) for the test set of 71 ternary oxides. The MAE is about an order of magnitude higher, *i.e.* 244 *vs.* 27 meV/atom, when using no *ab-initio* data as input, indicating that the information obtained from DFT is essential for CCE: predicting the stability of a compound only from nearest neighbor interactions (CCE@exp) is not a very good approximation. However, the DFT error for the formation enthalpy appears to be rather well reproduced from only nearest neighbor contributions.

## Appendix F: Tables with numerical data

TABLE A.V: **Structural data.** ICSD numbers, space group numbers and Pearson symbols for the structures of 79 binary and 71 ternary oxides. Space-groups and Pearson symbols are calculated with AFLOW-SYM [55].

| formula | ICSD # | space group # | Pearson symbol | formula | ICSD # | space group # | Pearson symbol |
|---|---|---|---|---|---|---|---|
| $Li_2O$ | 642216 | 225 | cF12 | $Ag_2CrO_4$ | 16298 | 62 | oP28 |
| $Li_2O_2$ | 152183 | 194 | hP8 | $Al_2SiO_5$ (kya.) | 85742 | 2 | aP32 |
| BeO | 62736 | 186 | hP4 | $Al_2SiO_5$ (and.) | 30679 | 58 | oP32 |
| $Na_2O$ | 60435 | 225 | cF12 | $BaAl_2O_4$ | 246027 | 173 | hP56 |
| $Na_2O_2$ | 25526 | 189 | hP12 | $BaMoO_4$ | 50821 | 88 | tI24 |
| $NaO_2$ | 87178 | 205 | cP12 | $BaTiO_3$ | 187292 | 221 | cP5 |
| MgO | 159378 | 225 | cF8 | $BaZrO_3$ | 27048 | 221 | cP5 |
| $K_2O$ | 180571 | 225 | cF12 | $BeAl_2O_4$ | 34806 | 62 | oP28 |
| $K_2O_2$ | 180559 | 64 | oS16 | $Be_2SiO_4$ | 64945 | 148 | hR42 |
| $KO_2$ | 38245 | 139 | tI6 | $CaAl_4O_7$ | 14270 | 15 | mS48 |
| CaO | 60704 | 225 | cF8 | $CaMoO_4$ | 417513 | 88 | tI24 |
| $Rb_2O$ | 77906 | 225 | cF12 | $Ca_3SiO_5$ | 81100 | 8 | mS54 |
| SrO | 181061 | 225 | cF8 | $CaSiO_3$ (wol.) | 201537 | 2 | aP30 |
| $SrO_2$ | 647474 | 139 | tI6 | $CaSiO_3$ (ps-wol.) | 87694 | 15 | mS60 |
| $Cs_2O$ | 27919 | 166 | hR3 | $CaTiO_3$ | 165801 | 62 | oP20 |
| $CsO_2$ | 38247 | 139 | tI6 | $Ca_2V_2O_7$ | 421266 | 2 | aP22 |
| BaO | 26961 | 225 | cF8 | $CaV_2O_6$ | 166516 | 12 | mS18 |

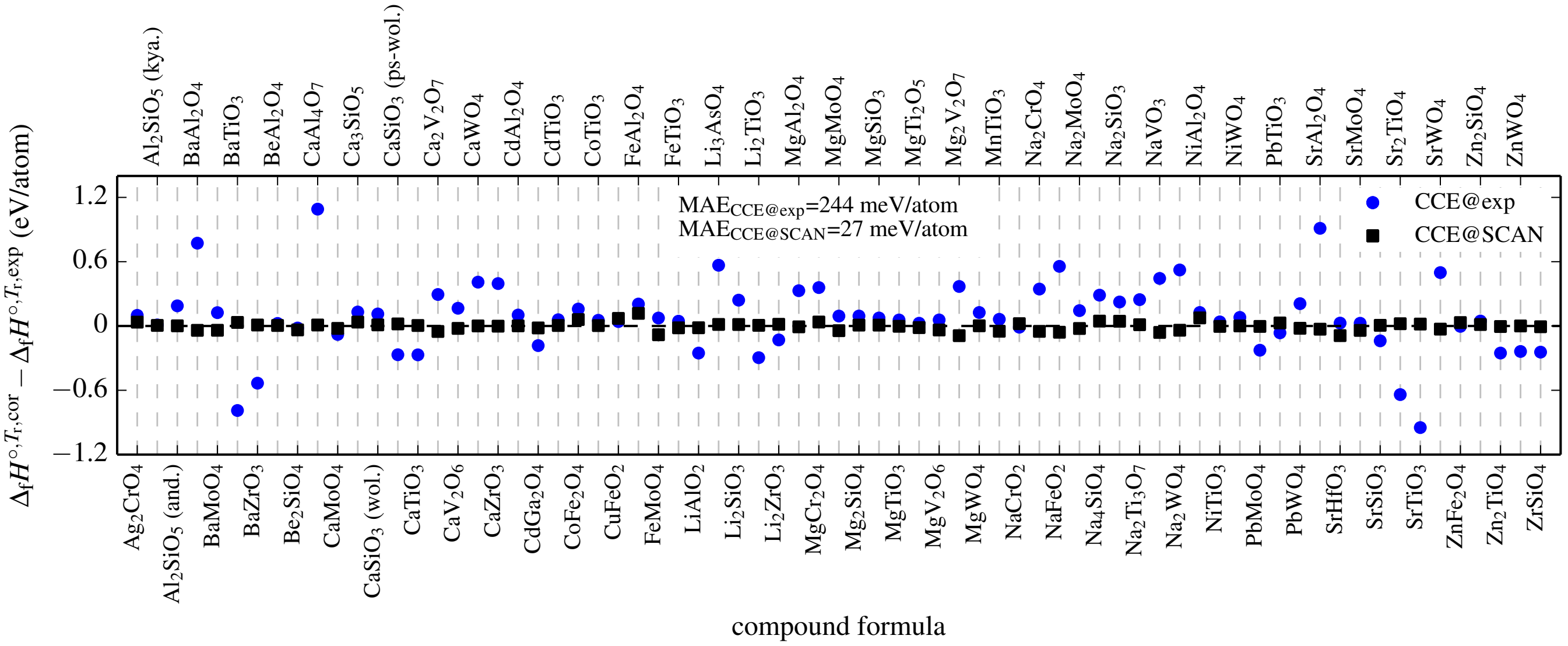

FIG. A.10. **Importance of input data for CCE.** Differences between corrected and experimental room temperature formation enthalpies for the test set of 71 ternary oxides if the CCE method is based on only experimental data for binary oxides (CCE@exp), or applied on top of calculated SCAN results (CCE@SCAN).

TABLE A.V. (*continued*)

| formula | ICSD # | space group # | Pearson symbol | formula | ICSD # | space group # | Pearson symbol |
|---|---|---|---|---|---|---|---|
| $BaO_2$ | 24248 | 139 | tI6 | $CaWO_4$ | 60547 | 88 | tI24 |
| $B_2O_3$ | 24649 | 144 | hP15 | $CaZrO_3$ | 97463 | 62 | oP20 |
| $Al_2O_3$ | 89664 | 167 | hR10 | $CdAl_2O_4$ | 183382 | 227 | cF56 |
| $SiO_2$ ($\alpha$-qua.) | 79635 | 152 | hP9 | $CdGa_2O_4$ | 159739 | 227 | cF56 |
| $SiO_2$ ($\alpha$-crist.) | 153886 | 92 | tP12 | $CdTiO_3$ | 15989 | 148 | hR10 |
| $Ga_2O_3$ | 83645 | 12 | mS20 | $CoFe_2O_4$ | 166200 | 227 | cF56 |
| $GeO_2$ | 158597 | 136 | tP6 | $CoTiO_3$ | 48107 | 148 | hR10 |
| $As_2O_5$ | 654040 | 19 | oP28 | $CuFeO_2$ | 246912 | 166 | hR4 |
| $SeO_2$ | 59712 | 135 | tP24 | $FeAl_2O_4$ | 56117 | 227 | cF56 |
| $In_2O_3$ | 33649 | 206 | cI80 | $FeMoO_4$ | 43012 | 12 | mS48 |
| SnO | 16481 | 129 | tP4 | $FeTiO_3$ | 187688 | 148 | hR10 |
| $SnO_2$ | 184420 | 136 | tP6 | $LiAlO_2$ | 28288 | 166 | hR4 |
| $Sb_2O_3$ | 31102 | 227 | cF80 | $Li_3AsO_4$ | 75927 | 31 | oP16 |
| $SbO_2$ | 4109 | 33 | oP24 | $Li_2SiO_3$ | 28192 | 36 | oS24 |
| $Sb_2O_5$ | 1422 | 15 | mS28 | $Li_2TiO_3$ | 162215 | 15 | mS48 |
| $TeO_2$ | 161691 | 92 | tP12 | $Li_2ZrO_3$ | 94893 | 15 | mS24 |
| $Tl_2O$ | 16220 | 166 | hR6 | $MgAl_2O_4$ | 24766 | 227 | cF56 |
| $Tl_2O_3$ | 26813 | 206 | cI80 | $MgCr_2O_4$ | 160953 | 227 | cF56 |
| PbO | 53927 | 129 | tP4 | $MgMoO_4$ | 20418 | 12 | mS48 |
| $Pb_3O_4$ | 29094 | 135 | tP28 | $Mg_2SiO_4$ | 83793 | 62 | oP28 |
| $PbO_2$ | 34234 | 136 | tP6 | $MgSiO_3$ | 30895 | 14 | mP40 |
| $Bi_2O_3$ | 168806 | 14 | mP20 | $MgTiO_3$ | 171792 | 148 | hR10 |
| $Sc_2O_3$ | 647398 | 206 | cI80 | $MgTi_2O_5$ | 37232 | 63 | oS32 |
| TiO | 56694 | 12 | mS20 | $MgV_2O_6$ | 10391 | 12 | mS18 |
| $Ti_2O_3$ | 647550 | 167 | hR10 | $Mg_2V_2O_7$ | 2321 | 2 | aP22 |
| $Ti_3O_5$ | 75193 | 12 | mS32 | $MgWO_4$ | 67903 | 13 | mP12 |
| $TiO_2$ (rut.) | 33837 | 136 | tP6 | $MnTiO_3$ | 171579 | 148 | hR10 |
| $TiO_2$ (ana.) | 154602 | 141 | tI12 | $NaCrO_2$ | 182235 | 166 | hR4 |
| VO | 647627 | 225 | cF8 | $Na_2CrO_4$ | 76001 | 63 | oS28 |
| $V_2O_3$ | 94768 | 167 | hR10 | $NaFeO_2$ | 186309 | 33 | oP16 |
| $VO_2$ | 647610 | 14 | mP12 | $Na_2MoO_4$ | 151970 | 227 | cF56 |
| $V_2O_5$ | 60767 | 59 | oP14 | $Na_4SiO_4$ | 15500 | 2 | aP18 |
| $Cr_2O_3$ | 167268 | 167 | hR10 | $Na_2SiO_3$ | 24664 | 36 | oS24 |
| $CrO_3$ | 16031 | 40 | oS16 | $Na_2Ti_3O_7$ | 15463 | 11 | mP24 |
| MnO | 53928 | 225 | cF8 | $NaVO_3$ | 2103 | 15 | mS40 |

TABLE A.V. (*continued*)

| formula | ICSD # | space group # | Pearson symbol | formula | ICSD # | space group # | Pearson symbol |
|---|---|---|---|---|---|---|---|
| $MnO_2$ | 20229 | 136 | tP6 | $Na_2WO_4$ | 44524 | 227 | cF56 |
| FeO | 31081 | 225 | cF8 | $NiAl_2O_4$ | 608815 | 227 | cF56 |
| $Fe_2O_3$ | 201097 | 167 | hR10 | $NiTiO_3$ | 33855 | 148 | hR10 |
| CoO | 9865 | 225 | cF8 | $NiWO_4$ | 16685 | 13 | mP12 |
| NiO | 166115 | 225 | cF8 | $PbMoO_4$ | 89034 | 88 | tI24 |
| $Cu_2O$ | 628619 | 224 | cP6 | $PbTiO_3$ | 162046 | 99 | tP5 |
| CuO | 92368 | 15 | mS8 | $PbWO_4$ | 93374 | 88 | tI24 |
| ZnO | 163380 | 186 | hP4 | $SrAl_2O_4$ | 160297 | 4 | mP28 |
| $Y_2O_3$ | 27772 | 206 | cI80 | $SrHfO_3$ | 84280 | 62 | oP20 |
| $ZrO_2$ | 80043 | 14 | mP12 | $SrMoO_4$ | 23700 | 88 | tI24 |
| NbO | 27574 | 221 | cP6 | $SrSiO_3$ | 32542 | 5 | mS60 |
| $MoO_2$ | 644064 | 14 | mP12 | $Sr_2TiO_4$ | 157402 | 139 | tI14 |
| $MoO_3$ | 151751 | 62 | oP16 | $SrTiO_3$ | 187296 | 221 | cP5 |
| $RuO_2$ | 647377 | 136 | tP6 | $SrWO_4$ | 23701 | 88 | tI24 |
| $Rh_2O_3$ | 181829 | 167 | hR10 | $ZnFe_2O_4$ | 91935 | 227 | cF56 |
| PdO | 26598 | 131 | tP4 | $Zn_2SiO_4$ | 20093 | 148 | hR42 |
| $Ag_2O$ | 174091 | 224 | cP6 | $Zn_2TiO_4$ | 109093 | 91 | tP28 |
| CdO | 181294 | 225 | cF8 | $ZnWO_4$ | 156482 | 13 | mP12 |
| $HfO_2$ | 173158 | 14 | mP12 | $ZrSiO_4$ | 71942 | 141 | tI24 |
| $WO_2$ | 8217 | 14 | mP12 | | | | |
| $WO_3$ | 84848 | 14 | mP16 | | | | |
| $ReO_2$ | 647349 | 14 | mP12 | | | | |
| $ReO_3$ | 16810 | 221 | cP4 | | | | |
| $OsO_2$ | 647244 | 136 | tP6 | | | | |
| $OsO_4$ | 23803 | 15 | mS20 | | | | |
| $IrO_2$ | 640887 | 136 | tP6 | | | | |
| HgO | 14124 | 62 | oP8 | | | | |

TABLE A.VI: **Structural data for halides.** ICSD numbers, space group numbers and Pearson symbols for the structures of 9 binary and 7 ternary halides. Space-groups and Pearson symbols are calculated with AFLOW-SYM [55].

| formula | ICSD # | space group # | Pearson symbol | formula | ICSD # | space group # | Pearson symbol |
|---|---|---|---|---|---|---|---|
| LiF | 44272 | 225 | cF8 | $KBF_4$ | 22260 | 62 | oP24 |
| NaF | 52238 | 225 | cF8 | $Li_3AlF_6$ | 85171 | 15 | mS120 |
| KF | 64686 | 225 | cF8 | $Li_2BeF_4$ | 72423 | 148 | hR42 |
| $BeF_2$ | 261194 | 152 | hP9 | $Na_3AlF_6$ | 30201 | 14 | mP20 |
| $AlF_3$ | 36034 | 167 | hR8 | $NaBF_4$ | 30435 | 63 | oS24 |
| NaCl | 162800 | 225 | cF8 | $Na_2SiF_6$ | 61274 | 150 | hP27 |
| KCl | 240527 | 225 | cF8 | $NaAlCl_4$ | 71159 | 19 | oP24 |
| $CaCl_2$ | 246416 | 58 | oP6 | | | | |
| $AlCl_3$ | 39566 | 12 | mS16 | | | | |

TABLE A.VII: **Formation enthalpies and CCE corrections for binary oxides.** Calculated and experimental room temperature formation enthalpies, corrections per bond $\delta H_{A-\mathrm{O}}^{A^{+\alpha}}$, number of cation-oxygen bonds $N_{A-\mathrm{O}}$ per formula unit and oxidation states $+\alpha$ of the cation for binary oxides. Experimental values from Kubaschewski *et al.* [15], NIST-JANAF [16] and Barin [17]. Values in brackets denote cases in which the corrections are applied to calculate corrected values (see section III B of the main text for details). The O-O bond correction for peroxides is obtained from $Li_2O_2$. The O-O bond correction for superoxides is derived from $KO_2$. The two values are listed in the table at the position of the respective compound. Enthalpies in eV/atom; corrections in eV/bond.

| formula | PBE+AGL | LDA+AGL | SCAN+AGL | Exp. | $\delta H_{A-\mathrm{O}}^{A^{+\alpha}}$ PBE+AGL | $\delta H_{A-\mathrm{O}}^{A^{+\alpha}}$ LDA+AGL | $\delta H_{A-\mathrm{O}}^{A^{+\alpha}}$ SCAN+AGL | $N_{A-\mathrm{O}}$ | $+\alpha$ |
|---|---|---|---|---|---|---|---|---|---|
| $Li_2O$ | −1.850 | −2.092 | −2.083 | −2.066 [15] | 0.0809 | −0.0100 | −0.0065 | 8 | +1 |
| $Li_2O_2$ | −1.426 | −1.704 | −1.605 | −1.643 [15] | −0.1050 | −0.1270 | 0.2270 | 12 | +1 |
| BeO | −2.751 | −3.122 | −3.126 | −3.158 [15] | 0.2035 | 0.0180 | 0.0160 | 4 | +2 |
| $Na_2O$ | −1.224 | −1.453 | −1.471 | −1.444 [16] | 0.0826 | −0.0033 | −0.0101 | 8 | +1 |
| $Na_2O_2$ | −1.083 | −1.350 | −1.285 | −1.329 [15] | (−1.304) | (−1.309) | (−1.311) | 12 | +1 |
| $NaO_2$ | −0.795 | −1.014 | −0.893 | −0.901 [15] | (−0.777) | (−0.918) | (−0.856) | 6 | +1 |
| MgO | −2.705 | −3.096 | −3.111 | −3.118 [15] | 0.1373 | 0.0072 | 0.0023 | 6 | +2 |
| $K_2O$ | −1.036 | −1.326 | −1.266 | −1.255 [15] | 0.0821 | −0.0269 | −0.0041 | 8 | +1 |
| $K_2O_2$ | −1.074 | −1.387 | −1.256 | −1.282 [15] | (−1.294) | (−1.275) | (−1.300) | 12 | +1 |

TABLE A.VII. (*continued*)

| formula | PBE+AGL | LDA+AGL | SCAN+AGL | Exp. | $\delta H^{A+\alpha}_{A-\mathrm{O}}$ PBE+AGL | $\delta H^{A+\alpha}_{A-\mathrm{O}}$ LDA+AGL | $\delta H^{A+\alpha}_{A-\mathrm{O}}$ SCAN+AGL | $N_{A-\mathrm{O}}$ | $+\alpha$ |
|---|---|---|---|---|---|---|---|---|---|
| $KO_2$ | −0.893 | −1.162 | −1.014 | −0.983 [15] | −0.5500 | −0.2680 | −0.0520 | 10 | +1 |
| CaO | −2.969 | −3.382 | −3.351 | −3.290 [15] | 0.1070 | −0.0308 | −0.0203 | 6 | +2 |
| $Rb_2O$ | −0.922 | −1.232 | −1.161 | −1.171 [15] | 0.0934 | −0.0229 | 0.0035 | 8 | +1 |
| SrO | −2.746 | −3.127 | −3.116 | −3.068 [15] | 0.1073 | −0.0195 | −0.0160 | 6 | +2 |
| $SrO_2$ | −1.902 | −2.314 | −2.177 | −2.189 [15] | (−2.224) | (−2.207) | (−2.199) | 10 | +2 |
| $Cs_2O$ | −0.999 | −1.313 | −1.210 | −1.195 [15] | 0.0983 | −0.0588 | −0.0073 | 6 | +1 |
| $CsO_2$ | −0.899 | −1.175 | −1.011 | −0.989 [15] | (−1.043) | (−0.890) | (−0.969) | 10 | +1 |
| BaO | −2.490 | −2.815 | −2.832 | −2.841 [15] | 0.1167 | 0.0085 | 0.0028 | 6 | +2 |
| $BaO_2$ | −1.866 | −2.255 | −2.128 | −2.191 [15] | (−2.220) | (−2.241) | (−2.213) | 10 | +2 |
| $B_2O_3$ | −2.396 | −2.708 | −2.683 | −2.640 [15] | 0.2030 | −0.0572 | −0.0357 | 6 | +3 |
| $Al_2O_3$ | −3.005 | −3.469 | −3.480 | −3.473 [15] | 0.1950 | 0.0020 | −0.0028 | 12 | +3 |
| $SiO_2$ ($\alpha$-qua.) | −2.794 | −3.160 | −3.156 | −3.147 [15] | 0.2648 | −0.0098 | −0.0070 | 4 | +4 |
| $SiO_2$ ($\alpha$-crist.) | −2.804 | −3.151 | −3.151 | −3.138 [15] | (−3.157) | (−3.138) | (−3.142) | 4 | +4 |
| $Ga_2O_3$ | −1.851 | −2.237 | −2.172 | −2.258 [15] | 0.2034 | 0.0105 | 0.0427 | 10 | +3 |
| $GeO_2$ | −1.598 | −2.064 | −1.912 | −2.004 [15] | 0.2030 | −0.0300 | 0.0457 | 6 | +4 |
| $As_2O_5$ | −1.071 | −1.448 | −1.326 | −1.362 [15] | 0.2039 | −0.0599 | 0.0251 | 10 | +5 |
| $SeO_2$ | −0.705 | −1.004 | −0.874 | −0.778 [15] | 0.0730 | −0.2267 | −0.0960 | 3 | +4 |
| $In_2O_3$ | −1.595 | −1.958 | −1.950 | −1.919 [15] | 0.1349 | −0.0163 | −0.0127 | 12 | +3 |
| SnO | −1.351 | −1.611 | −1.510 | −1.481 [17] | 0.0650 | −0.0653 | −0.0148 | 4 | +2 |
| $SnO_2$ | −1.704 | −2.095 | −2.033 | −2.007 [17] | 0.1512 | −0.0442 | −0.0133 | 6 | +4 |
| $Sb_2O_3$ | −1.343 | −1.622 | −1.511 | −1.484 [15] | 0.1177 | −0.1150 | −0.0218 | 6 | +3 |
| $SbO_2$ | −1.386 | −1.763 | −1.626 | −1.567 [15] | (−1.570) | (−1.556) | (−1.556) | 5 | +3, +5 |
| $Sb_2O_5$ | −1.258 | −1.663 | −1.533 | −1.439 [17] | 0.1056 | −0.1304 | −0.0551 | 12 | +5 |
| $TeO_2$ | −1.036 | −1.389 | −1.236 | −1.117 [17] | 0.0610 | −0.2035 | −0.0893 | 4 | +4 |
| $Tl_2O$ | −0.596 | −0.716 | −0.705 | −0.578 [15] | −0.0090 | −0.0687 | −0.0635 | 6 | +1 |
| $Tl_2O_3$ | −0.686 | −1.038 | −0.847 | −0.809 [15] | 0.0513 | −0.0955 | −0.0159 | 12 | +3 |
| PbO | −1.138 | −1.360 | −1.252 | −1.137 [15] | −0.0005 | −0.1115 | −0.0575 | 4 | +2 |
| $Pb_3O_4$ | −1.062 | −1.316 | −1.187 | −1.064 [15] | (−1.108) | (−1.112) | (−1.115) | 12 | +2, +4 |
| $PbO_2$ | −0.841 | −1.203 | −1.003 | −0.948 [15] | 0.0538 | −0.1272 | −0.0273 | 6 | +4 |
| $Bi_2O_3$ | −1.240 | −1.538 | −1.259 | −1.183 [15] | −0.0286 | −0.1775 | −0.0381 | 10 | +3 |
| $Sc_2O_3$ | −3.558 | −3.964 | −4.006 | −3.956 [15] | 0.1656 | −0.0034 | −0.0212 | 12 | +3 |
| TiO | −2.532 | −2.961 | −2.865 | −2.812 [15] | 0.1169 | −0.0619 | −0.0221 | 4.8 | +2 |
| $Ti_2O_3$ | −2.907 | −3.344 | −3.304 | −3.153 [15] | 0.1025 | −0.0796 | −0.0633 | 12 | +3 |
| $Ti_3O_5$ | −2.966 | −3.360 | −3.364 | −3.186 [15] | (−3.205) | (−3.174) | (−3.183) | 18 | +3, +4 |
| $TiO_2$ (rut.) | −3.034 | −3.438 | −3.491 | −3.261 [15] | 0.1138 | −0.0882 | −0.1150 | 6 | +4 |
| $TiO_2$ (ana.) | −3.067 | −3.449 | −3.502 | −3.243 [16] | (−3.295) | (−3.273) | (−3.272) | 6 | +4 |
| VO | −1.450 | −1.865 | −1.767 | −2.237 [15] | 0.2623 | 0.1240 | 0.1568 | 6 | +2 |
| $V_2O_3$ | −2.282 | −2.672 | −2.646 | −2.526 [15] | 0.1018 | −0.0608 | −0.0498 | 12 | +3 |
| $VO_2$ | −2.360 | −2.749 | −2.758 | −2.466 [15] | 0.0528 | −0.1413 | −0.1462 | 6 | +4 |
| $V_2O_5$ | −2.301 | −2.586 | −2.595 | −2.295 [15] | −0.0037 | −0.2033 | −0.2101 | 10 | +5 |
| $Cr_2O_3$ | −1.979 | −2.233 | −2.390 | −2.352 [15] | 0.1553 | 0.0495 | −0.0159 | 12 | +3 |
| $CrO_3$ | −1.651 | −1.819 | −1.796 | −1.521 [15] | −0.1305 | −0.2980 | −0.2745 | 4 | +6 |
| MnO | −1.247 | −1.190 | −2.095 | −1.994 [15] | 0.2492 | 0.2682 | −0.0333 | 6 | +2 |
| $MnO_2$ | −1.672 | −1.977 | −2.105 | −1.800 [15] | 0.0640 | −0.0885 | −0.1528 | 6 | +4 |
| FeO | −0.877 | −1.016 | −1.347 | −1.410 [16] | 0.1775 | 0.1312 | 0.0210 | 6 | +2 |
| $Fe_2O_3$ | −1.311 | −1.662 | −1.857 | −1.707 [15] | 0.1648 | 0.0187 | −0.0625 | 12 | +3 |
| CoO | −0.512 | −0.736 | −0.859 | −1.232 [15] | 0.2398 | 0.1655 | 0.1243 | 6 | +2 |
| NiO | −0.475 | −0.770 | −0.647 | −1.242 [15] | 0.2555 | 0.1572 | 0.1982 | 6 | +2 |
| $Cu_2O$ | −0.417 | −0.546 | −0.507 | −0.590 [16] | 0.1293 | 0.0328 | 0.0618 | 4 | +1 |
| CuO | −0.605 | −0.858 | −0.793 | −0.808 [16] | 0.1018 | −0.0245 | 0.0075 | 4 | +2 |
| ZnO | −1.445 | −1.730 | −1.723 | −1.817 [15] | 0.1858 | 0.0433 | 0.0468 | 4 | +2 |
| $Y_2O_3$ | −3.622 | −3.999 | −4.063 | −3.949 [15] | 0.1363 | −0.0209 | −0.0474 | 12 | +3 |
| $ZrO_2$ | −3.460 | −3.888 | −3.931 | −3.791 [16] | 0.1419 | −0.0416 | −0.0597 | 7 | +4 |
| NbO | −2.053 | −2.422 | −2.339 | −2.175 [15] | 0.0610 | −0.1235 | −0.0820 | 4 | +2 |
| $MoO_2$ | −1.966 | −2.390 | −2.233 | −2.031 [15] | 0.0327 | −0.1797 | −0.1008 | 6 | +4 |
| $MoO_3$ | −1.975 | −2.264 | −2.180 | −1.931 [15] | −0.0440 | −0.3335 | −0.2490 | 4 | +6 |
| $RuO_2$ | −1.059 | −1.458 | −1.271 | −1.054 [15] | −0.0027 | −0.2020 | −0.1085 | 6 | +4 |
| $Rh_2O_3$ | −0.710 | −1.060 | −0.874 | −0.737 [15] | 0.0115 | −0.1347 | −0.0572 | 12 | +3 |
| PdO | −0.485 | −0.757 | −0.648 | −0.599 [15] | 0.0568 | −0.0793 | −0.0250 | 4 | +2 |
| $Ag_2O$ | −0.123 | −0.183 | −0.198 | −0.107 [15] | −0.0115 | −0.0568 | −0.0683 | 4 | +1 |
| CdO | −1.028 | −1.292 | −1.324 | −1.339 [15] | 0.1037 | 0.0158 | 0.0050 | 6 | +2 |

TABLE A.VII. (*continued*)

| formula | PBE+AGL | LDA+AGL | SCAN+AGL | Exp. | $\delta H_{A-\mathrm{O}}^{A+\alpha}$ PBE+AGL | $\delta H_{A-\mathrm{O}}^{A+\alpha}$ LDA+AGL | $\delta H_{A-\mathrm{O}}^{A+\alpha}$ SCAN+AGL | $N_{A-\mathrm{O}}$ | $+\alpha$ |
|---|---|---|---|---|---|---|---|---|---|
| $HfO_2$ | −3.577 | −4.022 | −4.016 | −3.955 [17] | 0.1617 | −0.0290 | −0.0263 | 7 | +4 |
| $WO_2$ | −1.923 | −2.352 | −2.131 | −2.037 [15] | 0.0570 | −0.1575 | −0.0473 | 6 | +4 |
| $WO_3$ | −2.176 | −2.493 | −2.385 | −2.183 [15] | 0.0052 | −0.2063 | −0.1347 | 6 | +6 |
| $ReO_2$ | −1.371 | −1.797 | −1.507 | −1.551 [17] | 0.0898 | −0.1230 | 0.0218 | 6 | +4 |
| $ReO_3$ | −1.635 | −1.980 | −1.764 | −1.526 [17] | −0.0722 | −0.3027 | −0.1582 | 6 | +6 |
| $OsO_2$ | −0.895 | −1.305 | −1.015 | −1.018 [15] | 0.0613 | −0.1438 | 0.0015 | 6 | +4 |
| $OsO_4$ | −0.999 | −1.120 | −1.040 | −0.816 [15] | −0.2288 | −0.3805 | −0.2803 | 4 | +8 |
| $IrO_2$ | −0.799 | −1.200 | −0.803 | −0.838 [17] | 0.0198 | −0.1808 | 0.0177 | 6 | +4 |
| HgO | −0.319 | −0.557 | −0.433 | −0.470 [15] | 0.1515 | −0.0870 | 0.0375 | 2 | +2 |

TABLE A.VIII: **Formation enthalpies and CCE corrections for binary halides.** Calculated and experimental room temperature formation enthalpies, corrections per bond $\delta H_{A-X}^{+\alpha}$, number of cation-anion bonds $N_{A-X}$ per formula unit and oxidation states $+\alpha$ of the cation for binary halides. Experimental values from Kubaschewski *et al.* [15] and NIST-JANAF [16]. Note that $BF_3$ and $SiF_4$ are gaseous molecular systems. Enthalpies in eV/atom; corrections in eV/bond.

| formula | PBE+AGL | LDA+AGL | SCAN+AGL | Exp. | $\delta H_{A-X}^{A+\alpha}$ PBE+AGL | $\delta H_{A-X}^{A+\alpha}$ LDA+AGL | $\delta H_{A-X}^{A+\alpha}$ SCAN+AGL | $N_{A-X}$ | $+\alpha$ |
|---|---|---|---|---|---|---|---|---|---|
| LiF | −2.960 | −3.161 | −3.342 | −3.197 [16] | 0.0788 | 0.0120 | −0.0482 | 6 | +1 |
| NaF | −2.732 | −2.904 | −3.124 | −2.982 [16] | 0.0833 | 0.0258 | −0.0473 | 6 | +1 |
| KF | −2.731 | −2.922 | −3.088 | −2.946 [16] | 0.0718 | 0.0083 | −0.0472 | 6 | +1 |
| $BeF_2$ | −3.271 | −3.472 | −3.709 | −3.547 [15] | 0.2073 | 0.0563 | −0.1215 | 4 | +2 |
| $BF_3$(g) | −2.760 | −2.824 | −3.067 | −2.943 [15] | 0.2447 | 0.1587 | −0.1640 | 3 | +3 |
| $AlF_3$ | −3.540 | −3.828 | −4.095 | −3.913 [15] | 0.2488 | 0.0572 | −0.1208 | 6 | +3 |
| $SiF_4$(g) | −3.097 | −3.208 | −3.470 | −3.348 [15] | 0.3135 | 0.1750 | −0.1525 | 4 | +4 |
| NaCl | −1.832 | −1.961 | −2.077 | −2.131 [15] | 0.1000 | 0.0568 | 0.0180 | 6 | +1 |
| KCl | −1.981 | −2.116 | −2.212 | −2.263 [15] | 0.0938 | 0.0488 | 0.0168 | 6 | +1 |
| $CaCl_2$ | −2.425 | −2.598 | −2.701 | −2.747 [15] | 0.1608 | 0.0742 | 0.0227 | 6 | +2 |
| $AlCl_3$ | −1.538 | −1.753 | −1.755 | −1.828 [15] | 0.1933 | 0.0498 | 0.0485 | 6 | +3 |

TABLE A.IX: **Uncorrected and CCE corrected formation enthalpies for ternary oxides.** Calculated, coordination corrected and experimental room temperature formation enthalpies, number of cation-oxygen bonds $N_{1/2-\mathrm{O}}$ per formula unit and oxidation states $+\alpha$ of the cations for ternary oxides. For the cation-oxygen bonds- and oxidation numbers, the first (second) number in the column refers to the first (second) element in the formula. Experimental values from Kubaschewski *et al.* [15], NIST-JANAF [16] and Barin [17]. Enthalpies in eV/atom.

| formula | PBE+AGL | LDA+AGL | SCAN+AGL | PBE+AGL CCE | LDA+AGL CCE | SCAN+AGL CCE | Exp. | $N_{1/2-\mathrm{O}}$ | $+\alpha$ |
|---|---|---|---|---|---|---|---|---|---|
| $Ag_2CrO_4$ | −1.104 | −1.285 | −1.302 | −1.013 | −1.033 | −1.047 | −1.083 [17] | 10, 4 | +1, +6 |
| $Al_2SiO_5$ (kya.) | −2.917 | −3.365 | −3.363 | −3.342 | −3.364 | −3.356 | −3.361 [16] | 12, 4 | +3, +4 |
| $Al_2SiO_5$ (and.) | −2.940 | −3.354 | −3.364 | −3.340 | −3.352 | −3.357 | −3.358 [16] | 11, 4 | +3, +4 |
| $BaAl_2O_4$ | −3.065 | −3.402 | −3.482 | −3.413 | −3.413 | −3.482 | −3.441 [15] | 7.5, 8 | +2, +3 |
| $BaMoO_4$ | −2.616 | −2.872 | −2.876 | −2.742 | −2.661 | −2.714 | −2.674 [17] | 8, 4 | +2, +6 |
| $BaTiO_3$ | −3.090 | −3.526 | −3.538 | −3.506 | −3.440 | −3.407 | −3.441 [17] | 12, 6 | +2, +4 |
| $BaZrO_3$ | −3.302 | −3.696 | −3.744 | −3.752 | −3.667 | −3.679 | −3.689 [17] | 12, 6 | +2, +4 |
| $BeAl_2O_4$ | −2.953 | −3.385 | −3.398 | −3.403 | −3.399 | −3.402 | −3.407 [15] | 4, 12 | +2, +3 |
| $Be_2SiO_4$ | −2.778 | −3.161 | −3.157 | −3.162 | −3.176 | −3.171 | −3.134 [16] | 8, 4 | +2, +4 |
| $CaAl_4O_7$ | −3.064 | −3.413 | −3.480 | −3.368 | −3.403 | −3.468 | −3.477 [17] | 5, 16 | +2, +3 |
| $CaMoO_4$ | −2.594 | −2.915 | −2.886 | −2.707 | −2.652 | −2.693 | −2.671 [15] | 8, 4 | +2, +6 |
| $Ca_3SiO_5$ | −3.002 | −3.395 | −3.380 | −3.334 | −3.329 | −3.336 | −3.373 [15] | 18, 4 | +2, +4 |
| $CaSiO_3$ (wol.) | −3.038 | −3.412 | −3.410 | −3.385 | −3.366 | −3.379 | −3.389 [15] | $6.\overline{3}$, 4 | +2, +4 |
| $CaSiO_3$ (ps-wol.) | −3.018 | −3.391 | −3.393 | −3.401 | −3.334 | −3.355 | −3.375 [15] | 8, 4 | +2, +4 |
| $CaTiO_3$ | −3.163 | −3.591 | −3.608 | −3.471 | −3.436 | −3.438 | −3.442 [15] | 8, 6 | +2, +4 |
| $Ca_2V_2O_7$ | −2.804 | −3.118 | −3.152 | −2.928 | −2.915 | −2.956 | −2.905 [15] | 13, 9 | +2, +5 |
| $CaV_2O_6$ | −2.642 | −2.918 | −2.951 | −2.709 | −2.671 | −2.704 | −2.682 [15] | 6, 10 | +2, +5 |
| $CaWO_4$ | −2.661 | −2.984 | −2.958 | −2.807 | −2.805 | −2.841 | −2.842 [17] | 8, 4 | +2, +6 |
| $CaZrO_3$ | −3.329 | −3.727 | −3.759 | −3.628 | −3.640 | −3.663 | −3.662 [15] | 6, 6 | +2, +4 |
| $CdAl_2O_4$ | −2.438 | −2.834 | −2.838 | −2.831 | −2.846 | −2.837 | −2.838 [15] | 4, 12 | +2, +3 |
| $CdGa_2O_4$ | −1.645 | −1.999 | −1.949 | −2.053 | −2.026 | −2.025 | −2.007 [15] | 4, 12 | +2, +3 |
| $CdTiO_3$ | −2.289 | −2.629 | −2.678 | −2.550 | −2.542 | −2.546 | −2.551 [17] | 6, 6 | +2, +4 |
| $CoFe_2O_4$ | −1.143 | −1.458 | −1.588 | −1.562 | −1.585 | −1.552 | −1.612 [15] | 4, 12 | +2, +3 |

TABLE A.IX. (*continued*)

| formula | PBE+AGL | LDA+AGL | SCAN+AGL | PBE+AGL CCE | LDA+AGL CCE | SCAN+AGL CCE | Exp. | $N_{1/2-O}$ | $+\alpha$ |
|---|---|---|---|---|---|---|---|---|---|
| $CoTiO_3$ | −2.089 | −2.418 | −2.488 | −2.514 | −2.511 | −2.499 | −2.503 [15] | 6, 6 | +2, +4 |
| $CuFeO_2$ | −0.920 | −1.275 | −1.321 | −1.232 | −1.319 | −1.258 | −1.329 [15] | 2, 6 | +1, +3 |
| $FeAl_2O_4$ | −2.400 | −2.740 | −2.829 | −2.836 | −2.819 | −2.836 | −2.954 [17] | 4, 12 | +2, +3 |
| $FeMoO_4$ | −1.637 | −2.036 | −2.058 | −1.785 | −1.945 | −1.913 | −1.831 [15] | 6, 4 | +2, +6 |
| $FeTiO_3$ | −2.231 | −2.506 | −2.695 | −2.580 | −2.557 | −2.582 | −2.565 [15] | 6, 6 | +2, +4 |
| $LiAlO_2$ | −2.720 | −3.101 | −3.111 | −3.133 | −3.089 | −3.096 | −3.080 [15] | 6, 6 | +1, +3 |
| $Li_3AsO_4$ | −1.939 | −2.200 | −2.188 | −2.162 | −2.155 | −2.191 | −2.205 [17] | 12, 4 | +1, +5 |
| $Li_2SiO_3$ | −2.545 | −2.848 | −2.847 | −2.829 | −2.828 | −2.834 | −2.848 [15] | 8, 4 | +1, +4 |
| $Li_2TiO_3$ | −2.653 | −2.986 | −3.006 | −2.929 | −2.878 | −2.878 | −2.884 [15] | 12, 6 | +1, +4 |
| $Li_2ZrO_3$ | −2.759 | −3.076 | −3.101 | −3.063 | −3.015 | −3.028 | −3.044 [15] | 12, 6 | +1, +4 |
| $MgAl_2O_4$ | −2.976 | −3.393 | −3.415 | −3.388 | −3.401 | −3.412 | −3.404 [16] | 4, 12 | +2, +3 |
| $MgCr_2O_4$ | −2.225 | −2.458 | −2.621 | −2.569 | −2.547 | −2.595 | −2.632 [15] | 4, 12 | +2, +3 |
| $MgMoO_4$ | −2.371 | −2.634 | −2.626 | −2.479 | −2.419 | −2.463 | −2.419 [15] | 6, 4 | +2, +6 |
| $Mg_2SiO_4$ | −2.825 | −3.203 | −3.215 | −3.212 | −3.209 | −3.215 | −3.223 [15] | 12, 4 | +2, +4 |
| $MgSiO_3$ | −2.812 | −3.203 | −3.204 | −3.188 | −3.203 | −3.202 | −3.210 [15] | 6, 4 | +2, +4 |
| $MgTiO_3$ | −2.971 | −3.357 | −3.398 | −3.272 | −3.260 | −3.263 | −3.260 [15] | 6, 6 | +2, +4 |
| $MgTi_2O_5$ | −3.018 | −3.389 | −3.437 | −3.291 | −3.262 | −3.266 | −3.251 [16] | 6, 12 | +2, +4 |
| $MgV_2O_6$ | −2.479 | −2.783 | −2.804 | −2.567 | −2.562 | −2.572 | −2.534 [15] | 6, 10 | +2, +5 |
| $Mg_2V_2O_7$ | −2.573 | −2.866 | −2.911 | −2.720 | −2.726 | −2.761 | −2.671 [15] | 12, 8 | +2, +5 |
| $MgWO_4$ | −2.446 | −2.801 | −2.751 | −2.589 | −2.602 | −2.619 | −2.621 [15] | 6, 6 | +2, +6 |
| $MnTiO_3$ | −2.441 | −2.639 | −3.044 | −2.877 | −2.855 | −2.866 | −2.817 [15] | 6, 6 | +2, +4 |
| $NaCrO_2$ | −1.900 | −2.118 | −2.288 | −2.257 | −2.187 | −2.249 | −2.271 [15] | 6, 6 | +1, +3 |
| $Na_2CrO_4$ | −1.976 | −2.179 | −2.209 | −2.019 | −2.004 | −2.037 | −1.987 [17] | 10, 4 | +1, +6 |
| $NaFeO_2$ | −1.552 | −1.714 | −1.940 | −1.800 | −1.730 | −1.867 | −1.809 [15] | 4, 4 | +1, +3 |
| $Na_2MoO_4$ | −2.120 | −2.346 | −2.358 | −2.236 | −2.149 | −2.198 | −2.175 [17] | 12, 4 | +1, +6 |
| $Na_4SiO_4$ | −2.093 | −2.373 | −2.397 | −2.376 | −2.362 | −2.374 | −2.420 [15] | 18, 4 | +1, +4 |
| $Na_2SiO_3$ | −2.361 | −2.657 | −2.677 | −2.675 | −2.645 | −2.656 | −2.700 [15] | 10, 4 | +1, +4 |
| $Na_2Ti_3O_7$ | −2.787 | −3.116 | −3.167 | −3.017 | −2.989 | −2.995 | −3.007 [15] | 10, 17 | +1, +4 |
| $NaVO_3$ | −2.345 | −2.556 | −2.619 | −2.442 | −2.389 | −2.439 | −2.379 [17] | 6, 4 | +1, +5 |
| $Na_2WO_4$ | −2.175 | −2.397 | −2.417 | −2.319 | −2.273 | −2.322 | −2.283 [15] | 12, 4 | +1, +6 |
| $NiAl_2O_4$ | −2.279 | −2.641 | −2.659 | −2.759 | −2.735 | −2.768 | −2.844 [15] | 4, 12 | +2, +3 |
| $NiTiO_3$ | −2.070 | −2.398 | −2.396 | −2.513 | −2.481 | −2.496 | −2.492 [15] | 6, 6 | +2, +4 |
| $NiWO_4$ | −1.659 | −1.980 | −1.886 | −1.920 | −1.931 | −1.949 | −1.950 [15] | 6, 6 | +2, +6 |
| $PbMoO_4$ | −1.893 | −2.134 | −2.066 | −1.863 | −1.763 | −1.823 | −1.819 [15] | 8, 4 | +2, +6 |
| $PbTiO_3$ | −2.350 | −2.685 | −2.655 | −2.463 | −2.418 | −2.448 | −2.476 [15] | 8, 5 | +2, +4 |
| $PbWO_4$ | −1.948 | −2.187 | −2.125 | −1.951 | −1.901 | −1.958 | −1.937 [15] | 8, 4 | +2, +6 |
| $SrAl_2O_4$ | −3.074 | −3.419 | −3.489 | −3.389 | −3.405 | −3.472 | −3.442 [17] | 6, 8 | +2, +3 |
| $SrHfO_3$ | −3.416 | −3.826 | −3.844 | −3.782 | −3.760 | −3.787 | −3.697 [15] | 8, 6 | +2, +4 |
| $SrMoO_4$ | −2.624 | −2.918 | −2.905 | −2.738 | −2.670 | −2.717 | −2.676 [15] | 8, 4 | +2, +6 |
| $SrSiO_3$ | −3.039 | −3.401 | −3.411 | −3.422 | −3.362 | −3.380 | −3.386 [15] | 8, 4 | +2, +4 |
| $Sr_2TiO_4$ | −3.074 | −3.492 | −3.506 | −3.447 | −3.366 | −3.366 | −3.387 [15] | 18, 6 | +2, +4 |
| $SrTiO_3$ | −3.164 | −3.596 | −3.622 | −3.558 | −3.444 | −3.445 | −3.463 [15] | 12, 6 | +2, +4 |
| $SrWO_4$ | −2.688 | −2.981 | −2.973 | −2.834 | −2.818 | −2.862 | −2.832 [17] | 8, 4 | +2, +6 |
| $ZnFe_2O_4$ | −1.288 | −1.669 | −1.783 | −1.676 | −1.726 | −1.703 | −1.735 [17] | 4, 12 | +2, +3 |
| $Zn_2SiO_4$ | −2.054 | −2.376 | −2.370 | −2.417 | −2.420 | −2.419 | −2.433 [15] | 8, 4 | +2, +4 |
| $Zn_2TiO_4$ | −2.115 | −2.457 | −2.479 | −2.478 | −2.444 | −2.448 | −2.443 [15] | 10, 6 | +2, +4 |
| $ZnWO_4$ | −1.937 | −2.282 | −2.214 | −2.128 | −2.119 | −2.126 | −2.126 [15] | 6, 6 | +2, +6 |
| $ZrSiO_4$ | −3.146 | −3.579 | −3.587 | −3.511 | −3.518 | −3.503 | −3.496 [16] | 8, 4 | +4, +4 |

TABLE A.X: **Uncorrected and CCE corrected formation enthalpies for ternary halides.** Calculated, coordination corrected and experimental room temperature formation enthalpies, number of cation-anion bonds $N_{1/2-X}$ per formula unit and oxidation states $+\alpha$ of the cations for ternary halides. For the cation-oxygen bonds- and oxidation numbers, the first (second) number in the column refers to the first (second) element in the formula. Experimental values from Kubaschewski *et al.* [15] and NIST-JANAF [16]. Enthalpies in eV/atom.

| formula | PBE+AGL | LDA+AGL | SCAN+AGL | PBE+AGL CCE | LDA+AGL CCE | SCAN+AGL CCE | Exp. | $N_{1/2-X}$ | $+\alpha$ |
|---|---|---|---|---|---|---|---|---|---|
| $KBF_4$ | −3.035 | −3.235 | −3.425 | −3.318 | −3.354 | −3.237 | −3.255 [15] | 10, 4 | +1, +3 |
| $Li_3AlF_6$ | −3.231 | −3.443 | −3.663 | −3.485 | −3.494 | −3.526 | −3.507 [16] | $13.\overline{3}$, 6 | +1, +3 |
| $Li_2BeF_4$ | −3.135 | −3.299 | −3.512 | −3.344 | −3.345 | −3.388 | −3.365 [15] | 8, 4 | +1, +2 |
| $Na_3AlF_6$ | −3.079 | −3.267 | −3.508 | −3.361 | −3.343 | −3.359 | −3.433 [15] | 16, 6 | +1, +3 |

TABLE A.X. (*continued*)

| formula | PBE+AGL | LDA+AGL | SCAN+AGL | PBE+AGL CCE | LDA+AGL CCE | SCAN+AGL CCE | Exp. | $N_{1/2-X}$ | $+\alpha$ |
|---|---|---|---|---|---|---|---|---|---|
| $NaBF_4$ | −2.961 | −3.150 | −3.349 | −3.235 | −3.290 | −3.177 | −3.187 [15] | 8, 4 | +1, +3 |
| $Na_2SiF_6$ | −3.088 | −3.316 | −3.518 | −3.408 | −3.467 | −3.353 | −3.354 [15] | 12, 6 | +1, +4 |
| $NaAlCl_4$ | −1.696 | −1.816 | −1.904 | −1.909 | −1.896 | −1.952 | −1.973 [16] | 5, 4 | +1, +3 |

TABLE A.XI: **Uncorrected and quasi-FERE corrected formation enthalpies for ternary oxides.** Calculated, quasi-FERE corrected and experimental room temperature formation enthalpies for ternary oxides. Experimental values from Kubaschewski *et al.* [15], NIST-JANAF [16] and Barin [17]. Enthalpies in eV/atom.

| formula | PBE+AGL | LDA+AGL | SCAN+AGL | PBE+AGL quasi-FERE corr. | LDA+AGL quasi-FERE corr. | SCAN+AGL quasi-FERE corr. | Exp. |
|---|---|---|---|---|---|---|---|
| $Ag_2CrO_4$ | −1.104 | −1.285 | −1.302 | −1.153 | −1.167 | −1.163 | −1.083 [17] |
| $Al_2SiO_5$ (kya.) | −2.917 | −3.365 | −3.363 | −3.339 | −3.364 | −3.355 | −3.361 [16] |
| $Al_2SiO_5$ (and.) | −2.940 | −3.354 | −3.364 | −3.361 | −3.352 | −3.355 | −3.358 [16] |
| $BaAl_2O_4$ | −3.065 | −3.402 | −3.482 | −3.534 | −3.431 | −3.516 | −3.441 [15] |
| $BaMoO_4$ | −2.616 | −2.872 | −2.876 | −2.757 | −2.660 | −2.760 | −2.674 [17] |
| $BaTiO_3$ | −3.090 | −3.526 | −3.538 | −3.381 | −3.430 | −3.476 | −3.441 [17] |
| $BaZrO_3$ | −3.302 | −3.696 | −3.744 | −3.689 | −3.674 | −3.715 | −3.689 [17] |
| $BeAl_2O_4$ | −2.953 | −3.385 | −3.398 | −3.403 | −3.399 | −3.402 | −3.407 [15] |
| $Be_2SiO_4$ | −2.778 | −3.161 | −3.157 | −3.158 | −3.176 | −3.170 | −3.134 [16] |
| $CaAl_4O_7$ | −3.064 | −3.413 | −3.480 | −3.507 | −3.402 | −3.464 | −3.477 [17] |
| $CaMoO_4$ | −2.594 | −2.915 | −2.886 | −2.684 | −2.642 | −2.704 | −2.671 [15] |
| $Ca_3SiO_5$ | −3.002 | −3.395 | −3.380 | −3.330 | −3.329 | −3.336 | −3.373 [15] |
| $CaSiO_3$ (wol.) | −3.038 | −3.412 | −3.410 | −3.372 | −3.368 | −3.379 | −3.389 [15] |
| $CaSiO_3$ (ps-wol.) | −3.018 | −3.391 | −3.393 | −3.353 | −3.346 | −3.362 | −3.375 [15] |
| $CaTiO_3$ | −3.163 | −3.591 | −3.608 | −3.394 | −3.423 | −3.467 | −3.442 [15] |
| $Ca_2V_2O_7$ | −2.804 | −3.118 | −3.152 | −2.976 | −2.928 | −2.996 | −2.905 [15] |
| $CaV_2O_6$ | −2.642 | −2.918 | −2.951 | −2.780 | −2.706 | −2.774 | −2.682 [15] |
| $CaWO_4$ | −2.661 | −2.984 | −2.958 | −2.781 | −2.730 | −2.819 | −2.842 [17] |
| $CaZrO_3$ | −3.329 | −3.727 | −3.759 | −3.656 | −3.632 | −3.651 | −3.662 [15] |
| $CdAl_2O_4$ | −2.438 | −2.834 | −2.838 | −2.861 | −2.851 | −2.838 | −2.838 [15] |
| $CdGa_2O_4$ | −1.645 | −1.999 | −1.949 | −2.024 | −2.027 | −2.015 | −2.007 [15] |
| $CdTiO_3$ | −2.289 | −2.629 | −2.678 | −2.517 | −2.517 | −2.567 | −2.551 [17] |
| $CoFe_2O_4$ | −1.143 | −1.458 | −1.588 | −1.629 | −1.657 | −1.607 | −1.612 [15] |
| $CoTiO_3$ | −2.089 | −2.418 | −2.488 | −2.480 | −2.486 | −2.520 | −2.503 [15] |
| $CuFeO_2$ | −0.920 | −1.275 | −1.321 | −1.244 | −1.345 | −1.279 | −1.329 [15] |
| $FeAl_2O_4$ | −2.400 | −2.740 | −2.829 | −2.890 | −2.807 | −2.804 | −2.954 [17] |
| $FeMoO_4$ | −1.637 | −2.036 | −2.058 | −1.802 | −1.867 | −1.873 | −1.831 [15] |
| $FeTiO_3$ | −2.231 | −2.506 | −2.695 | −2.551 | −2.463 | −2.550 | −2.565 [15] |
| $LiAlO_2$ | −2.720 | −3.101 | −3.111 | −3.120 | −3.115 | −3.134 | −3.080 [15] |
| $Li_3AsO_4$ | −1.939 | −2.200 | −2.188 | −2.228 | −2.178 | −2.245 | −2.205 [17] |
| $Li_2SiO_3$ | −2.545 | −2.848 | −2.847 | −2.860 | −2.855 | −2.877 | −2.848 [15] |
| $Li_2TiO_3$ | −2.653 | −2.986 | −3.006 | −2.883 | −2.891 | −2.945 | −2.884 [15] |
| $Li_2ZrO_3$ | −2.759 | −3.076 | −3.101 | −3.069 | −3.042 | −3.068 | −3.044 [15] |
| $MgAl_2O_4$ | −2.976 | −3.393 | −3.415 | −3.428 | −3.403 | −3.413 | −3.404 [16] |
| $MgCr_2O_4$ | −2.225 | −2.458 | −2.621 | −2.544 | −2.507 | −2.568 | −2.632 [15] |
| $MgMoO_4$ | −2.371 | −2.634 | −2.626 | −2.492 | −2.399 | −2.467 | −2.419 [15] |
| $Mg_2SiO_4$ | −2.825 | −3.203 | −3.215 | −3.208 | −3.209 | −3.214 | −3.223 [15] |
| $MgSiO_3$ | −2.812 | −3.203 | −3.204 | −3.183 | −3.203 | −3.200 | −3.210 [15] |
| $MgTiO_3$ | −2.971 | −3.357 | −3.398 | −3.239 | −3.234 | −3.283 | −3.260 [15] |
| $MgTi_2O_5$ | −3.018 | −3.389 | −3.437 | −3.249 | −3.230 | −3.292 | −3.251 [16] |
| $MgV_2O_6$ | −2.479 | −2.783 | −2.804 | −2.638 | −2.596 | −2.642 | −2.534 [15] |
| $Mg_2V_2O_7$ | −2.573 | −2.866 | −2.911 | −2.777 | −2.718 | −2.780 | −2.671 [15] |
| $MgWO_4$ | −2.446 | −2.801 | −2.751 | −2.597 | −2.585 | −2.634 | −2.621 [15] |
| $MnTiO_3$ | −2.441 | −2.639 | −3.044 | −2.753 | −2.665 | −2.849 | −2.817 [15] |
| $NaCrO_2$ | −1.900 | −2.118 | −2.288 | −2.197 | −2.187 | −2.278 | −2.271 [15] |
| $Na_2CrO_4$ | −1.976 | −2.179 | −2.209 | −2.169 | −2.131 | −2.153 | −1.987 [17] |
| $NaFeO_2$ | −1.552 | −1.714 | −1.940 | −1.919 | −1.796 | −1.902 | −1.809 [15] |
| $Na_2MoO_4$ | −2.120 | −2.346 | −2.358 | −2.244 | −2.175 | −2.264 | −2.175 [17] |
| $Na_4SiO_4$ | −2.093 | −2.373 | −2.397 | −2.422 | −2.427 | −2.463 | −2.420 [15] |
| $Na_2SiO_3$ | −2.361 | −2.657 | −2.677 | −2.694 | −2.694 | −2.723 | −2.700 [15] |
| $Na_2Ti_3O_7$ | −2.787 | −3.116 | −3.167 | −2.997 | −2.974 | −3.046 | −3.007 [15] |

TABLE A.XI. (*continued*)

| formula | PBE+AGL | LDA+AGL | SCAN+AGL | PBE+AGL quasi-FERE corr. | LDA+AGL quasi-FERE corr. | SCAN+AGL quasi-FERE corr. | Exp. |
|---|---|---|---|---|---|---|---|
| $NaVO_3$ | −2.345 | −2.556 | −2.619 | −2.502 | −2.410 | −2.503 | −2.379 [17] |
| $Na_2WO_4$ | −2.175 | −2.397 | −2.417 | −2.324 | −2.243 | −2.359 | −2.283 [15] |
| $NiAl_2O_4$ | −2.279 | −2.641 | −2.659 | −2.832 | −2.779 | −2.825 | −2.844 [15] |
| $NiTiO_3$ | −2.070 | −2.398 | −2.396 | −2.480 | −2.455 | −2.516 | −2.492 [15] |
| $NiWO_4$ | −1.659 | −1.980 | −1.886 | −1.928 | −1.914 | −1.965 | −1.950 [15] |
| $PbMoO_4$ | −1.893 | −2.134 | −2.066 | −1.895 | −1.823 | −1.880 | −1.819 [15] |
| $PbTiO_3$ | −2.350 | −2.685 | −2.655 | −2.475 | −2.471 | −2.508 | −2.476 [15] |
| $PbWO_4$ | −1.948 | −2.187 | −2.125 | −1.980 | −1.895 | −1.981 | −1.937 [15] |
| $SrAl_2O_4$ | −3.074 | −3.419 | −3.489 | −3.531 | −3.423 | −3.504 | −3.442 [17] |
| $SrHfO_3$ | −3.416 | −3.826 | −3.844 | −3.814 | −3.786 | −3.835 | −3.697 [15] |
| $SrMoO_4$ | −2.624 | −2.918 | −2.905 | −2.751 | −2.676 | −2.766 | −2.676 [15] |
| $SrSiO_3$ | −3.039 | −3.401 | −3.411 | −3.416 | −3.394 | −3.432 | −3.386 [15] |
| $Sr_2TiO_4$ | −3.074 | −3.492 | −3.506 | −3.393 | −3.398 | −3.462 | −3.387 [15] |
| $SrTiO_3$ | −3.164 | −3.596 | −3.622 | −3.439 | −3.465 | −3.532 | −3.463 [15] |
| $SrWO_4$ | −2.688 | −2.981 | −2.973 | −2.844 | −2.758 | −2.877 | −2.832 [17] |
| $ZnFe_2O_4$ | −1.288 | −1.669 | −1.783 | −1.675 | −1.751 | −1.722 | −1.735 [17] |
| $Zn_2SiO_4$ | −2.054 | −2.376 | −2.370 | −2.413 | −2.420 | −2.418 | −2.433 [15] |
| $Zn_2TiO_4$ | −2.115 | −2.457 | −2.479 | −2.401 | −2.413 | −2.449 | −2.443 [15] |
| $ZnWO_4$ | −1.937 | −2.282 | −2.214 | −2.074 | −2.087 | −2.126 | −2.126 [15] |
| $ZrSiO_4$ | −3.146 | −3.579 | −3.587 | −3.483 | −3.524 | −3.512 | −3.496 [16] |

TABLE A.XII: **AGL contributions to the formation enthalpies for binary oxides.** Total vibrational (TVC), zero-point (ZPC) and thermal (TC) contributions to the calculated formation enthalpies obtained from AGL [36–40] for each functional for binary oxides. The sum of ZPC and TC might not match exactly the total AGL contribution listed due to rounding. All values in eV/atom.

| formula | PBE | | | LDA | | | SCAN | | |
|---|---|---|---|---|---|---|---|---|---|
| | AGL-TVC | AGL-ZPC | AGL-TC | AGL-TVC | AGL-ZPC | AGL-TC | AGL-TVC | AGL-ZPC | AGL-TC |
| $Li_2O$ | 0.011 | 0.028 | −0.017 | 0.014 | 0.033 | −0.018 | 0.014 | 0.033 | −0.018 |
| $Li_2O_2$ | 0.010 | 0.027 | −0.017 | 0.014 | 0.033 | −0.019 | 0.013 | 0.031 | −0.019 |
| BeO | 0.017 | 0.032 | −0.016 | 0.019 | 0.036 | −0.016 | 0.020 | 0.036 | −0.016 |
| $Na_2O$ | 0.001 | 0.013 | −0.012 | 0.002 | 0.017 | −0.014 | 0.003 | 0.018 | −0.015 |
| $Na_2O_2$ | 0.001 | 0.015 | −0.014 | 0.003 | 0.020 | −0.017 | 0.003 | 0.020 | −0.017 |
| $NaO_2$ | −0.001 | 0.011 | −0.012 | 0.002 | 0.017 | −0.015 | 0.001 | 0.016 | −0.015 |
| MgO | 0.012 | 0.031 | −0.019 | 0.014 | 0.034 | −0.020 | 0.014 | 0.034 | −0.020 |
| $K_2O$ | −0.003 | 0.005 | −0.007 | −0.001 | 0.009 | −0.010 | −0.002 | 0.008 | −0.010 |
| $K_2O_2$ | −0.003 | 0.006 | −0.009 | −0.002 | 0.011 | −0.012 | −0.002 | 0.009 | −0.011 |
| $KO_2$ | −0.005 | 0.004 | −0.009 | −0.002 | 0.011 | −0.013 | −0.003 | 0.008 | −0.011 |
| CaO | 0.004 | 0.020 | −0.016 | 0.007 | 0.026 | −0.019 | 0.005 | 0.023 | −0.018 |
| $Rb_2O$ | −0.004 | −0.002 | −0.003 | −0.004 | 0.002 | −0.005 | −0.004 | 0.001 | −0.005 |
| SrO | −0.002 | 0.008 | −0.010 | −0.001 | 0.011 | −0.012 | −0.001 | 0.010 | −0.012 |
| $SrO_2$ | −0.003 | 0.007 | −0.010 | −0.001 | 0.012 | −0.013 | −0.002 | 0.010 | −0.012 |
| $Cs_2O$ | −0.005 | −0.005 | 0 | −0.004 | −0.001 | −0.003 | −0.005 | −0.003 | −0.002 |
| $CsO_2$ | −0.009 | −0.010 | 0.001 | −0.008 | −0.005 | −0.003 | −0.008 | −0.007 | −0.001 |
| BaO | −0.005 | 0.001 | −0.005 | −0.004 | 0.003 | −0.007 | −0.004 | 0.002 | −0.006 |
| $BaO_2$ | −0.006 | 0 | −0.006 | −0.004 | 0.004 | −0.008 | −0.005 | 0.002 | −0.007 |
| $B_2O_3$ | 0.009 | 0.025 | −0.015 | 0.015 | 0.032 | −0.017 | 0.014 | 0.031 | −0.017 |
| $Al_2O_3$ | 0.019 | 0.041 | −0.022 | 0.022 | 0.045 | −0.023 | 0.023 | 0.047 | −0.023 |
| $SiO_2$ ($\alpha$-qua.) | 0.016 | 0.036 | −0.020 | 0.018 | 0.039 | −0.021 | 0.018 | 0.040 | −0.021 |
| $SiO_2$ ($\alpha$-crist.) | 0.013 | 0.032 | −0.019 | 0.015 | 0.035 | −0.020 | 0.016 | 0.036 | −0.020 |
| $Ga_2O_3$ | 0.005 | 0.022 | −0.017 | 0.007 | 0.025 | −0.018 | 0.008 | 0.028 | −0.019 |
| $GeO_2$ | 0.008 | 0.027 | −0.019 | 0.011 | 0.032 | −0.021 | 0.012 | 0.033 | −0.021 |
| $As_2O_5$ | 0.003 | 0.017 | −0.015 | 0.005 | 0.022 | −0.017 | 0.006 | 0.023 | −0.017 |
| $SeO_2$ | −0.002 | 0.007 | −0.009 | 0.001 | 0.013 | −0.012 | 0 | 0.012 | −0.012 |
| $In_2O_3$ | −0.001 | 0.011 | −0.012 | 0.001 | 0.014 | −0.013 | 0.001 | 0.014 | −0.013 |
| SnO | −0.004 | 0 | −0.004 | −0.003 | 0.003 | −0.006 | −0.003 | 0.002 | −0.005 |
| $SnO_2$ | 0.001 | 0.016 | −0.014 | 0.003 | 0.020 | −0.016 | 0.004 | 0.021 | −0.017 |
| $Sb_2O_3$ | −0.004 | 0.003 | −0.006 | −0.002 | 0.007 | −0.009 | −0.002 | 0.006 | −0.008 |
| $SbO_2$ | −0.001 | 0.011 | −0.012 | 0.001 | 0.015 | −0.014 | 0.001 | 0.015 | −0.014 |
| $Sb_2O_5$ | 0.001 | 0.015 | −0.014 | 0.003 | 0.019 | −0.016 | 0.004 | 0.020 | −0.017 |
| $TeO_2$ | −0.002 | 0.007 | −0.010 | 0 | 0.012 | −0.012 | −0.001 | 0.010 | −0.011 |

TABLE A.XII. (*continued*)

| formula | PBE | | | LDA | | | SCAN | | |
|---|---|---|---|---|---|---|---|---|---|
| | AGL-TVC | AGL-ZPC | AGL-TC | AGL-TVC | AGL-ZPC | AGL-TC | AGL-TVC | AGL-ZPC | AGL-TC |
| $Tl_2O$ | −0.005 | −0.007 | 0.002 | −0.004 | −0.005 | 0.001 | −0.005 | −0.006 | 0.001 |
| $Tl_2O_3$ | −0.006 | −0.001 | −0.004 | −0.005 | 0.002 | −0.006 | −0.005 | 0.002 | −0.006 |
| PbO | −0.006 | −0.007 | 0.001 | −0.006 | −0.005 | −0.001 | −0.006 | −0.005 | −0.001 |
| $Pb_3O_4$ | −0.006 | −0.006 | −0.001 | −0.006 | −0.003 | −0.003 | −0.006 | −0.003 | −0.002 |
| $PbO_2$ | −0.006 | −0.001 | −0.005 | −0.004 | 0.002 | −0.007 | −0.005 | 0.002 | −0.007 |
| $Bi_2O_3$ | −0.006 | −0.002 | −0.004 | −0.005 | 0.001 | −0.005 | −0.005 | 0 | −0.005 |
| $Sc_2O_3$ | 0.009 | 0.028 | −0.018 | 0.012 | 0.032 | −0.020 | 0.011 | 0.030 | −0.020 |
| TiO | 0.009 | 0.024 | −0.015 | 0.012 | 0.029 | −0.017 | 0.010 | 0.026 | −0.016 |
| $Ti_2O_3$ | 0.011 | 0.029 | −0.018 | 0.014 | 0.034 | −0.019 | 0.013 | 0.032 | −0.019 |
| $Ti_3O_5$ | 0.009 | 0.026 | −0.017 | 0.013 | 0.032 | −0.019 | 0.012 | 0.03 | −0.018 |
| $TiO_2$ (rut.) | 0.013 | 0.033 | −0.020 | 0.016 | 0.038 | −0.021 | 0.017 | 0.039 | −0.022 |
| $TiO_2$ (ana.) | 0.012 | 0.030 | −0.019 | 0.014 | 0.035 | −0.020 | 0.015 | 0.036 | −0.021 |
| VO | −0.004 | 0.001 | −0.005 | 0.011 | 0.027 | −0.015 | 0 | 0.006 | −0.007 |
| $V_2O_3$ | 0.008 | 0.024 | −0.015 | 0.013 | 0.031 | −0.018 | 0.009 | 0.025 | −0.016 |
| $VO_2$ | 0.012 | 0.031 | −0.018 | 0.017 | 0.037 | −0.020 | 0.015 | 0.035 | −0.020 |
| $V_2O_5$ | 0.006 | 0.022 | −0.015 | 0.012 | 0.031 | −0.019 | 0.011 | 0.029 | −0.018 |
| $Cr_2O_3$ | 0.006 | 0.019 | −0.013 | 0.010 | 0.025 | −0.015 | 0.007 | 0.021 | −0.014 |
| $CrO_3$ | 0.002 | 0.014 | −0.012 | 0.007 | 0.023 | −0.016 | 0.007 | 0.022 | −0.016 |
| MnO | −0.006 | −0.006 | 0 | −0.005 | −0.003 | −0.002 | −0.002 | 0.002 | −0.005 |
| $MnO_2$ | 0.008 | 0.023 | −0.015 | 0.012 | 0.029 | −0.017 | 0.011 | 0.028 | −0.017 |
| FeO | 0.004 | 0.014 | −0.010 | 0 | 0.007 | −0.007 | 0.006 | 0.020 | −0.014 |
| $Fe_2O_3$ | 0.004 | 0.015 | −0.011 | 0.014 | 0.032 | −0.018 | 0.007 | 0.022 | −0.015 |
| CoO | 0.001 | 0.008 | −0.007 | −0.002 | 0.003 | −0.005 | 0.004 | 0.015 | −0.011 |
| NiO | −0.001 | 0.006 | −0.006 | 0.006 | 0.018 | −0.012 | 0.001 | 0.008 | −0.007 |
| $Cu_2O$ | −0.002 | 0 | −0.002 | −0.002 | 0.002 | −0.003 | −0.002 | 0 | −0.002 |
| CuO | 0.001 | 0.009 | −0.009 | 0.003 | 0.013 | −0.010 | 0.002 | 0.011 | −0.009 |
| ZnO | 0.001 | 0.011 | −0.010 | 0.002 | 0.013 | −0.011 | 0.003 | 0.014 | −0.012 |
| $Y_2O_3$ | 0.001 | 0.014 | −0.013 | 0.003 | 0.017 | −0.015 | 0.002 | 0.016 | −0.014 |
| $ZrO_2$ | 0.006 | 0.023 | −0.017 | 0.008 | 0.027 | −0.018 | 0.008 | 0.026 | −0.018 |
| NbO | 0.004 | 0.016 | −0.012 | 0.005 | 0.018 | −0.013 | 0.005 | 0.018 | −0.013 |
| $MoO_2$ | 0.007 | 0.022 | −0.015 | 0.009 | 0.026 | −0.017 | 0.009 | 0.026 | −0.017 |
| $MoO_3$ | 0.003 | 0.016 | −0.013 | 0.007 | 0.024 | −0.017 | 0.007 | 0.023 | −0.016 |
| $RuO_2$ | 0.004 | 0.018 | −0.013 | 0.007 | 0.023 | −0.015 | 0.006 | 0.021 | −0.015 |
| $Rh_2O_3$ | −0.006 | 0.012 | −0.018 | 0.004 | 0.016 | −0.012 | 0.003 | 0.014 | −0.011 |
| PdO | −0.002 | 0.004 | −0.006 | 0 | 0.007 | −0.007 | −0.001 | 0.006 | −0.007 |
| $Ag_2O$ | −0.004 | −0.006 | 0.002 | −0.004 | −0.004 | 0 | −0.004 | −0.005 | 0.001 |
| CdO | −0.002 | 0.007 | −0.009 | −0.001 | 0.009 | −0.009 | −0.001 | 0.008 | −0.009 |
| $HfO_2$ | 0 | 0.012 | −0.012 | 0.001 | 0.015 | −0.013 | 0.001 | 0.015 | −0.013 |
| $WO_2$ | 0.001 | 0.012 | −0.011 | 0.002 | 0.015 | −0.012 | 0.002 | 0.014 | −0.012 |
| $WO_3$ | 0 | 0.011 | −0.011 | 0.002 | 0.015 | −0.013 | 0.002 | 0.015 | −0.013 |
| $ReO_2$ | 0 | 0.011 | −0.010 | 0.002 | 0.014 | −0.012 | 0.002 | 0.013 | −0.011 |
| $ReO_3$ | 0.001 | 0.012 | −0.011 | 0.002 | 0.015 | −0.013 | 0.002 | 0.015 | −0.013 |
| $OsO_2$ | −0.001 | 0.009 | −0.009 | 0.001 | 0.012 | −0.011 | 0.001 | 0.011 | −0.010 |
| $OsO_4$ | −0.005 | 0.001 | −0.006 | −0.001 | 0.009 | −0.010 | −0.002 | 0.006 | −0.009 |
| $IrO_2$ | −0.001 | 0.008 | −0.009 | 0.001 | 0.012 | −0.011 | −0.001 | 0.008 | −0.009 |
| HgO | −0.019 | −0.001 | −0.017 | −0.022 | −0.001 | −0.022 | −0.020 | 0 | −0.020 |

TABLE A.XIII: **AGL contributions to the formation enthalpies for binary halides.** Total vibrational (TVC), zero-point (ZPC) and thermal (TC) contributions to the calculated formation enthalpies obtained from AGL [36–40] for each functional for binary halides. Note that $BF_3$ and $SiF_4$ are gaseous molecular systems. The sum of ZPC and TC might not match exactly the total AGL contribution listed due to rounding. All values in eV/atom.

| formula | PBE | | | LDA | | | SCAN | | |
|---|---|---|---|---|---|---|---|---|---|
| | AGL-TVC | AGL-ZPC | AGL-TC | AGL-TVC | AGL-ZPC | AGL-TC | AGL-TVC | AGL-ZPC | AGL-TC |
| LiF | 0.012 | 0.023 | −0.011 | 0.015 | 0.029 | −0.014 | 0.015 | 0.028 | −0.013 |
| NaF | 0.008 | 0.017 | −0.009 | 0.010 | 0.022 | −0.013 | 0.009 | 0.022 | −0.013 |
| KF | 0.005 | 0.010 | −0.005 | 0.007 | 0.016 | −0.009 | 0.006 | 0.014 | −0.008 |
| $BeF_2$ | 0.008 | 0.016 | −0.007 | 0.011 | 0.021 | −0.010 | 0.011 | 0.021 | −0.009 |
| $BF_3$(g) | 0.026 | 0.034 | −0.007 | 0.026 | 0.033 | −0.007 | 0.026 | 0.033 | −0.007 |
| $AlF_3$ | 0.020 | 0.036 | −0.015 | 0.023 | 0.041 | −0.018 | 0.024 | 0.042 | −0.018 |
| $SiF_4$(g) | 0.024 | 0.036 | −0.012 | 0.024 | 0.036 | −0.012 | 0.024 | 0.035 | −0.011 |

TABLE A.XIII. (*continued*)

| formula | PBE | | | LDA | | | SCAN | | |
|---|---|---|---|---|---|---|---|---|---|
| | AGL-TVC | AGL-ZPC | AGL-TC | AGL-TVC | AGL-ZPC | AGL-TC | AGL-TVC | AGL-ZPC | AGL-TC |
| NaCl | 0.008 | 0.009 | −0.001 | 0.009 | 0.013 | −0.004 | 0.009 | 0.011 | −0.002 |
| KCl | 0.008 | 0.007 | 0 | 0.008 | 0.011 | −0.003 | 0.008 | 0.010 | −0.002 |
| $CaCl_2$ | 0.011 | 0.012 | 0 | 0.013 | 0.016 | −0.003 | 0.012 | 0.014 | −0.002 |
| $AlCl_3$ | 0.013 | 0.014 | −0.001 | 0.015 | 0.019 | −0.004 | 0.015 | 0.018 | −0.003 |

TABLE A.XIV: **AGL contributions to the formation enthalpies for ternary oxides.** Total vibrational (TVC), zero-point (ZPC) and thermal (TC) contributions to the calculated formation enthalpies obtained from AGL [36–40] for each functional for ternary oxides. The sum of ZPC and TC might not match exactly the total AGL contribution listed due to rounding. All values in eV/atom.

| formula | PBE | | | LDA | | | SCAN | | |
|---|---|---|---|---|---|---|---|---|---|
| | AGL-TVC | AGL-ZPC | AGL-TC | AGL-TVC | AGL-ZPC | AGL-TC | AGL-TVC | AGL-ZPC | AGL-TC |
| $Ag_2CrO_4$ | −0.003 | 0.002 | −0.006 | −0.001 | 0.007 | −0.008 | −0.002 | 0.006 | −0.008 |
| $Al_2SiO_5$ (kya.) | 0.019 | 0.041 | −0.022 | 0.022 | 0.045 | −0.023 | 0.023 | 0.046 | −0.023 |
| $Al_2SiO_5$ (and.) | 0.016 | 0.036 | −0.020 | 0.019 | 0.040 | −0.021 | 0.019 | 0.041 | −0.022 |
| $BaAl_2O_4$ | 0 | 0.010 | −0.010 | 0.001 | 0.013 | −0.012 | 0.001 | 0.013 | −0.012 |
| $BaMoO_4$ | −0.003 | 0.006 | −0.009 | −0.001 | 0.009 | −0.011 | −0.002 | 0.008 | −0.010 |
| $BaTiO_3$ | 0 | 0.013 | −0.013 | 0.002 | 0.017 | −0.015 | 0.002 | 0.017 | −0.015 |
| $BaZrO_3$ | −0.001 | 0.011 | −0.012 | 0.001 | 0.014 | −0.014 | 0 | 0.014 | −0.013 |
| $BeAl_2O_4$ | 0.017 | 0.037 | −0.020 | 0.020 | 0.041 | −0.021 | 0.021 | 0.042 | −0.021 |
| $Be_2SiO_4$ | 0.016 | 0.033 | −0.018 | 0.018 | 0.036 | −0.018 | 0.019 | 0.037 | −0.019 |
| $CaAl_4O_7$ | 0.010 | 0.027 | −0.017 | 0.012 | 0.031 | −0.019 | 0.012 | 0.031 | −0.019 |
| $CaMoO_4$ | 0.004 | 0.019 | −0.015 | 0.007 | 0.024 | −0.017 | 0.006 | 0.023 | −0.016 |
| $Ca_3SiO_5$ | 0.006 | 0.022 | −0.017 | 0.008 | 0.028 | −0.019 | 0.008 | 0.026 | −0.018 |
| $CaSiO_3$ (wol.) | 0.009 | 0.027 | −0.018 | 0.011 | 0.031 | −0.020 | 0.010 | 0.030 | −0.019 |
| $CaSiO_3$ (ps-wol.) | 0.008 | 0.026 | −0.018 | 0.011 | 0.030 | −0.019 | 0.010 | 0.029 | −0.019 |
| $CaTiO_3$ | 0.010 | 0.029 | −0.019 | 0.013 | 0.034 | −0.021 | 0.012 | 0.033 | −0.021 |
| $Ca_2V_2O_7$ | 0.006 | 0.022 | −0.016 | 0.009 | 0.027 | −0.018 | 0.008 | 0.026 | −0.018 |
| $CaV_2O_6$ | 0.005 | 0.020 | −0.015 | 0.009 | 0.026 | −0.017 | 0.007 | 0.024 | −0.017 |
| $CaWO_4$ | 0.001 | 0.013 | −0.012 | 0.003 | 0.017 | −0.014 | 0.002 | 0.016 | −0.014 |
| $CaZrO_3$ | 0.005 | 0.022 | −0.017 | 0.007 | 0.026 | −0.018 | 0.007 | 0.025 | −0.018 |
| $CdAl_2O_4$ | 0.005 | 0.020 | −0.015 | 0.007 | 0.023 | −0.016 | 0.007 | 0.023 | −0.016 |
| $CdGa_2O_4$ | 0.002 | 0.015 | −0.014 | 0.003 | 0.018 | −0.015 | 0.003 | 0.019 | −0.015 |
| $CdTiO_3$ | 0.002 | 0.016 | −0.013 | 0.004 | 0.019 | −0.015 | 0.004 | 0.019 | −0.015 |
| $CoFe_2O_4$ | 0.001 | 0.009 | −0.008 | 0.011 | 0.026 | −0.016 | 0.005 | 0.017 | −0.012 |
| $CoTiO_3$ | 0.008 | 0.024 | −0.016 | 0.013 | 0.031 | −0.018 | 0.011 | 0.028 | −0.017 |
| $CuFeO_2$ | 0.005 | 0.017 | −0.012 | 0.006 | 0.019 | −0.013 | 0.003 | 0.014 | −0.011 |
| $FeAl_2O_4$ | 0.010 | 0.027 | −0.017 | 0.013 | 0.030 | −0.017 | 0.014 | 0.032 | −0.019 |
| $FeMoO_4$ | 0.006 | 0.021 | −0.015 | 0.009 | 0.025 | −0.016 | 0.006 | 0.021 | −0.015 |
| $FeTiO_3$ | 0.008 | 0.024 | −0.016 | 0.014 | 0.033 | −0.019 | 0.012 | 0.029 | −0.018 |
| $LiAlO_2$ | 0.018 | 0.039 | −0.021 | 0.021 | 0.044 | −0.022 | 0.022 | 0.045 | −0.023 |
| $Li_3AsO_4$ | 0.007 | 0.022 | −0.015 | 0.009 | 0.026 | −0.017 | 0.010 | 0.027 | −0.017 |
| $Li_2SiO_3$ | 0.016 | 0.036 | −0.020 | 0.019 | 0.040 | −0.021 | 0.019 | 0.040 | −0.021 |
| $Li_2TiO_3$ | 0.015 | 0.035 | −0.020 | 0.019 | 0.040 | −0.021 | 0.019 | 0.040 | −0.021 |
| $Li_2ZrO_3$ | 0.009 | 0.025 | −0.017 | 0.011 | 0.029 | −0.018 | 0.011 | 0.029 | −0.018 |
| $MgAl_2O_4$ | 0.015 | 0.036 | −0.020 | 0.018 | 0.039 | −0.021 | 0.018 | 0.040 | −0.022 |
| $MgCr_2O_4$ | 0.006 | 0.020 | −0.014 | 0.009 | 0.025 | −0.016 | 0.008 | 0.023 | −0.015 |
| $MgMoO_4$ | 0.004 | 0.018 | −0.014 | 0.007 | 0.022 | −0.015 | 0.006 | 0.021 | −0.015 |
| $Mg_2SiO_4$ | 0.013 | 0.032 | −0.019 | 0.015 | 0.036 | −0.021 | 0.015 | 0.036 | −0.021 |
| $MgSiO_3$ | 0.015 | 0.035 | −0.020 | 0.018 | 0.039 | −0.021 | 0.018 | 0.039 | −0.021 |
| $MgTiO_3$ | 0.012 | 0.032 | −0.019 | 0.015 | 0.036 | −0.021 | 0.015 | 0.036 | −0.021 |
| $MgTi_2O_5$ | 0.011 | 0.029 | −0.018 | 0.014 | 0.034 | −0.020 | 0.014 | 0.034 | −0.020 |
| $MgV_2O_6$ | 0.007 | 0.023 | −0.016 | 0.014 | 0.033 | −0.020 | 0.013 | 0.032 | −0.019 |
| $Mg_2V_2O_7$ | 0.008 | 0.025 | −0.017 | 0.011 | 0.029 | −0.018 | 0.011 | 0.029 | −0.018 |
| $MgWO_4$ | 0.003 | 0.017 | −0.014 | 0.005 | 0.020 | −0.015 | 0.005 | 0.020 | −0.015 |
| $MnTiO_3$ | 0.005 | 0.019 | −0.013 | 0.009 | 0.024 | −0.015 | 0.008 | 0.023 | −0.015 |
| $NaCrO_2$ | 0.005 | 0.021 | −0.016 | 0.009 | 0.027 | −0.018 | 0.008 | 0.025 | −0.017 |
| $Na_2CrO_4$ | 0.004 | 0.019 | −0.016 | 0.007 | 0.024 | −0.018 | 0.007 | 0.024 | −0.018 |
| $NaFeO_2$ | 0.002 | 0.013 | −0.011 | 0.002 | 0.015 | −0.012 | 0.003 | 0.017 | −0.014 |
| $Na_2MoO_4$ | 0.004 | 0.019 | −0.015 | 0.006 | 0.023 | −0.017 | 0.006 | 0.023 | −0.017 |
| $Na_4SiO_4$ | 0.006 | 0.023 | −0.018 | 0.008 | 0.027 | −0.019 | 0.008 | 0.028 | −0.020 |

TABLE A.XIV. (*continued*)

| formula | PBE | | | LDA | | | SCAN | | |
|---|---|---|---|---|---|---|---|---|---|
| | AGL-TVC | AGL-ZPC | AGL-TC | AGL-TVC | AGL-ZPC | AGL-TC | AGL-TVC | AGL-ZPC | AGL-TC |
| $Na_2SiO_3$ | 0.008 | 0.026 | −0.019 | 0.010 | 0.030 | −0.020 | 0.010 | 0.030 | −0.020 |
| $Na_2Ti_3O_7$ | 0.009 | 0.027 | −0.019 | 0.012 | 0.032 | −0.020 | 0.012 | 0.032 | −0.020 |
| $NaVO_3$ | 0.005 | 0.020 | −0.015 | 0.008 | 0.025 | −0.017 | 0.007 | 0.025 | −0.017 |
| $Na_2WO_4$ | 0.001 | 0.013 | −0.012 | 0.003 | 0.017 | −0.014 | 0.003 | 0.017 | −0.014 |
| $NiAl_2O_4$ | 0.010 | 0.026 | −0.016 | 0.013 | 0.030 | −0.018 | 0.013 | 0.031 | −0.018 |
| $NiTiO_3$ | 0.008 | 0.023 | −0.015 | 0.011 | 0.028 | −0.017 | 0.010 | 0.027 | −0.017 |
| $NiWO_4$ | 0.002 | 0.013 | −0.012 | 0.004 | 0.017 | −0.013 | 0.004 | 0.017 | −0.013 |
| $PbMoO_4$ | −0.004 | 0.002 | −0.006 | −0.003 | 0.005 | −0.008 | −0.003 | 0.005 | −0.008 |
| $PbTiO_3$ | −0.003 | 0.004 | −0.007 | −0.001 | 0.009 | −0.010 | −0.001 | 0.008 | −0.009 |
| $PbWO_4$ | −0.005 | 0.001 | −0.006 | −0.003 | 0.004 | −0.007 | −0.004 | 0.004 | −0.007 |
| $SrAl_2O_4$ | 0.003 | 0.017 | −0.014 | 0.005 | 0.020 | −0.015 | 0.005 | 0.020 | −0.015 |
| $SrHfO_3$ | −0.001 | 0.011 | −0.012 | 0 | 0.014 | −0.013 | 0 | 0.013 | −0.013 |
| $SrMoO_4$ | 0 | 0.012 | −0.012 | 0.002 | 0.016 | −0.014 | 0.001 | 0.015 | −0.014 |
| $SrSiO_3$ | 0.001 | 0.014 | −0.013 | 0.003 | 0.018 | −0.015 | 0.003 | 0.017 | −0.014 |
| $Sr_2TiO_4$ | 0.001 | 0.015 | −0.014 | 0.003 | 0.018 | −0.015 | 0.003 | 0.018 | −0.015 |
| $SrTiO_3$ | 0.004 | 0.020 | −0.016 | 0.006 | 0.024 | −0.018 | 0.006 | 0.024 | −0.018 |
| $SrWO_4$ | −0.002 | 0.008 | −0.010 | 0 | 0.012 | −0.012 | 0 | 0.011 | −0.012 |
| $ZnFe_2O_4$ | −0.003 | 0 | −0.004 | 0.009 | 0.025 | −0.016 | 0.005 | 0.019 | −0.014 |
| $Zn_2SiO_4$ | 0.004 | 0.017 | −0.013 | 0.006 | 0.020 | −0.014 | 0.006 | 0.021 | −0.015 |
| $Zn_2TiO_4$ | 0.005 | 0.020 | −0.015 | 0.007 | 0.023 | −0.016 | 0.008 | 0.024 | −0.016 |
| $ZnWO_4$ | 0.002 | 0.015 | −0.013 | 0.004 | 0.018 | −0.014 | 0.004 | 0.018 | −0.014 |
| $ZrSiO_4$ | 0.011 | 0.029 | −0.019 | 0.013 | 0.033 | −0.020 | 0.013 | 0.033 | −0.020 |

TABLE A.XV: **AGL contributions to the formation enthalpies for ternary halides.** Total vibrational (TVC), zero-point (ZPC) and thermal (TC) contributions to the calculated formation enthalpies obtained from AGL [36–40] for each functional for ternary halides. The sum of ZPC and TC might not match exactly the total AGL contribution listed due to rounding. All values in eV/atom.

| formula | PBE | | | LDA | | | SCAN | | |
|---|---|---|---|---|---|---|---|---|---|
| | AGL-TVC | AGL-ZPC | AGL-TC | AGL-TVC | AGL-ZPC | AGL-TC | AGL-TVC | AGL-ZPC | AGL-TC |
| $KBF_4$ | 0.005 | 0.013 | −0.008 | 0.008 | 0.020 | −0.012 | 0.007 | 0.018 | −0.011 |
| $Li_3AlF_6$ | 0.016 | 0.029 | −0.013 | 0.019 | 0.034 | −0.016 | 0.019 | 0.034 | −0.015 |
| $Li_2BeF_4$ | 0.009 | 0.018 | −0.008 | 0.012 | 0.022 | −0.010 | 0.011 | 0.022 | −0.010 |
| $Na_3AlF_6$ | 0.011 | 0.022 | −0.011 | 0.013 | 0.026 | −0.013 | 0.013 | 0.026 | −0.013 |
| $NaBF_4$ | 0.008 | 0.019 | −0.010 | 0.011 | 0.025 | −0.013 | 0.011 | 0.024 | −0.013 |
| $Na_2SiF_6$ | 0.015 | 0.028 | −0.014 | 0.017 | 0.033 | −0.016 | 0.017 | 0.033 | −0.016 |
| $NaAlCl_4$ | 0.010 | 0.009 | 0.001 | 0.012 | 0.015 | −0.003 | 0.012 | 0.014 | −0.002 |

TABLE A.XVI: **Formation energies and CCE corrections for binary oxides.** DFT formation energies (without vibrational contribution) and experimental room temperature formation enthalpies, corrections per bond $\delta H_{A-\mathrm{O}}^{A^{+\alpha}}$, number of cation-oxygen bonds $N_{A-\mathrm{O}}$ per formula unit and oxidation states $+\alpha$ of the cation for binary oxides. Experimental values from Kubaschewski *et al.* [15], NIST-JANAF [16] and Barin [17]. Values in brackets denote cases in which the corrections are applied to calculate corrected values (see section III B of the main text for details). The O-O bond correction for peroxides is obtained from $Li_2O_2$. The O-O bond correction for superoxides is derived from $KO_2$. The two values are listed in the table at the position of the respective compound. Energies and enthalpies in eV/atom; corrections in eV/bond.

| formula | PBE | LDA | SCAN | Exp. | $\delta H_{A-\mathrm{O}}^{A^{+\alpha}}$ PBE | $\delta H_{A-\mathrm{O}}^{A^{+\alpha}}$ LDA | $\delta H_{A-\mathrm{O}}^{A^{+\alpha}}$ SCAN | $N_{A-\mathrm{O}}$ | $+\alpha$ |
|---|---|---|---|---|---|---|---|---|---|
| $Li_2O$ | −1.861 | −2.107 | −2.097 | −2.066 [15] | 0.0766 | −0.0154 | −0.0118 | 8 | +1 |
| $Li_2O_2$ | −1.436 | −1.718 | −1.618 | −1.643 [15] | −0.0930 | −0.1160 | 0.2400 | 12 | +1 |
| BeO | −2.768 | −3.142 | −3.146 | −3.158 [15] | 0.1953 | 0.0083 | 0.0060 | 4 | +2 |
| $Na_2O$ | −1.225 | −1.455 | −1.474 | −1.444 [16] | 0.0823 | −0.0043 | −0.0113 | 8 | +1 |
| $Na_2O_2$ | −1.084 | −1.354 | −1.288 | −1.329 [15] | (−1.307) | (−1.312) | (−1.314) | 12 | +1 |
| $NaO_2$ | −0.794 | −1.016 | −0.894 | −0.901 [15] | (−0.776) | (−0.918) | (−0.855) | 6 | +1 |
| MgO | −2.717 | −3.110 | −3.124 | −3.118 [15] | 0.1335 | 0.0025 | −0.0023 | 6 | +2 |
| $K_2O$ | −1.033 | −1.325 | −1.264 | −1.255 [15] | 0.0830 | −0.0263 | −0.0035 | 8 | +1 |
| $K_2O_2$ | −1.071 | −1.386 | −1.254 | −1.282 [15] | (−1.297) | (−1.278) | (−1.303) | 12 | +1 |
| $KO_2$ | −0.888 | −1.160 | −1.011 | −0.983 [15] | −0.5450 | −0.2690 | −0.0490 | 10 | +1 |
| CaO | −2.973 | −3.389 | −3.357 | −3.290 [15] | 0.1057 | −0.0332 | −0.0222 | 6 | +2 |
| $Rb_2O$ | −0.917 | −1.228 | −1.158 | −1.171 [15] | 0.0950 | −0.0215 | 0.0049 | 8 | +1 |

TABLE A.XVI. (*continued*)

| formula | PBE | LDA | SCAN | Exp. | $\delta H_{A-\mathrm{O}}^{A^{+\alpha}}$ PBE | $\delta H_{A-\mathrm{O}}^{A^{+\alpha}}$ LDA | $\delta H_{A-\mathrm{O}}^{A^{+\alpha}}$ SCAN | $N_{A-\mathrm{O}}$ | $+\alpha$ |
|---|---|---|---|---|---|---|---|---|---|
| SrO | −2.744 | −3.126 | −3.115 | −3.068 [15] | 0.1082 | −0.0192 | −0.0155 | 6 | +2 |
| $SrO_2$ | −1.898 | −2.313 | −2.174 | −2.189 [15] | (−2.228) | (−2.210) | (−2.203) | 10 | +2 |
| $Cs_2O$ | −0.994 | −1.309 | −1.205 | −1.195 [15] | 0.1008 | −0.0567 | −0.0050 | 6 | +1 |
| $CsO_2$ | −0.890 | −1.168 | −1.003 | −0.989 [15] | (−1.045) | (−0.889) | (−0.970) | 10 | +1 |
| BaO | −2.486 | −2.811 | −2.828 | −2.841 [15] | 0.1183 | 0.0098 | 0.0042 | 6 | +2 |
| $BaO_2$ | −1.860 | −2.250 | −2.123 | −2.191 [15] | (−2.223) | (−2.244) | (−2.217) | 10 | +2 |
| $B_2O_3$ | −2.406 | −2.723 | −2.696 | −2.640 [15] | 0.1952 | −0.0693 | −0.0472 | 6 | +3 |
| $Al_2O_3$ | −3.025 | −3.491 | −3.503 | −3.473 [15] | 0.1869 | −0.0073 | −0.0124 | 12 | +3 |
| $SiO_2$ ($\alpha$-qua.) | −2.810 | −3.178 | −3.175 | −3.147 [15] | 0.2530 | −0.0233 | −0.0208 | 4 | +4 |
| $SiO_2$ ($\alpha$-crist.) | −2.817 | −3.167 | −3.167 | −3.138 [15] | (−3.154) | (−3.136) | (−3.139) | 4 | +4 |
| $Ga_2O_3$ | −1.856 | −2.244 | −2.180 | −2.258 [15] | 0.2009 | 0.0068 | 0.0386 | 10 | +3 |
| $GeO_2$ | −1.605 | −2.075 | −1.924 | −2.004 [15] | 0.1992 | −0.0357 | 0.0397 | 6 | +4 |
| $As_2O_5$ | −1.073 | −1.453 | −1.332 | −1.362 [15] | 0.2022 | −0.0636 | 0.0212 | 10 | +5 |
| $SeO_2$ | −0.703 | −1.005 | −0.874 | −0.778 [15] | 0.0750 | −0.2277 | −0.0963 | 3 | +4 |
| $In_2O_3$ | −1.595 | −1.959 | −1.950 | −1.919 [15] | 0.1353 | −0.0167 | −0.0130 | 12 | +3 |
| SnO | −1.347 | −1.609 | −1.507 | −1.481 [17] | 0.0670 | −0.0638 | −0.0130 | 4 | +2 |
| $SnO_2$ | −1.706 | −2.099 | −2.037 | −2.007 [17] | 0.1505 | −0.0460 | −0.0153 | 6 | +4 |
| $Sb_2O_3$ | −1.340 | −1.621 | −1.508 | −1.484 [15] | 0.1207 | −0.1135 | −0.0198 | 6 | +3 |
| $SbO_2$ | −1.385 | −1.764 | −1.627 | −1.567 [15] | (−1.571) | (−1.556) | (−1.557) | 5 | +3, +5 |
| $Sb_2O_5$ | −1.259 | −1.666 | −1.537 | −1.439 [17] | 0.1052 | −0.1323 | −0.0573 | 12 | +5 |
| $TeO_2$ | −1.033 | −1.388 | −1.235 | −1.117 [17] | 0.0630 | −0.2033 | −0.0885 | 4 | +4 |
| $Tl_2O$ | −0.591 | −0.711 | −0.701 | −0.578 [15] | −0.0065 | −0.0665 | −0.0612 | 6 | +1 |
| $Tl_2O_3$ | −0.681 | −1.034 | −0.843 | −0.809 [15] | 0.0536 | −0.0936 | −0.0140 | 12 | +3 |
| PbO | −1.131 | −1.355 | −1.246 | −1.137 [15] | 0.0028 | −0.1088 | −0.0548 | 4 | +2 |
| $Pb_3O_4$ | −1.056 | −1.311 | −1.182 | −1.064 [15] | (−1.107) | (−1.111) | (−1.113) | 12 | +2, +4 |
| $PbO_2$ | −0.835 | −1.198 | −0.998 | −0.948 [15] | 0.0568 | −0.1250 | −0.0250 | 6 | +4 |
| $Bi_2O_3$ | −1.235 | −1.533 | −1.254 | −1.183 [15] | −0.02580 | −0.1752 | −0.0356 | 10 | +3 |
| $Sc_2O_3$ | −3.567 | −3.976 | −4.017 | −3.956 [15] | 0.1618 | −0.0083 | −0.0257 | 12 | +3 |
| TiO | −2.541 | −2.973 | −2.876 | −2.812 [15] | 0.1131 | −0.0667 | −0.0265 | 4.8 | +2 |
| $Ti_2O_3$ | −2.917 | −3.358 | −3.318 | −3.153 [15] | 0.0980 | −0.0855 | −0.0688 | 12 | +3 |
| $Ti_3O_5$ | −2.975 | −3.373 | −3.375 | −3.186 [15] | (−3.202) | (−3.172) | (−3.179) | 18 | +3, +4 |
| $TiO_2$ (rut.) | −3.047 | −3.454 | −3.509 | −3.261 [15] | 0.1072 | −0.0965 | −0.1237 | 6 | +4 |
| $TiO_2$ (ana.) | −3.078 | −3.464 | −3.517 | −3.243 [16] | (−3.293) | (−3.271) | (−3.270) | 6 | +4 |
| VO | −1.446 | −1.877 | −1.767 | −2.237 [15] | 0.2637 | 0.1203 | 0.1568 | 6 | +2 |
| $V_2O_3$ | −2.290 | −2.685 | −2.654 | −2.526 [15] | 0.0984 | −0.0661 | −0.0531 | 12 | +3 |
| $VO_2$ | −2.373 | −2.765 | −2.773 | −2.466 [15] | 0.0467 | −0.1497 | −0.1537 | 6 | +4 |
| $V_2O_5$ | −2.307 | −2.598 | −2.607 | −2.295 [15] | −0.0082 | −0.2118 | −0.2179 | 10 | +5 |
| $Cr_2O_3$ | −1.985 | −2.243 | −2.397 | −2.352 [15] | 0.1528 | 0.0454 | −0.0189 | 12 | +3 |
| $CrO_3$ | −1.653 | −1.826 | −1.802 | −1.521 [15] | −0.1323 | −0.3053 | −0.2813 | 4 | +6 |
| MnO | −1.240 | −1.185 | −2.092 | −1.994 [15] | 0.2513 | 0.2700 | −0.0325 | 6 | +2 |
| $MnO_2$ | −1.680 | −1.988 | −2.116 | −1.800 [15] | 0.0600 | −0.0943 | −0.1582 | 6 | +4 |
| FeO | −0.881 | −1.017 | −1.353 | −1.410 [16] | 0.1763 | 0.1310 | 0.0188 | 6 | +2 |
| $Fe_2O_3$ | −1.315 | −1.676 | −1.864 | −1.707 [15] | 0.1633 | 0.0130 | −0.0655 | 12 | +3 |
| CoO | −0.513 | −0.734 | −0.863 | −1.232 [15] | 0.2397 | 0.1662 | 0.1230 | 6 | +2 |
| NiO | −0.474 | −0.776 | −0.648 | −1.242 [15] | 0.2558 | 0.1552 | 0.1980 | 6 | +2 |
| $Cu_2O$ | −0.415 | −0.544 | −0.505 | −0.590 [16] | 0.1310 | 0.0340 | 0.0635 | 4 | +1 |
| CuO | −0.606 | −0.860 | −0.795 | −0.808 [16] | 0.1015 | −0.0258 | 0.0068 | 4 | +2 |
| ZnO | −1.446 | −1.732 | −1.726 | −1.817 [15] | 0.1853 | 0.0423 | 0.0455 | 4 | +2 |
| $Y_2O_3$ | −3.623 | −4.001 | −4.065 | −3.949 [15] | 0.1358 | −0.0219 | −0.0483 | 12 | +3 |
| $ZrO_2$ | −3.466 | −3.897 | −3.939 | −3.791 [16] | 0.1393 | −0.0451 | −0.0631 | 7 | +4 |
| NbO | −2.057 | −2.428 | −2.344 | −2.175 [15] | 0.0593 | −0.1263 | −0.0845 | 4 | +2 |
| $MoO_2$ | −1.973 | −2.400 | −2.242 | −2.031 [15] | 0.0292 | −0.1843 | −0.1053 | 6 | +4 |
| $MoO_3$ | −1.978 | −2.272 | −2.187 | −1.931 [15] | −0.0470 | −0.3410 | −0.2558 | 4 | +6 |
| $RuO_2$ | −1.063 | −1.465 | −1.277 | −1.054 [15] | −0.0048 | −0.2057 | −0.1118 | 6 | +4 |
| $Rh_2O_3$ | −0.703 | −1.064 | −0.877 | −0.737 [15] | 0.0141 | −0.1363 | −0.0583 | 12 | +3 |
| PdO | −0.483 | −0.757 | −0.647 | −0.599 [15] | 0.0578 | −0.0793 | −0.0245 | 4 | +2 |
| $Ag_2O$ | −0.118 | −0.179 | −0.194 | −0.107 [15] | −0.0083 | −0.0540 | −0.0653 | 4 | +1 |
| CdO | −1.026 | −1.290 | −1.323 | −1.339 [15] | 0.1042 | 0.0162 | 0.0053 | 6 | +2 |
| $HfO_2$ | −3.577 | −4.024 | −4.017 | −3.955 [17] | 0.1617 | −0.0296 | −0.0269 | 7 | +4 |
| $WO_2$ | −1.923 | −2.354 | −2.133 | −2.037 [15] | 0.0567 | −0.1587 | −0.0483 | 6 | +4 |
| $WO_3$ | −2.176 | −2.495 | −2.388 | −2.183 [15] | 0.0050 | −0.2078 | −0.1362 | 6 | +6 |

TABLE A.XVI. (*continued*)

| formula | PBE | LDA | SCAN | Exp. | $\delta H_{A-\mathrm{O}}^{A+\alpha}$ PBE | $\delta H_{A-\mathrm{O}}^{A+\alpha}$ LDA | $\delta H_{A-\mathrm{O}}^{A+\alpha}$ SCAN | $N_{A-\mathrm{O}}$ | $+\alpha$ |
|---|---|---|---|---|---|---|---|---|---|
| $ReO_2$ | −1.372 | −1.799 | −1.509 | −1.551 [17] | 0.0897 | −0.1242 | 0.0212 | 6 | +4 |
| $ReO_3$ | −1.635 | −1.982 | −1.766 | −1.526 [17] | −0.0727 | −0.3040 | −0.1597 | 6 | +6 |
| $OsO_2$ | −0.894 | −1.306 | −1.015 | −1.018 [15] | 0.0617 | −0.1443 | 0.0012 | 6 | +4 |
| $OsO_4$ | −0.994 | −1.119 | −1.038 | −0.816 [15] | −0.2225 | −0.3793 | −0.2773 | 4 | +8 |
| $IrO_2$ | −0.798 | −1.201 | −0.802 | −0.838 [17] | 0.0202 | −0.1813 | 0.0180 | 6 | +4 |
| HgO | −0.300 | −0.535 | −0.412 | −0.470 [15] | 0.1700 | −0.0650 | 0.0580 | 2 | +2 |

TABLE A.XVII: **Formation energies and CCE corrections for binary halides.** DFT formation energies (without vibrational contribution) and experimental room temperature formation enthalpies, corrections per bond $\delta H_{A-X}^{A+\alpha}$, number of cation-anion bonds $N_{A-X}$ per formula unit and oxidation states $+\alpha$ of the cation for binary halides. Experimental values from Kubaschewski *et al.* [15] and NIST-JANAF [16]. Note that $BF_3$ and $SiF_4$ are gaseous molecular systems. Energies and enthalpies in eV/atom; corrections in eV/bond.

| formula | PBE | LDA | SCAN | Exp. | $\delta H_{A-X}^{A+\alpha}$ PBE | $\delta H_{A-X}^{A+\alpha}$ LDA | $\delta H_{A-X}^{A+\alpha}$ SCAN | $N_{A-X}$ | $+\alpha$ |
|---|---|---|---|---|---|---|---|---|---|
| LiF | −2.972 | −3.176 | −3.356 | −3.197 [16] | 0.0748 | 0.0070 | −0.0532 | 6 | +1 |
| NaF | −2.739 | −2.914 | −3.133 | −2.982 [16] | 0.0807 | 0.0225 | −0.0503 | 6 | +1 |
| KF | −2.736 | −2.928 | −3.094 | −2.946 [16] | 0.0702 | 0.0060 | −0.0490 | 6 | +1 |
| $BeF_2$ | −3.280 | −3.483 | −3.721 | −3.547 [15] | 0.2008 | 0.0480 | −0.1300 | 4 | +2 |
| $BF_3$(g) | −2.786 | −2.850 | −3.092 | −2.943 [15] | 0.2093 | 0.1247 | −0.1987 | 3 | +3 |
| $AlF_3$ | −3.560 | −3.851 | −4.118 | −3.913 [15] | 0.2353 | 0.0415 | −0.1367 | 6 | +3 |
| $SiF_4$(g) | −3.121 | −3.232 | −3.494 | −3.348 [15] | 0.2833 | 0.1450 | −0.1825 | 4 | +4 |
| NaCl | −1.840 | −1.970 | −2.086 | −2.131 [15] | 0.0972 | 0.0537 | 0.0152 | 6 | +1 |
| KCl | −1.988 | −2.124 | −2.220 | −2.263 [15] | 0.0913 | 0.0460 | 0.0142 | 6 | +1 |
| $CaCl_2$ | −2.436 | −2.611 | −2.713 | −2.747 [15] | 0.1552 | 0.0680 | 0.0167 | 6 | +2 |
| $AlCl_3$ | −1.551 | −1.768 | −1.770 | −1.828 [15] | 0.1845 | 0.0400 | 0.0388 | 6 | +3 |

TABLE A.XVIII: **Uncorrected and CCE corrected formation energies for ternary oxides.** DFT formation energies (without vibrational contribution), coordination corrected values (using corrections from Table A.XVI) and experimental room temperature formation enthalpies, number of cation-oxygen bonds $N_{1/2-\mathrm{O}}$ per formula unit and oxidation states $+\alpha$ of the cations for ternary oxides. For the cation-oxygen bonds- and oxidation numbers, the first (second) number in the column refers to the first (second) element in the formula. Experimental values from Kubaschewski *et al.* [15], NIST-JANAF [16] and Barin [17]. Energies and enthalpies in eV/atom.

| formula | PBE | LDA | SCAN | PBE CCE | LDA CCE | SCAN CCE | Exp. | $N_{1/2-\mathrm{O}}$ | $+\alpha$ |
|---|---|---|---|---|---|---|---|---|---|
| $Ag_2CrO_4$ | −1.100 | −1.284 | −1.300 | −1.013 | −1.032 | −1.046 | −1.083 [17] | 10, 4 | +1, +6 |
| $Al_2SiO_5$ (kya.) | −2.937 | −3.388 | −3.386 | −3.343 | −3.365 | −3.357 | −3.361 [16] | 12, 4 | +3, +4 |
| $Al_2SiO_5$ (and.) | −2.956 | −3.373 | −3.383 | −3.339 | −3.351 | −3.356 | −3.358 [16] | 11, 4 | +3, +4 |
| $BaAl_2O_4$ | −3.064 | −3.403 | −3.483 | −3.405 | −3.405 | −3.473 | −3.441 [15] | 7.5, 8 | +2, +3 |
| $BaMoO_4$ | −2.613 | −2.871 | −2.874 | −2.739 | −2.657 | −2.709 | −2.674 [17] | 8, 4 | +2, +6 |
| $BaTiO_3$ | −3.090 | −3.528 | −3.540 | −3.503 | −3.436 | −3.402 | −3.441 [17] | 12, 6 | +2, +4 |
| $BaZrO_3$ | −3.301 | −3.697 | −3.745 | −3.752 | −3.666 | −3.679 | −3.689 [17] | 12, 6 | +2, +4 |
| $BeAl_2O_4$ | −2.970 | −3.405 | −3.419 | −3.402 | −3.397 | −3.401 | −3.407 [15] | 4, 12 | +2, +3 |
| $Be_2SiO_4$ | −2.794 | −3.179 | −3.175 | −3.162 | −3.175 | −3.170 | −3.134 [16] | 8, 4 | +2, +4 |
| $CaAl_4O_7$ | −3.074 | −3.425 | −3.492 | −3.367 | −3.402 | −3.466 | −3.477 [17] | 5, 16 | +2, +3 |
| $CaMoO_4$ | −2.598 | −2.922 | −2.892 | −2.707 | −2.650 | −2.692 | −2.671 [15] | 8, 4 | +2, +6 |
| $Ca_3SiO_5$ | −3.007 | −3.403 | −3.388 | −3.331 | −3.327 | −3.334 | −3.373 [15] | 18, 4 | +2, +4 |
| $CaSiO_3$ (wol.) | −3.047 | −3.423 | −3.421 | −3.383 | −3.363 | −3.376 | −3.389 [15] | $6.\overline{3}$, 4 | +2, +4 |
| $CaSiO_3$ (ps-wol.) | −3.026 | −3.401 | −3.404 | −3.398 | −3.330 | −3.352 | −3.375 [15] | 8, 4 | +2, +4 |
| $CaTiO_3$ | −3.173 | −3.604 | −3.621 | −3.470 | −3.435 | −3.437 | −3.442 [15] | 8, 6 | +2, +4 |
| $Ca_2V_2O_7$ | −2.810 | −3.127 | −3.160 | −2.928 | −2.914 | −2.955 | −2.905 [15] | 13, 9 | +2, +5 |
| $CaV_2O_6$ | −2.647 | −2.926 | −2.959 | −2.708 | −2.669 | −2.702 | −2.682 [15] | 6, 10 | +2, +5 |
| $CaWO_4$ | −2.662 | −2.987 | −2.960 | −2.806 | −2.804 | −2.840 | −2.842 [17] | 8, 4 | +2, +6 |
| $CaZrO_3$ | −3.335 | −3.734 | −3.766 | −3.629 | −3.640 | −3.664 | −3.662 [15] | 6, 6 | +2, +4 |
| $CdAl_2O_4$ | −2.443 | −2.841 | −2.846 | −2.823 | −2.838 | −2.827 | −2.838 [15] | 4, 12 | +2, +3 |
| $CdGa_2O_4$ | −1.646 | −2.002 | −1.953 | −2.050 | −2.023 | −2.022 | −2.007 [15] | 4, 12 | +2, +3 |
| $CdTiO_3$ | −2.291 | −2.634 | −2.683 | −2.545 | −2.537 | −2.541 | −2.551 [17] | 6, 6 | +2, +4 |
| $CoFe_2O_4$ | −1.144 | −1.465 | −1.593 | −1.561 | −1.582 | −1.551 | −1.612 [15] | 4, 12 | +2, +3 |
| $CoTiO_3$ | −2.097 | −2.431 | −2.499 | −2.513 | −2.515 | −2.498 | −2.503 [15] | 6, 6 | +2, +4 |

TABLE A.XVIII. (*continued*)

| formula | PBE | LDA | SCAN | PBE CCE | LDA CCE | SCAN CCE | Exp. | $N_{1/2-\mathrm{O}}$ | $+\alpha$ |
|---|---|---|---|---|---|---|---|---|---|
| $CuFeO_2$ | −0.925 | −1.281 | −1.325 | −1.236 | −1.318 | −1.258 | −1.329 [15] | 2, 6 | +1, +3 |
| $FeAl_2O_4$ | −2.411 | −2.753 | −2.843 | −2.832 | −2.815 | −2.832 | −2.954 [17] | 4, 12 | +2, +3 |
| $FeMoO_4$ | −1.643 | −2.044 | −2.065 | −1.788 | −1.948 | −1.913 | −1.831 [15] | 6, 4 | +2, +6 |
| $FeTiO_3$ | −2.238 | −2.520 | −2.707 | −2.579 | −2.561 | −2.581 | −2.565 [15] | 6, 6 | +2, +4 |
| $LiAlO_2$ | −2.738 | −3.123 | −3.132 | −3.133 | −3.088 | −3.096 | −3.080 [15] | 6, 6 | +1, +3 |
| $Li_3AsO_4$ | −1.946 | −2.209 | −2.198 | −2.162 | −2.154 | −2.191 | −2.205 [17] | 12, 4 | +1, +5 |
| $Li_2SiO_3$ | −2.561 | −2.867 | −2.866 | −2.832 | −2.831 | −2.836 | −2.848 [15] | 8, 4 | +1, +4 |
| $Li_2TiO_3$ | −2.668 | −3.005 | −3.025 | −2.929 | −2.878 | −2.878 | −2.884 [15] | 12, 6 | +1, +4 |
| $Li_2ZrO_3$ | −2.768 | −3.088 | −3.112 | −3.061 | −3.012 | −3.025 | −3.044 [15] | 12, 6 | +1, +4 |
| $MgAl_2O_4$ | −2.991 | −3.411 | −3.434 | −3.388 | −3.400 | −3.411 | −3.404 [16] | 4, 12 | +2, +3 |
| $MgCr_2O_4$ | −2.231 | −2.468 | −2.629 | −2.569 | −2.547 | −2.595 | −2.632 [15] | 4, 12 | +2, +3 |
| $MgMoO_4$ | −2.375 | −2.640 | −2.633 | −2.477 | −2.416 | −2.460 | −2.419 [15] | 6, 4 | +2, +6 |
| $Mg_2SiO_4$ | −2.838 | −3.218 | −3.230 | −3.211 | −3.209 | −3.214 | −3.223 [15] | 12, 4 | +2, +4 |
| $MgSiO_3$ | −2.826 | −3.220 | −3.222 | −3.189 | −3.205 | −3.203 | −3.210 [15] | 6, 4 | +2, +4 |
| $MgTiO_3$ | −2.983 | −3.372 | −3.413 | −3.272 | −3.259 | −3.262 | −3.260 [15] | 6, 6 | +2, +4 |
| $MgTi_2O_5$ | −3.029 | −3.402 | −3.451 | −3.289 | −3.259 | −3.264 | −3.251 [16] | 6, 12 | +2, +4 |
| $MgV_2O_6$ | −2.487 | −2.796 | −2.817 | −2.566 | −2.563 | −2.573 | −2.534 [15] | 6, 10 | +2, +5 |
| $Mg_2V_2O_7$ | −2.581 | −2.878 | −2.922 | −2.721 | −2.726 | −2.761 | −2.671 [15] | 12, 8 | +2, +5 |
| $MgWO_4$ | −2.450 | −2.806 | −2.756 | −2.588 | −2.601 | −2.618 | −2.621 [15] | 6, 6 | +2, +6 |
| $MnTiO_3$ | −2.447 | −2.648 | −3.052 | −2.877 | −2.856 | −2.865 | −2.817 [15] | 6, 6 | +2, +4 |
| $NaCrO_2$ | −1.905 | −2.127 | −2.296 | −2.258 | −2.188 | −2.251 | −2.271 [15] | 6, 6 | +1, +3 |
| $Na_2CrO_4$ | −1.979 | −2.186 | −2.215 | −2.021 | −2.005 | −2.038 | −1.987 [17] | 10, 4 | +1, +6 |
| $NaFeO_2$ | −1.554 | −1.717 | −1.944 | −1.799 | −1.726 | −1.867 | −1.809 [15] | 4, 4 | +1, +3 |
| $Na_2MoO_4$ | −2.123 | −2.352 | −2.364 | −2.238 | −2.149 | −2.198 | −2.175 [17] | 12, 4 | +1, +6 |
| $Na_4SiO_4$ | −2.098 | −2.381 | −2.406 | −2.375 | −2.362 | −2.374 | −2.420 [15] | 18, 4 | +1, +4 |
| $Na_2SiO_3$ | −2.369 | −2.667 | −2.687 | −2.674 | −2.644 | −2.655 | −2.700 [15] | 10, 4 | +1, +4 |
| $Na_2Ti_3O_7$ | −2.796 | −3.128 | −3.178 | −3.016 | −2.988 | −2.993 | −3.007 [15] | 10, 17 | +1, +4 |
| $NaVO_3$ | −2.350 | −2.563 | −2.626 | −2.442 | −2.389 | −2.439 | −2.379 [17] | 6, 4 | +1, +5 |
| $Na_2WO_4$ | −2.175 | −2.399 | −2.419 | −2.319 | −2.273 | −2.322 | −2.283 [15] | 12, 4 | +1, +6 |
| $NiAl_2O_4$ | −2.289 | −2.654 | −2.672 | −2.755 | −2.730 | −2.764 | −2.844 [15] | 4, 12 | +2, +3 |
| $NiTiO_3$ | −2.077 | −2.408 | −2.406 | −2.513 | −2.479 | −2.495 | −2.492 [15] | 6, 6 | +2, +4 |
| $NiWO_4$ | −1.661 | −1.984 | −1.889 | −1.922 | −1.931 | −1.951 | −1.950 [15] | 6, 6 | +2, +6 |
| $PbMoO_4$ | −1.888 | −2.131 | −2.063 | −1.861 | −1.759 | −1.819 | −1.819 [15] | 8, 4 | +2, +6 |
| $PbTiO_3$ | −2.346 | −2.684 | −2.654 | −2.458 | −2.413 | −2.443 | −2.476 [15] | 8, 5 | +2, +4 |
| $PbWO_4$ | −1.944 | −2.184 | −2.121 | −1.950 | −1.900 | −1.957 | −1.937 [15] | 8, 4 | +2, +6 |
| $SrAl_2O_4$ | −3.077 | −3.424 | −3.494 | −3.383 | −3.399 | −3.466 | −3.442 [17] | 6, 8 | +2, +3 |
| $SrHfO_3$ | −3.415 | −3.827 | −3.844 | −3.782 | −3.760 | −3.787 | −3.697 [15] | 8, 6 | +2, +4 |
| $SrMoO_4$ | −2.624 | −2.920 | −2.906 | −2.737 | −2.667 | −2.715 | −2.676 [15] | 8, 4 | +2, +6 |
| $SrSiO_3$ | −3.040 | −3.405 | −3.414 | −3.415 | −3.355 | −3.372 | −3.386 [15] | 8, 4 | +2, +4 |
| $Sr_2TiO_4$ | −3.075 | −3.495 | −3.508 | −3.445 | −3.363 | −3.363 | −3.387 [15] | 18, 6 | +2, +4 |
| $SrTiO_3$ | −3.168 | −3.603 | −3.628 | −3.556 | −3.441 | −3.442 | −3.463 [15] | 12, 6 | +2, +4 |
| $SrWO_4$ | −2.686 | −2.981 | −2.972 | −2.833 | −2.817 | −2.861 | −2.832 [17] | 8, 4 | +2, +6 |
| $ZnFe_2O_4$ | −1.284 | −1.678 | −1.788 | −1.670 | −1.725 | −1.702 | −1.735 [17] | 4, 12 | +2, +3 |
| $Zn_2SiO_4$ | −2.057 | −2.382 | −2.376 | −2.414 | −2.417 | −2.416 | −2.433 [15] | 8, 4 | +2, +4 |
| $Zn_2TiO_4$ | −2.120 | −2.465 | −2.487 | −2.476 | −2.442 | −2.446 | −2.443 [15] | 10, 6 | +2, +4 |
| $ZnWO_4$ | −1.939 | −2.286 | −2.217 | −2.129 | −2.120 | −2.127 | −2.126 [15] | 6, 6 | +2, +6 |
| $ZrSiO_4$ | −3.156 | −3.592 | −3.600 | −3.510 | −3.517 | −3.502 | −3.496 [16] | 8, 4 | +4, +4 |

TABLE A.XIX: **Uncorrected and CCE corrected formation energies for ternary halides.** DFT formation energies (without vibrational contribution), coordination corrected values (using corrections from Table A.XVII) and experimental room temperature formation enthalpies, number of cation-anion bonds $N_{1/2-X}$ per formula unit and oxidation states $+\alpha$ of the cations for ternary halides. For the cation-oxygen bonds- and oxidation numbers, the first (second) number in the column refers to the first (second) element in the formula. Experimental values from Kubaschewski *et al.* [15] and NIST-JANAF [16]. Energies and enthalpies in eV/atom.

| formula | PBE | LDA | SCAN | PBE CCE | LDA CCE | SCAN CCE | Exp. | $N_{1/2-X}$ | $+\alpha$ |
|---|---|---|---|---|---|---|---|---|---|
| $KBF_4$ | −3.040 | −3.243 | −3.433 | −3.297 | −3.336 | −3.219 | −3.255 [15] | 10, 4 | +1, +3 |
| $Li_3AlF_6$ | −3.246 | −3.462 | −3.682 | −3.487 | −3.496 | −3.529 | −3.507 [16] | $13.\overline{3}$, 6 | +1, +3 |
| $Li_2BeF_4$ | −3.145 | −3.310 | −3.524 | −3.345 | −3.346 | −3.389 | −3.365 [15] | 8, 4 | +1, +2 |
| $Na_3AlF_6$ | −3.090 | −3.280 | −3.520 | −3.360 | −3.341 | −3.358 | −3.433 [15] | 16, 6 | +1, +3 |

TABLE A.XIX. (*continued*)

| formula | PBE | LDA | SCAN | PBE CCE | LDA CCE | SCAN CCE | Exp. | $N_{1/2-X}$ | $+\alpha$ |
|---|---|---|---|---|---|---|---|---|---|
| $NaBF_4$ | −2.969 | −3.162 | −3.360 | −3.216 | −3.275 | −3.161 | −3.187 [15] | 8, 4 | +1, +3 |
| $Na_2SiF_6$ | −3.103 | −3.333 | −3.535 | −3.399 | −3.459 | −3.346 | −3.354 [15] | 12, 6 | +1, +4 |
| $NaAlCl_4$ | −1.707 | −1.828 | −1.916 | −1.911 | −1.899 | −1.955 | −1.973 [16] | 5, 4 | +1, +3 |